\documentclass[twocolumn]{aastex701}

\newcommand{\oi}{[O\,{\scriptsize I}]}
\newcommand{\neii}{[Ne\,{\scriptsize II}]}
\newcommand{\arii}{[Ar\,{\scriptsize II}]}
\newcommand{\hi}{H\,{\scriptsize I}}
\newcommand{\feii}{[Fe\,{\scriptsize II}]}

\usepackage{caption}
\usepackage{comment}
\usepackage{xcolor}
\usepackage{longtable}

\begin{document}

\title{JWST/MIRI Reveals the Evolution from Molecular to Atomic Disk Winds}
\shorttitle{JWST/MIRI Reveals the Evolution from Molecular to Atomic Disk Winds}

\author[orcid=0000-0003-3401-1704,sname='Bajaj']{Naman S. Bajaj}
\affiliation{Lunar and Planetary Laboratory, The University of Arizona, Tucson, AZ 85721, USA}
\email[show]{namanbajaj@arizona.edu}  

\author[0000-0001-7962-1683]{Ilaria Pascucci}
\affiliation{Lunar and Planetary Laboratory, The University of Arizona, Tucson, AZ 85721, USA}
\email[]{pascucci@arizona.edu}

\author[0000-0002-1593-3693]{Sylvie Cabrit}
\affiliation{Observatoire de Paris - PSL University, Sorbonne Université, LERMA, CNRS, Paris, France}
\affiliation{Univ. Grenoble Alpes, CNRS, IPAG, Grenoble, France}
\email[]{sylvie.cabrit@obspm.fr}

\author[0000-0002-3232-665X]{Suzan Edwards}
\affiliation{Five College Astronomy Department, Smith College, Northampton, MA, USA}
\email[]{sedwards@smith.edu}

\author[0000-0001-7255-3251]{Gabriele Cugno}
\affiliation{Department of Astrophysics, University of Zürich, Winterthurerstrasse 190, 8057 Zürich, Switzerland}
\email[]{gabriele.cugno@uzh.ch}

\author[0000-0003-0330-1506]{Andrew D. Sellek}
\affiliation{Leiden Observatory, Leiden University, 2300 RA Leiden, The Netherlands}
\email[]{sellek@strw.leidenuniv.nl}

\author[0000-0002-5758-150X]{Joan R. Najita}
\affiliation{NSF NOIRLab, 950 N. Cherry Avenue, Tucson, AZ  85719, USA}
\email[]{joan.najita@noirlab.edu}

\author[0000-0002-0661-7517]{Ke Zhang}
\affiliation{Department of Astronomy, University of Wisconsin-Madison, Madison, WI 53706, USA}
\email[]{ke.zhang@wisc.edu}

\author[0000-0001-6410-2899]{Richard Alexander}
\affiliation{School of Physics and Astronomy, University of Leicester, University Rd, Leicester, LE1 7RH, UK}
\email[]{richard.alexander@leicester.ac.uk}

\author[0000-0002-7154-6065]{Gregory J. Herczeg}
\affiliation{Kavli Institute for Astronomy and Astrophysics, Peking University, Beijing 100871, People’s Republic of China}
\affiliation{Department of Astronomy, Peking University, Beijing 100871, People’s Republic of China}
\email[]{gherczeg1@gmail.com}

\author[0000-0002-3311-5918]{Uma Gorti}
\affiliation{Carl Sagan Center, SETI Institute, Mountain View, CA, USA}
\affiliation{Ames Research Center, NASA, Moffett Field, CA, USA}
\email[]{ugorti@seti.org}

\author[0009-0007-0600-6461]{Sophie C. Clark}
\affiliation{Lunar and Planetary Laboratory, The University of Arizona, Tucson, AZ 85721, USA}
\email[]{sophieclark@arizona.edu}

\author[0000-0002-6881-0574]{Tracy L. Beck}
\affiliation{The Space Telescope Science Institute, 3700 San Martin Drive, Baltimore, MD 21218, USA}
\email[]{tbeck@stsci.edu}

\begin{abstract}

The evolution and dispersal of protoplanetary disks--governed by accretion, magnetically launched jets and winds, and photoevaporative winds--fundamentally shape planetary systems. Determining how these mass-loss processes co-evolve is crucial for constraining planet formation pathways. We analyze archival JWST/MIRI/IFU data of 72 inclined (\textit{i}~$>$~40$^{\circ}$) mostly Class~II disks to identify and characterize spatially resolved jets and winds, focusing on \neii{} and H$_2$ lines. Extended emission in H$_2$ S(1), S(3), S(5), S(7) and/or \neii{} is detected toward 66 disks, revealing diverse morphologies. We develop a framework to identify conical H$_2$ winds and high-velocity \neii{} jets perpendicular to the disk, detecting them toward 46 and 40 disks, respectively. All sources with \neii{} jets exhibit a corresponding wind traced in either H$_2$ (85\%) or \oi{}, establishing a connection between jets and winds. The detection fractions of \neii{}-jets and H$_2$-winds correlate positively with mass accretion rate, with no dependence on disk inclination or stellar mass. Conversely, marginally resolved low-velocity \neii{} winds are found preferentially toward lower accretors. Among sources with H$_2$ winds, detection of hotter winds traced by S(7) and S(5) declines more rapidly with decreasing accretion rate than the colder S(1) component. Comparison with high-resolution \oi{}~6300\text{\AA} spectroscopy reveals \oi{} LVC and extended H$_2$ wind detections preferentially toward moderate-to-high accretors ($\gtrsim~10^{-8.5}~\dot{M}_{\odot}~yr^{-1}$), whereas lower accretors exhibit only \oi{} and \neii{} winds. Together, these results indicate that atomic jets and atomic+molecular winds, consistent with an MHD disk-wind origin, dominate during early, actively accreting disk phases, while at lower accretion rates, jets weaken and winds become predominantly atomic.

\end{abstract}


\section{Introduction} 
\label{sec:intro}

Protoplanetary disks are a natural outcome of star formation and provide the reservoir of material from which planets form \citep[see][for a review]{Pineda2023}. Their evolution is governed by the redistribution of mass and angular momentum, which regulates both stellar accretion and the surface density structure that sets the initial conditions for planet formation \citep[see][for reviews]{Williams2011,Andrews2020,Manara2023}. While classic viscous evolving models invoke turbulence, such as that driven by the magnetorotational instability \citep[MRI,][]{Balbus1991}, its efficiency may be limited by low ionization levels across much of the planet-forming region \citep[e.g.,][]{Gammie1996,Dzyurkevich2013,Okuzumi2015}. In this context, magnetocentrifugal (MHD) disk winds have emerged as a compelling alternative, capable of driving accretion by removing angular momentum along large-scale magnetic field lines \citep[][and \cite{Pascucci2023} for the latest review]{Blandford1982,Bai2013,Gressel2015}. 

At later stages, disk dispersal is thought to be controlled by photoevaporative (PE) winds, in which high-energy stellar radiation heats and unbinds the gas, leading to rapid clearing once accretion rates fall below the PE wind mass-loss rate \citep[][]{Alexander2014,Pascucci2020,Sellek2024b}. Together, MHD and PE winds likely regulate both the evolution and dispersal of protoplanetary disks, and are therefore central to understanding the timeline and outcome of planet formation.

Forbidden atomic lines provide a powerful diagnostic of outflowing gas launched from the disk. The \oi{} $\lambda$6300 line is commonly detected in Class~II T-Tauri stars \citep{Natta2014,McGinnis2018,Nisini2018} and can be decomposed into low-velocity ($<$30~km~s$^{-1}$) and high-velocity ($>$30~km~s$^{-1}$) components \citep[e.g.,][]{Simon2016}, with the former proposed as a tracer of MHD disk winds \citep[see][for an alternative explanation for the narrow low-velocity component]{Weber2020} and the latter tracing collimated jets \citep[e.g.,][]{Fang2018,Banzatti2019,Pascucci2020,Fang2023}. Similarly, high-resolution spectroscopy found \neii{}~12.81~\micron{} to have two components \citep{Pascucci2020}. The low-velocity is consistent with PE winds while the high-velocity component traces jets \citep[e.g.,][]{Pascucci2009,vanBoekel2009,Alexander2014,Pascucci2020}. \cite{Gangi2020} examined the H$_2$ 2.12~\micron{} line in several T~Tauri disks utilizing high-resolution spectroscopy and identified blueshifted low-velocity components ($<$-20~km~s$^{-1}$) toward 7 of 17 targets with H$_2$ detections, tracing molecular disk winds. \cite{Beck2019} investigated ground-based IFU (Integral Field Unit) data of 10 disks in the same H$_2$ line and revealed spatially resolved wide-angle morphologies toward 5 disks, suggestive of disk winds. 

\cite{Pascucci2020}, in a study of 31 Class~II disks, analyzed both the \oi{}~$\lambda$6300 and \neii{}~12.81~\micron{} lines using high-resolution ($\Delta$v $\leq$ 10~km~s$^{-1}$) spectroscopy. They found \neii{} tracing either high-velocity jets or low-velocity winds (but never together) with the jets detected preferentially towards higher accretors ($\dot{M}_acc$ $>$ 10$^{-8}$~M$_{\odot}$~yr$^{-1}$). Importantly, they found that as the inner dust disk depletes (higher n$_{13-31}$, associated with sources with overall lower accretion rates), the \neii{} luminosity increases while the \oi{} weakens, which led them to propose the existence of mostly molecular inner MHD disk winds that are dense enough to screen hard X-rays in the higher accretors. As the disks evolve, these inner molecular winds weaken, allowing increasingly more X-ray photons to penetrate and ionize neon further out and drive a PE wind. In this work, we aim to test this hypothesis by detecting and characterizing spatially resolved molecular winds along with jets towards a large number of Class~II disks.

Recent observations with the James Webb Space Telescope (JWST) Mid-Infrared Instrument (MIRI) Medium-resolution spectroscopy (MRS) IFU have demonstrated the unprecedented capability  of this instrument to spatially resolve jets and winds simultaneously in multiple atomic and molecular tracers \citep[e.g.,][]{Arulanantham2024,Bajaj2024,Delabrosse2024,Tychoniec2024,Schwarz2025,Francis2026}. For inclined disks, spatially resolved H$_2$ emission often exhibits a conical morphology with an axis perpendicular to the disk position angle (PA). Recently, \citet{Narang2026b}, reported 10 such H$_2$ cone-like morphologies towards moderate-to-high inclination disks (\textit{i} $\gtrsim$ 50$^{\circ}$). This morphology is characteristic of a slow molecular wind launched from the disk \citep[e.g.,][]{Federman2024,Nisini2024b,Narang2025,Navarro2025,Francis2026,Narang2026b,Narang2026a}. In one particular study, with JWST NIRSpec, similar conical H$_2$ flows detected towards 3 Class~II and 1 Class~I edge-on disks were shown to be most likely MHD-disk-winds based on their launch radius and their nested morphology with the \feii{} jets \citep{Pascucci2025}. 

At the same time, \neii{} 12.81~\micron{} is among the most commonly detected forbidden atomic lines in mid-IR spectra of protostellar disks \citep[e.g.,][]{Rigliaco2015,Arulanantham2025}. For the handful of objects observed (and published) with JWST, the \neii{} emission is observed either as collimated high-velocity jets \citep[e.g.,][]{Arulanantham2024,Tychoniec2024,Schwarz2025} or as modestly extended low-velocity emission consistent with winds \citep{Bajaj2024}. This dichotomy is consistent with high-resolution spectroscopic studies of Class~II sources described earlier, which detect either a low-velocity or a high-velocity component in \neii{}, but not both simultaneously, suggesting that it traces either jets or winds depending on the stage of disk evolution \citep{Pascucci2020}.

Here we analyze a large sample of 72 inclined disks (\textit{i} $>$ 40$^{\circ}$), approximately four times larger than that analyzed by \cite{Narang2026b} for inclined disks, and present a robust framework to identify jets and winds. Details of the sample selection, calibration, and continuum-subtraction procedures are provided in Section \ref{sec:sample_details}. We outline our strategy for identifying extended emission, classifying H$_2$ emission as wind-like or non-wind-like, and \neii{} emission as jet-like or non-jet-like in Section \ref{sec:identification_classification}. A tabular summary of line detections, spatial extent, and outflow classifications for each source is presented in Figure \ref{fig:det_ext_out_stat}. Next, we briefly discuss any intrinsically asymmetric flows before exploring correlations of sources exhibiting H$_2$ and \neii{} outflows with stellar and disk properties, as well as with previous studies on outflows traced in \oi{}~6300\AA{} in Section \ref{sec:results}. Finally, we discuss the mechanism likely responsible for the observed conical H$_2$ wind emission, the empirical evolutionary sequence of outflows, and how the results of this study connect with those from Class~0/I outflows in Section \ref{sec:Discussion} before summarizing our conclusions in Section \ref{sec:conclusion}. Line and continuum maps for all sources are provided in Section \ref{sec:line_cont_maps_all}, with brief notes on selected interesting systems in Section \ref{sec:individual_sources}.

\section{Sample details, calibration, and continuum subtraction} \label{sec:sample_details}

\subsection{Sample Details}

\begin{figure}
    \centering
    \includegraphics[width=\linewidth]{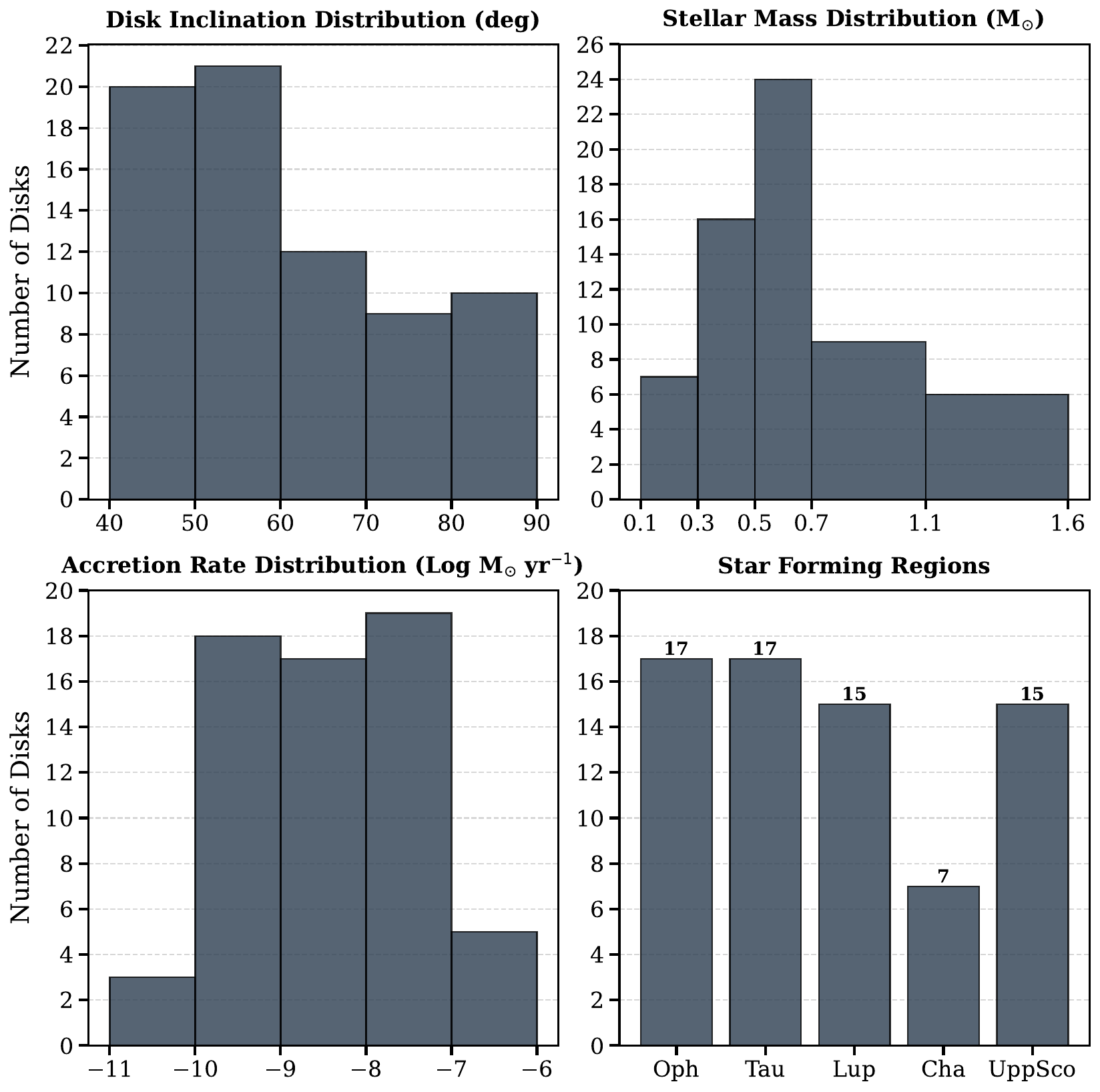}
    \caption{Histograms showing distribution of various properties for our sample of inclined (40$^{\circ}$-- 90$^{\circ}$) disks. 5 of the 10 highly inclined disks ($\geq$80$^{\circ}$) lack information on accretion rate.}
    \label{fig:statistics}
\end{figure}

The goal of this study is to test the evolution of inner winds from molecular to atomic as proposed by \cite{Pascucci2020} based on a sample of Class~II disks. To this end, we compiled a list of all ($>$100) T Tauri stars \citep[][]{Manara2023} in the well-studied star-forming regions of Taurus, Chamaeleon, Upper Sco, Ophiuchus, and Lupus. These targets were observed in cycles 1, 2, and 3, and the data for all targets are publicly available. The observations were carried out as part of the following programs: GTO~1282 \citep{Henning2017}, ERS~1309 \citep{McClure2017}, GO~1584 \citep{Salyk2021}, 1640 \citep{Banzatti2021}, 1751 \citep{McClure2021}, 2260 \citep{Pascucci2021}, 2970 \citep{Pascucci2023}, 3034 \citep{Zhang2023}, and 4201 \citep{vanthoff2023}.

These targets were observed with comparably deep exposures, optimized to detect faint molecular emission at the level of only a few~\% of the $\sim$15~$\mu$m continuum, thereby ensuring relatively uniform sensitivity. Because disk emission is likely to contaminate outflow emission at low disk inclinations, i.e., it is hard to spatially separate the two components, we restrict our analysis to disks with inclinations $>40^{\circ}$ (Figure~\ref{fig:statistics}), as determined from resolved ALMA continuum observations \citep[e.g.,][]{Long2019,Carpenter2025,Deng2025}. This selection yields a final sample of 72 inclined disks, of which nine are part of multiple star systems with companions within the MIRI FOV, and ten of them have inclinations between 80 and 90$^{\circ}$, with accretion rates unknown for many of the latter (see Tables \ref{tab:source_properties} and \ref{tab:source_properties_references}). The resulting sample contains 67 Class~II disks and 5 disks classified as Class~I in the literature (however, 4 have inclinations $>$ 70$^o$, making the classification less certain). Our sample has a stellar mass distribution that peaks at $\sim$0.5-0.6 M$_{\odot}$ with minimum and maximum stellar masses of 0.1 and 1.6~M$_{\odot}$, respectively (see Figure \ref{fig:statistics}). In addition, the disks are approximately evenly distributed on either side of $\dot{M}_{\mathrm{acc}} \sim 3 \times 10^{-9}$~M$_{\odot}$~yr$^{-1}$, providing a balanced sample of high- and low-accretion-rate systems, which serve as proxies for disks at earlier and later evolutionary stages \citep[e.g.,][]{Fang2023}. While a comparable number of targets are drawn from most star-forming regions, we note that the current MIRI MRS observations, and consequently our sample, are significantly underrepresented in the Chamaeleon~I star-forming region. Finally, all disks were observed with 4 dithers without any mosaicing, except IRAS04302. Further details are available in the publications describing the respective JWST programs \citep[e.g.,][]{Henning2024,Arulanantham2025,Raul2026,Xie2026}.

\begin{longtable}{p{1.6cm}lp{1.3cm}p{0.7cm}p{0.7cm}p{0.8cm}p{0.9cm}p{0.9cm}p{0.7cm}p{0.7cm}p{0.7cm}p{0.7cm}p{1.7cm}}
\caption{Source properties collected from literature} \\
\hline
Source &
2MASS Name &
Region &
Dist (pc) &
SpT &
M$_{star}$ (solar mass) &
log $\dot M_{\rm acc}$ &
Rad-Vel (Km/s) (Helio) &
Disk incl (Deg) &
Incl Err (Deg)$^a$ &
PA (Deg) &
PA Err (Deg) &
{[}O I{]} high-res data\\ 
\hline
\endfirsthead

\multicolumn{13}{c}
{{\bfseries Table \thetable\ continued}} \\ 
\hline
Source &
2MASS Name &
Region &
Dist (pc) &
SpT &
M$_{star}$ (solar mass) &
log $\dot M_{\rm acc}$ &
Rad-Vel (Km/s) (Helio) &
Disk incl (Deg) &
Incl Err (Deg)$^a$ &
PA (Deg) &
PA Err (Deg) &
{[}O I{]} high-res data\\ 
\hline
\endhead

ESOHA569 & J11111083-7641574 & ChamI & 190.0 & M2.5 & 0.35 &  & -11.4 & 83.0 & 4.8 & 144.0 & 1.0 & \\
$**$HH48NE & J11042275-7718080 & ChamI &  & K7 & 1.2 &  & -11.4 & 82.3 & 1.0 & 75.0 & 1.0 & \\
J11085090-7625135 & J11085090-7625135 & ChamI & 192.0 & M5.25 & 0.1 & -10.27 & 14.7 & 45.0 & 35.0 &  &  & \\
SYCha & J10563044-7711393 & ChamI & 181.0 & K7 & 0.67 & -9.18 & 15.69 & 51.1 & 0.5 & -12.65 & 1.9 & \\
TCha & J11571348-7921313 & ChamI & 103.0 & G8 & 1.5 & -8.4 & 15.8 & 73.0 & 5.0 & 113.0 &  & LVC BC\\
$**$VWCha & J11080148-7742288 & ChamI & 191.0 & K7 & 0.7 & -7.38 & 14.4 & 44.0 & 17.0 &  &  & HVC + LVC BC\\
$**$WXCha & J11095873-7737088 & ChamI & 190.0 & M0.5 & 0.49 & -6.73 & -11.4 & 87.0 & 31 &  &  & \\
GQLup & J15491210-3539051 & Lupus & 154.0 & K5 & 0.61 & -7.44 & -3.13 & 60.5 & 0.5 & -11.6 & 0.5 & LVC BC\\
HKLup & J16082249-3904464 & Lupus & 156.0 & K7 & 0.55 & -7.44 & -1.18 & 47.1 & 0.7 & 111.6 & 0.1 & HVC + LVC NC\\
$**$HTLup & J15451286-3417305 & Lupus & 158.0 & K2 & 1.32 & -8.12 & -2.15 & 48.1 & 4.5 & 166.1 & 6.0 & \\
IMLup & J15560921-3756057 & Lupus & 156.0 & M0 & 0.72 & -7.85 & -1.82 & 47.5 & 0.3 & 144.5 & 0.5 & LVC BC\\
J16085324-3914401 & J16085324-3914401 & Lupus & 161.5 & M3 & 0.29 & -10.0 & 0.9 & 52.8 & 1.0 & 256.47 & 1.0 & \\
J16124373-3815031 & J16124373-3815031 & Lupus & 160.0 & M1 & 0.47 & -9.08 & -2.3 & 51.9 & 0.7 & 16.3 & 1.0 & \\
MYLup & J16004452-4155310 & Lupus & 158.0 & K0 & 1.2 & -8.0 & 4.4 & 73.2 & 0.1 & 58.8 & 0.1 & \\
RYLup & J15592838-4021513 & Lupus & 158.0 &  & 1.27 & -8.05 & 0.8 & 67.0 & 5.0 & 109.0 & 5.0 & LVC NC\\
Sz65 & J15392776-3446171 & Lupus & 154.0 & M7 & 0.61 & -9.48 & -3.45 & 61.6 & 1.0 & 289.58 & 1.0 & LVC BC\\
Sz66 & J15392828-3446180 & Lupus & 156.0 & M2 & 0.29 & -8.51 & 2.4 & 52.8 & 1.0 & 256.47 & 1.0 & \\
Sz77 & J15514695-3556440 & Lupus & 155.0 & M5.5 & 0.67 & -8.69 & 2.4 & 47.4 & 5.3 & 110.8 & 5.5 & \\
Sz95 & J16075230-3858059 & Lupus & 160.0 & M3 & 0.29 & -9.38 & -2.8 & 63.3 & 2.9 & 21.1 & 3.0 & \\
Sz131 & J16004943-4130038 & Lupus & 161.0 & M3 & 0.3 & -9.17 & 2.4 & 42.9 & 1.0 & 342.14 & 1.0 & \\
V1094SCO & J16083617-3923024 & Lupus & 158.0 & K6 & 0.64 & -7.88 & 2.2 & 49.9 & 1.0 & 106.1 & 1.0 & \\
BBRCG39 & J16271838-2439146 & $\rho$ Oph &  & M0 & 0.48 &  & -6.85 & 70.5 & 0.2 & 53.2 & 0.3 & \\
DoAr25 & J16262367-2443138 & $\rho$ Oph & 139.0 & K5 & 0.62 & -8.94 & -6.51 & 67.4 & 0.2 & 110.6 & 0.2 & \\
DoAr33 & J16273901-2358187 & $\rho$ Oph & 141.0 & K4 & 0.69 & -9.6 & -6.55 & 41.8 & 0.8 & 81.1 & 1.2 & \\
Elia2-32 & J16272844-2427210 & $\rho$ Oph & 139.0 & M6.5 & 0.55 & -7.98 & -4.85 & 53.2 & 7.5 & 90.3 & 15.0 & \\
Elias2-20 & J16261886-2428196 & $\rho$ Oph & 137.0 & M0 & 0.59 & -6.71 & -3.26 & 49.0 & 1.0 & 153.2 & 1.3 & \\
Elias2-27 & J16264502-2423077 & $\rho$ Oph & 110.0 & M0 & 0.56 & -7.38 & -7.67 & 56.2 & 0.8 & 118.8 & 0.7 & \\
GY92-21 & J16262357-2424394 & $\rho$ Oph & 139.4 & K7 & 0.69 & -8.5 & -7.9 & 73.8 & 0.2 & 48.2 & 0.3 & \\
GY92-312 & J16273894-2440206 & $\rho$ Oph & 139.0 & M2.5 & 0.39 & -8.95 & -6.85 & 68.1 & 0.4 & 168.4 & 0.6 & \\
GY92-33 & J16262753-2441535 & $\rho$ Oph &  & M2 & 0.36 & -8.71 & -3.67 & 72.0 & 3.3 & 155.2 & 3.0 & \\
HD163296 & J17562128-2157218 & $\rho$ Oph & 104.0 & A0 & 2.3 & -6.34 & -4.0 & 46.7 & 0.1 & 133.3 & 0.15 & \\
IRAS16285-2355 & J16313565-2401294 & $\rho$ Oph &  & K6 & 0.65 & -7.3 & -6.85 & 46.2 & 0.4 & 149.3 & 1.7 & \\
J16230544-2302566 & J16230544-2302566 & $\rho$ Oph &  & M1 & 0.4 &  & -6.85 & 66.1 & 2.3 & 73.2 & 2.1 & \\
J16270359-2420054 & J16270359-2420054 & $\rho$ Oph &  & M1.5 & 0.36 & -9.55 & -6.85 & 74.5 & 3.0 & 48.7 & 0.4 & \\
J16281370-2431391 & J16281370-2431391 & $\rho$ Oph &  &  & 0.57 &  & -6.85 & 87.0 & 1.0 & 93.6 & 0.4 & \\
WaOph6 & J16484562-1416359 & $\rho$ Oph &  & K & 0.9 & -6.6 & -7.6 & 47.3 & 0.7 & 174.2 & 0.8 & HVC + LVC NC\\
WL3 & J16271921-2428438 & $\rho$ Oph &  & K7 & 0.64 &  & -6.9 & 69.0 & 3.8 & 94.1 & 3.5 & \\
WSB52 & J16273942-2439155 & $\rho$ Oph & 135.0 & M1 & 0.55 & -7.89 & -5.5 & 54.4 & 0.3 & 138.4 & 0.3 & \\
AATau & J04345542+2428531 & Taurus & 140.0 & M0.6 & 0.49 & -7.64 & 15.4 & 59.1 & 0.3 & 93.0 & 1.2 & HVC + LVC BC + NC\\
CITau & J04335200+2250301 & Taurus & 160.0 & K5.5 & 0.65 & -7.28 & 17.06 & 50.0 & 0.3 & 11.2 & 0.4 & HVC + LVC NC\\
CXTau & J04144786+2648110 & Taurus & 127.0 & M2.5 & 0.33 & -9.15 & 18.75 & 55.1 & 1.0 & 66.2 & 1.4 & LVC NC\\
$**$DFTau & J04270280+2542223 & Taurus &  & M2.7 & 0.32 & -7.55 & 14.8 & 41.0 & 13.0 & 40.0 & 20.0 & HVC + LVC BC + NC\\
DLTau & J04333906+2520382 & Taurus & 160.0 & K5.5 & 0.66 & -7.19 & 16.56 & 45.0 & 0.2 & 52.1 & 0.4 & HVC + LVC BC\\
FTTau & J04233919+2456141 & Taurus & 130.0 & M2.8 & 0.3 & -7.52 & 16.65 & 40.5 & 0.5 & 121.8 & 0.7 & HVC + LVC NC\\
GKTau & J04333456+2421058 & Taurus & 129.0 & K6.5 & 0.58 & -8.31 & 20.96 & 40.2 & 6.2 & 119.9 & 9.1 & HVC + LVC BC + NC\\
GOTau & J04430309+2520187 & Taurus & 142.0 & M2.3 & 0.34 & -9.52 & 17.1 & 53.9 & 0.5 & 20.9 & 0.6 & LVC BC\\
HH30 & IRAS 04368+2557 & Taurus &  &  & 0.45 &  & 18.0 & $>$85.0 & & 1.5 & 0.2 & \\
$**$HKTauA & J04315056+2424180 & Taurus & 131.0 & M1.5 & 0.4 & -8.7 & 16.22 & 56.9 & 0.5 & 174.9 & 0.5 & \\
$**$HVTauC & J04383528+2610386 & Taurus &  & K6 & 1.33 &  & 18.0 & $>$80.0 &  & 108.0 & 1.0 & \\
IQTau & J04295156+2606448 & Taurus & 140.0 & M1.1 & 0.42 & -8.54 & 17.5 & 62.1 & 0.5 & 42.4 & 0.6 & HVC + LVC BC + NC\\
IRAS-04385 & J04413882+2556267 & Taurus & 140.0 & M0 & 0.64 & -8.11 & 16.97 & 52.0 & 5.0 & 142.0 & 1.0 & \\
IRAS04302 & J04331650+2253204 & Taurus &  &  & 1.5 & -7 & 18.0 & $>$84.0 &  & 175.0 & 1.0 & \\
J04381486+ 2611399 & J04381486+2611399 & Taurus & 140.0 & M7.25 & 0.045 & -10.8 & 21.24 & 70.0 &  &  &  & \\
LkCa15 & J04391779+2221034 & Taurus & 157.0 & K5.5 & 0.7 & -7.94 & 19.43 & 50.7 & 0.1 & 61.7 & 0.1 & LVC BC\\
$**$RWAurA & J05074953+3024050 & Taurus & 140.0 & K0 & 1.48 & -7.07 & 27.7 & 55.1 & 0.5 & 41.1 & 0.6 & HVC\\
Tau042021 & J04202144+2813491 & Taurus &  &  & 0.4 & -7.1 & 18.0 & $>$85.0 &  & -16.0 & 1.0 & \\
PDS70 & J14081015-4123525 & UpperCent & 112.0 & K7 & 0.82 & -10.0 & 0.73 & 51.0 & 1.1 & 158.2 & 2.0 & \\
J15582981-2310077 & J15582981-2310077 & UpperSco & 141.0 & M3 & 0.144 & -9.1 & -6.0 & 80.5 & 32.7 & 53.2 & 71.0 & LVC BC + NC\\
J16035767-2031055 & J16035767-2031055 & UpperSco & 143.0 & K5.1 & 0.8 & -8.98 & -6.4 & 74.6 & 13.7 & 6.4 & 13.8 & LVC SC\\
J16054540-2023088 & J16054540-2023088 & UpperSco & 139.0 & M2 & 0.15 & -9.36 & -6.43 & 51.0 & 1.0 & 63.0 & 1.0 & \\
J16064385-1908056 & J16064385-1908056 & UpperSco & 145.0 & K8 & 0.78 & -9.54 & -5.7 & 66.4 & 16.0 & 71.3 & 41.0 & Not detected\\
J16064794-1841437 & J16064794-1841437 & UpperSco & 153.0 & K9 & 0.56 & -9.42 & -3.3 & 55.5 & 0.1 & 20.0 & 1.0 & Not detected\\
J16075796-2040087 & J16075796-2040087 & UpperSco &  & K4 & 0.71 & -8.96 & 0.6 & 65.8 & 15.0 & 1.4 & 7.7 & HVC + LVC BC + NC\\
J16082324-1930009 & J16082324-1930009 & UpperSco & 138.0 & K9 & 0.65 & -9.14 & -6.19 & 71.0 & 1.0 & 123.0 & 1.0 & \\
J16090075-1908526 & J16090075-1908526 & UpperSco & 137.0 & M1 & 0.65 & -8.81 & -6.7 & 50.3 & 1.0 & 155.3 & 1.0 & LVC SC\\
J16111534-1757214 & J16111534-1757214 & UpperSco & 135.0 & M1.2 & 0.4 & -9.7 & -8.2 & 41.4 &  &  &  & LVC SC\\
J16111742-1918285 & J16111742-1918285 & UpperSco & 137.0 & M0.25 & 0.5 &  & -6.7 & 80.6 & 1.0 & 70.9 & 1.0 & \\
J16123916-1859284 & J16123916-1859284 & UpperSco & 135.0 & M2 & 0.52 & -9.38 & -4.8 & 51.0 & 36.0 & 46.0 & 27.0 & LVC SC\\
$**$J16153456-2242421 & J16153456-2242421 & UpperSco &  & M0.2 & 0.48 & -8.68 & -3.2 & 49.4 & 19.2 & 161.7 & 24.0 & LVC SC\\
J16163345-2521505 & J16163345-2521505 & UpperSco & 158.0 & M0.1 & 0.56 & -10.91 & -2.8 & 62.0 & 1.0 & 62.0 & 1.0 & Not detected\\
J16202863-2442087 & J16202863-2442087 & UpperSco & 153.0 & M2 & 0.8 &  & -5.52 & 44.4 & 4.8 & 179.1 & 4.0 & \\
J16221532-2511349 & J16221532-2511349 & UpperSco & 139.0 & M3 & 0.29 &  & -16.17 & 52.0 & 1.0 & 20.0 & 1.0 & \\ \hline
\multicolumn{13}{p{18.5cm}}{NOTE: The table is available in a digital format in the published version of this manuscript. Sources whose names are prefixed with `$**$’ denote multiple systems with companion(s) within the same FOV. Each of these nine sources is discussed further in Section \ref{sec:individual_sources}.}\\
\multicolumn{13}{p{18.5cm}}{$^a$ Some ALMA studies report extremely small inclination uncertainties ($\sim$0.1$^{\circ}$), but these likely underestimate the true errors because they account mainly for SNR and $uv$-coverage while neglecting systematics such as phase calibration uncertainties and limitations of the adopted disk model \citep[see the discussion in][]{Miley2024}. These effects are particularly important for compact or marginally resolved disks, for which true uncertainties are likely on the order of a few degrees.}
\label{tab:source_properties}
\end{longtable}

\begin{longtable}{p{1.4cm}lp{2cm}p{0.4cm}p{0.2cm}p{1.5cm}p{1.5cm}p{0.5cm}p{0.5cm}p{0.5cm}p{0.5cm}p{1.7cm}} 
\caption{References for Table \ref{tab:source_properties}} \\
\hline
Source &
2MASS Name &
JWST ID / PI &
SpT &
M$_{star}$ &
log $\dot M_{\rm acc}$ &
Rad-Vel &
Disk incl &
Incl Err &
PA &
PA Err &
[O I] high-res data \\
\hline
\endfirsthead

\multicolumn{12}{c}
{{\bfseries Table \thetable\ continued}} \\
\hline
Source &
2MASS Name &
JWST ID / PI &
SpT &
M$_{star}$ &
log $\dot M_{\rm acc}$ &
Rad-Vel &
Disk incl &
Incl Err &
PA &
PA Err &
[O I] high-res data \\
\hline
\endhead

\hline
\endfoot

ESOHA569 & J11111083-7641574 & 1751/McClure  & \multicolumn{2}{l}{Villenave2020} & \multicolumn{1}{l}{} & Biazzo2012 & \multicolumn{4}{c}{Wolff2017, Villenave2020} &   \\
$**$HH48NE  & J11042275-7718080   & ERS1309/ McClure & \multicolumn{2}{l}{Villenave2020} & \multicolumn{1}{l}{} & Biazzo2012 & \multicolumn{4}{c}{Villenave2020} &  \\
J11085090-7625135 & J11085090-7625135   & 1282/Henning  & \multicolumn{2}{l}{Manara2023} & Manara2023    & Gutierrez Albarran2020 & \multicolumn{4}{c}{Greenwood2017}  &  \\
SYCha & J10563044-7711393   & 1282/Henning  & \multicolumn{2}{l}{Manara2023}  & Manara2023                & Sacco2017   & \multicolumn{4}{c}{Orihara2023} &    \\
TCha & J11571348-7921313   & 2260/Pascucci  & \multicolumn{2}{l}{Cieza2011} & Cahill2019      & Pascucci2020 & \multicolumn{4}{c}{Hendler2018} & Pascucci2020 \\
\nodata & \nodata & \nodata & \multicolumn{2}{l}{\nodata} & \nodata & \nodata & \multicolumn{4}{c}{\nodata} & \nodata \\
\nodata & \nodata & \nodata & \multicolumn{2}{l}{\nodata} & \nodata & \nodata & \multicolumn{4}{c}{\nodata} & \nodata \\
\hline
\multicolumn{12}{p{18.5cm}}{NOTE: The full version of this table is available in a digital format in the published version of this manuscript.}\\
\multicolumn{12}{p{18.5cm}}{References: \cite{Hartmann1998,Bacciotti1999,Bary2003,White2004,Eisner2005,Gontcharov2006,Natta2006,Pety2006,Riaud2006,Luhman2007,Schaefer2009,Cieza2010,Dahm2010,Pontoppidan2010,Ricci2010,Cieza2011,Biazzo2012,Garufi2014,Banzatti2015,Manara2015,Rigliaco+2016,Simon2016,Aso2017,Banzatti2017,Barenfeld2017,Dutrey2017,Frasca2017,Greenwood2017,Henning2017,Loomis2017,McClure2017,vanDishoeck2017,Wolff2017,Fang2018,Hendler2018,Huang2018a,Huang2018b,Kurtovic2018,Louvet2018,Mendigutia2018,Banzatti2019,Beck2019,Cahill2019,Facchini2019,Keppler2019,Kiman2019,Long2019,Sullivan2019,Ansdell2020,Francis2020,Gangi2020,Gutierrezalbarran2020,Hendler2020,Jonsson2020,Manara2020,Pascucci2020,Sanchis2020,Thanathibodee2020,Villenave2020,Banzatti2021,McClure2021,Pascucci2021,Salyk2021,Abdurrouf2022,Gangi2022,Fang2023,Flores2023,Hourihare2023,Lin2023,Manara2023,Orihara2023,Pascucci2023,Sturm2023,vanthoff2023,Zhang2023,Duchene2024,Nisini2024a,Agurto-gangas2025,Bajaj2025,Carpenter2025,Deng2025,Gardner2025,Guerra-Alvarado2025,Kutra2025,Miley2025,Ruiz-Rodriguez2025}}
\label{tab:source_properties_references}
\end{longtable}

\subsection{Calibration} \label{sec:calibration}

All observations were calibrated locally using the standard JWST pipeline. We first downloaded the raw \texttt{uncal} files from the MAST (Barbara A. Mikulski Archive for Space Telescopes) server and processed them using version~2.0.1 of the JWST calibration pipeline \citep{Bushouse2026}, which was the most recent release available at the time of submission (July~06,~2026). We adopted CRDS version~13.1.16 together with the context \texttt{jwst\_1535.pmap} for the selection of reference files. To maintain consistency across the sample, the same pipeline version and CRDS configuration were applied to all observations.

As the first step of the calibration process, we run the \texttt{Detector1Pipeline}, using mostly default parameters. The only modification is in the jump detection step, where we set the \texttt{expand\_large\_events} and \texttt{find\_showers} options to \texttt{True}. These settings improve the identification of large cosmic-ray events on the MIRI detector by detecting showers and flagging an expanded region of affected pixels around each event.

We then run the \texttt{calwebb\_spec2} pipeline. In this step, we enable the \texttt{pixel\_replace} step and use the \texttt{mingrad} algorithm, which is optimized for MIRI MRS data, to replace NaN values in the detector images. We skip background subtraction, residual fringe correction, cube building, and 1D spectral extraction at this stage. Background subtraction is omitted for consistency, as not all targets in our sample have dedicated background observations. Cube building and spectral extraction are deferred because they are not required at this point and are computationally expensive. We also postpone residual fringe correction, as this step is particularly costly and is more effective for MIRI MRS data when applied later in the \texttt{calwebb\_spec3} pipeline.

As a final step, we run the \texttt{calwebb\_spec3} pipeline, which combines the calibrated data from multiple exposures into a single final product. Outlier detection is performed using the default kernel size and threshold. As in the previous steps, we skip background subtraction (\texttt{master\_background}) for consistency and replace any remaining NaN pixels using the \texttt{mingrad} algorithm. We construct the data cubes in the \texttt{ifualign} mode to avoid rotational interpolation in the image plane, which can otherwise suppress faint, spatially extended structures. Spectra are only used for line detection and SNR calculation in this work (Section \ref{sec:SNR_calculation}) and are extracted over the full field of view, since our analysis focuses on extended emission. During spectral extraction, residual fringe correction and adaptive trace model steps are applied. This procedure results in 12 three-dimensional data cubes and 12 one-dimensional spectra, corresponding to the three sub-bands in each of the four MIRI MRS channels.

\subsection{Continuum Subtraction}
\label{sec:cont_sub}

\begin{figure*}
    \centering`
    \includegraphics[width=\linewidth]{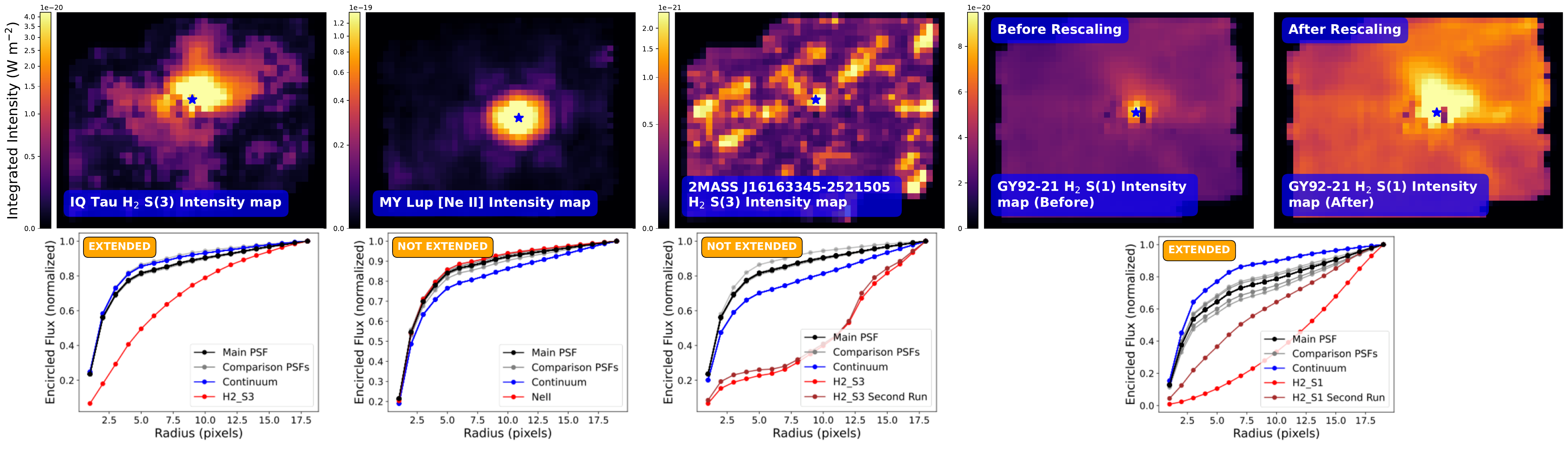}
    \caption{Examples of the methodology employed to identify spatially extended emission with respect to the MIRI PSF. On top are the spaxel-by-spaxel continuum-subtracted intensity-integrated maps of the mentioned lines, where the blue stars represent the centroids of the corresponding continuum. For GY92-21, two maps are shown, one before rescaling the emission and one after (see Section \ref{sec:extended_emission_id} for details). On the bottom are the encircled flux curves as described in the main text, with the conclusion about the emission's extent written on top left.}
    \label{fig:extended_emission_method}
\end{figure*}

We follow the approach adopted in previous studies from our group \citep[e.g.,][]{Bajaj2025,Pascucci2025} to construct spaxel-by-spaxel, continuum-subtracted, line-integrated intensity maps (e.g., the line maps shown in the top row of Figure~\ref{fig:extended_emission_method}). For each line in a spaxel, the continuum is estimated locally by considering only the spectral region within $\pm$0.2~\micron{} of the line rest wavelength. The continuum underlying the line is determined by fitting a straight line to the spectral regions [line\_center$-0.2$\micron{}, line\_center$-0.02$\micron{}] and [line\_center$+0.02$\micron{}, line\_center$+0.2$\micron{}], and subtracting this fit from the data. We find that this range provides enough points to get a good fit on the continuum while simultaneously avoiding contamination from nearby lines. We then fit a Gaussian profile to the continuum-subtracted line, using 16 different combinations of initial parameter guesses. Each initialization converges to a $\chi^2$-minimized solution, and we adopt the best-fit Gaussian with the lowest $\chi^2$ as the final fit. This process is repeated for each spaxel in the cube.

To assess the significance of the line detection in each spaxel, we compare the amplitude of the best-fit Gaussian to the standard deviation of the data points around the emission line. If the amplitude exceeds three times this standard deviation, we compute the integrated line intensity by evaluating the area under the best-fit Gaussian. Otherwise, we assign a 3$\sigma$ upper limit, where $\sigma$ is taken as the standard deviation over the full 0.4~\micron{} spectral window, including the line wavelengths. The result of this process is a 2D intensity map with each pixel representing the integrated line intensity (or upper limit) in the corresponding spaxel. We note that the effect of water contamination on both H$_2$ and \neii{} 2D line maps is insignificant since the water emission is spatially unresolved in most cases, whereas our interest lies in the broader extended structure. 

We construct pixel-by-pixel \neii{} velocity maps for the \neii{} emission by converting the wavelength axis to velocity (in km~s$^{-1}$) using the radio Doppler convention. The velocities are referenced (zero-velocity) to the rest wavelength of the \neii{} transition \citep[12.813548~\micron{}][]{NIST_ASD_2024}. For each spaxel in the cube, we then determine the velocity corresponding to the centroid of the best-fit Gaussian profile. These centroid velocities are subsequently corrected for systematic offsets and for the source radial velocity and disk inclination, as described in Section~\ref{sec:jet_classification}, before using them to identify outflows.

\section{Extended emission identification and outflow classification} \label{sec:identification_classification}

We outline our methodology to identify extended emission in Section \ref{sec:extended_emission_id}, followed by a technique to classify the extended H$_2$ emission as wind-like or not wind-like (Section \ref{sec:wind_classification}) and the \neii{} as jet-like or not jet-like (Section \ref{sec:jet_classification}). The assessment of disk winds is based on the morphology of four ortho-H$_2$ lines: $\nu$0-0 S(1), S(3), S(5), and S(7). The assessment of jets is based on both morphology and inclination-corrected velocity maps of \neii{}. We focus only on ortho-H$_2$ transitions in this work as we find them to be consistently brighter than the para-H$_2$ transitions, in agreement with ortho-para-ratio$\sim$3 in disks and outflows \citep[e.g.,][]{Neufeld1998,Rosenthal2000,Schwarz2025,vanDishoeck2025}.

\subsection{Extended emission identification}
\label{sec:extended_emission_id}

To identify extended emission, we compare the spaxel-by-spaxel continuum-subtracted and intensity-integrated line maps against a group of standard calibration stars whose emission represents the instrument's point spread function (PSF). The standard calibration stars (or PSFs) used here were observed as part of various STScI calibration programs. Of them, we picked the ones that are bright across all MIRI wavelengths and have exposure times comparable (or higher) to those of the sources analyzed in this work ($\sim$2 hours). In total, we select five observations corresponding to three stars, names and observation details of which are as follows: HD~159222 (PID: 1050, Obs: 3 and 5), delta~UMi (PID: 1524, Obs: 1 and PID: 1536, Obs: 24), and 16 Cyg~B (PID: 1538, Obs: 1).

To establish if the line emission is extended beyond the PSF, we plot the encircled flux as a function of increasing radius; that is, we sum the flux within a series of concentric circles centered on the stellar position (obtained as the centroid of the 2D Gaussian fit on the continuum at the same wavelength). The radius of successive circles differs by 1 pixel. The resulting intensity sum as a function of radius is shown in Figure \ref{fig:extended_emission_method} for four representative sources. We construct these curves for the line emission (red), the continuum near the line (blue), the main PSF (16 Cyg~B; black; selected at random)\footnote{Since all three stars have similar fluxes at $\sim$13~\micron{} ($\sim$0.5~Jy), we select the main PSF at random and confirm that changing this selection does not affect our results.}, and the comparison PSFs (HD~159222 and delta~UMi; gray). 

For each curve, we compute a $\chi^{2}$ difference relative to the main PSF, defined as $\Sigma (F_s(r) - F_{main\_psf}(r))^2$. To estimate the typical PSF-to-PSF variation, we compute the mean and standard deviation of the $\chi^{2}$ values obtained by comparing each of the comparison PSFs with the main PSF. The typical PSF variation is then estimated as the mean $\chi^{2}$ plus three times the standard deviation. If the $\chi^{2}$ value of the line-emission curve relative to the main PSF exceeds this typical PSF variation, we classify the emission as spatially extended. The same criterion is applied to the continuum emission. Figure \ref{fig:extended_emission_method} shows IQTau as an example of extended emission and MYLup as an example of compact PSF-like emission.

We also identify cases in which the encircled-flux curve is convex rather than concave (e.g., 2MASS J16163345-2521505 and GY92-21 in Figure \ref{fig:extended_emission_method}). A concave (convex) curve corresponds to a negative (positive) second derivative of the encircled flux with radius. A concave shape indicates that most of the emission is concentrated near the centroid, consistent with PSF-like (e.g., MYLup) or centrally peaked wind emission (e.g., IQTau). In contrast, a convex curve reflects relatively weak central emission and a gradual flux increase with radius, which may arise from faint extended emission (e.g., GY92-21) or, in low–signal-to-noise cases, from the cumulative integration of noise rather than a clearly resolved structure (e.g., 2MASS~J16163345-2521505). To distinguish between these two possibilities, we rescale the emission to enhance the contrast between the source emission and the background and generate an additional encircled-flux curve (shown in dark red). The rescaling is performed by linearly normalizing positive-valued pixels between the 5th and 99.5th percentiles of the pixel distribution, clipping the result to the range [0, 1], and then renormalizing by the maximum pixel value. This approach is similar to background subtraction, without requiring assumptions about the location of background emission in the maps.

If the rescaled curve remains predominantly convex, we classify the emission as compact (unresolved). Otherwise, we classify it as extended. This last step is particularly helpful for the H$_2$~S(1) line, where the background (or diffuse contamination) can be very bright, leading to misclassification. For example, GY92-21 in Figure \ref{fig:extended_emission_method} initially follows a convex curve. However, after scaling the line emission, the curve becomes concave, leading to its classification as extended, which is consistent with the visual presence of extended emission with wind-like morphology (Section \ref{sec:wind_classification}). We also notice an artefact in the H$_2$ S(5) line maps of some sources, in which the leftward-pointing wing of the MIRI PSF exhibits enhanced emission (for details, see Section \ref{sec:chan1_psf_artifact}). Although such cases are identified by this method as extended emission, we visually detect them (see Section \ref{sec:wind_classification} for an example) and classify them as compact.

Overall, we find that this approach recovers all the extended and compact line emissions identified by visual inspection. Following this, we identified spatially extended emission in the four ortho-H$_2$ lines, found in $\sim$67\% (48/72) of all targets in H$_2$~S(1),  $\sim$76\% (55/72) in H$_2$~S(3), $\sim$82\% (59/72) in H$_2$~S(5), and $\sim$46\% in H$_2$~S(7). We adopt the same approach for \neii{} and identify extended emission towards $\sim$79\% (57/72) of the targets.

\subsection{Wind Classification in Extended H$_2$ Emission}
\label{sec:wind_classification}

\begin{figure*}
    \centering
    \includegraphics[width=0.7\linewidth]{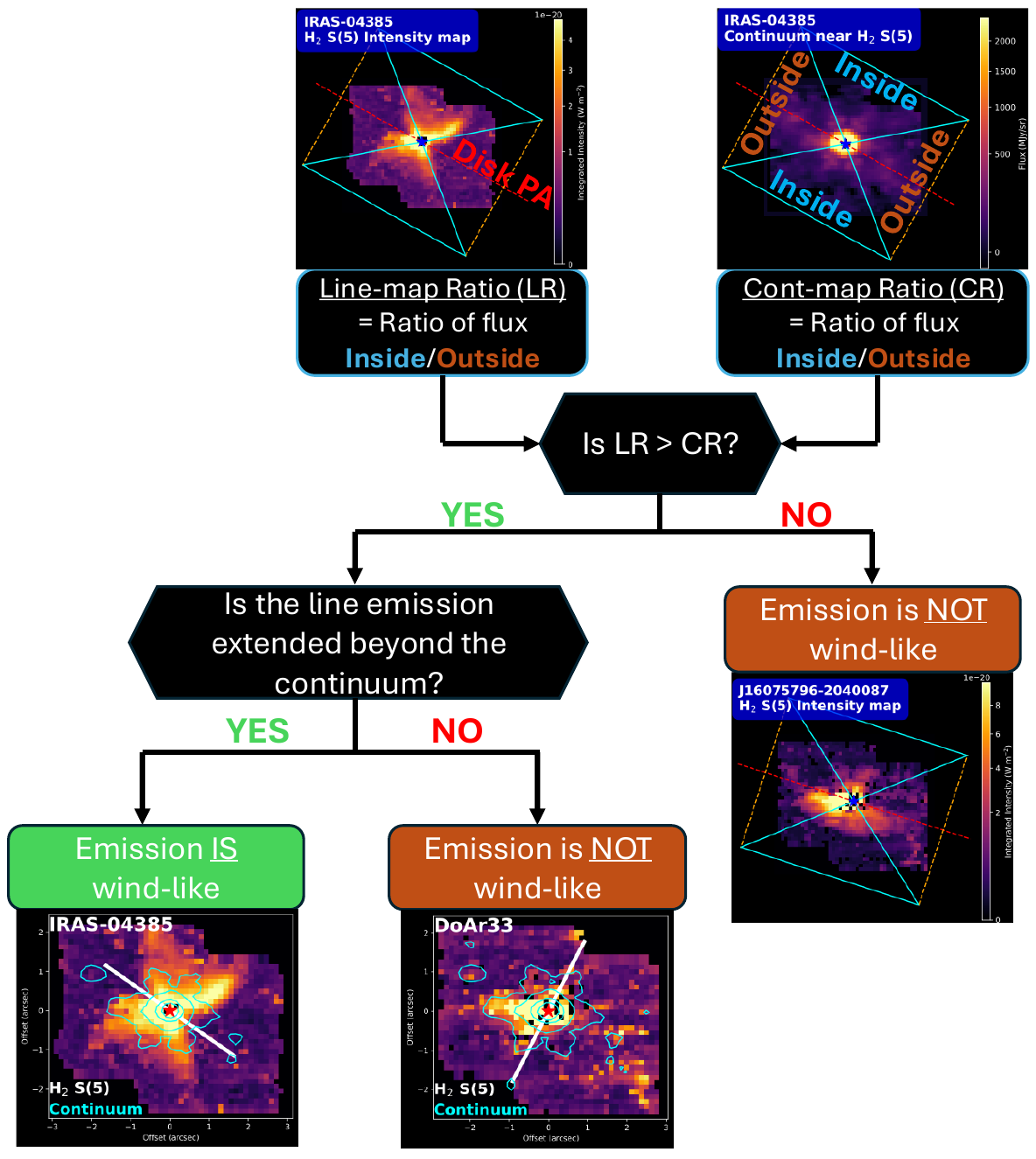}
    \caption{This flowchart outlines our strategy to identify wind-like H$_2$ emission as described in Section \ref{sec:wind_classification}. Only line maps that appear spatially resolved relative to the PSF in the flux curves (see Figure \ref{fig:extended_emission_method}) are investigated for wind emission using this flowchart. Every line map shown in the flowchart is subtracted for continuum and integrated under a 1D Gaussian in every spaxel.}
    \label{fig:wind_classification_flowchart}
\end{figure*}

\begin{figure}
    \centering
    \includegraphics[width=0.8\linewidth]{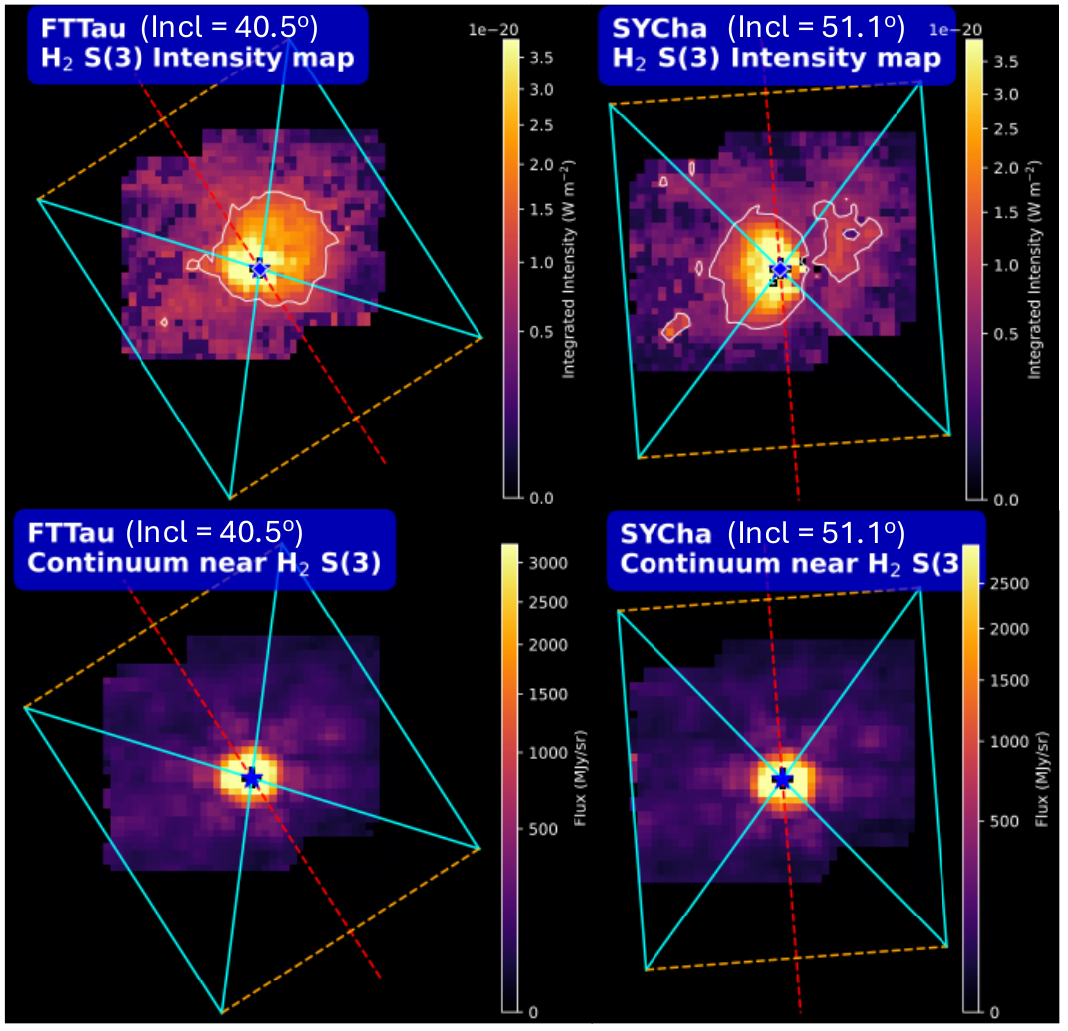}
    \caption{H$_2$ S(3) line intensity and continuum maps for FTTau (incl = 40.5$^{\circ}$) and SYCha (incl = 51.1$^{\circ}$). The blue stars show the continuum centroid, and the red dashed line shows the known disk position angle from literature ALMA results (Table \ref{tab:source_properties}).}
    \label{fig:sycha_fttau_outliers}
\end{figure}

\begin{figure*}
    \centering
    \includegraphics[width=0.8\linewidth]{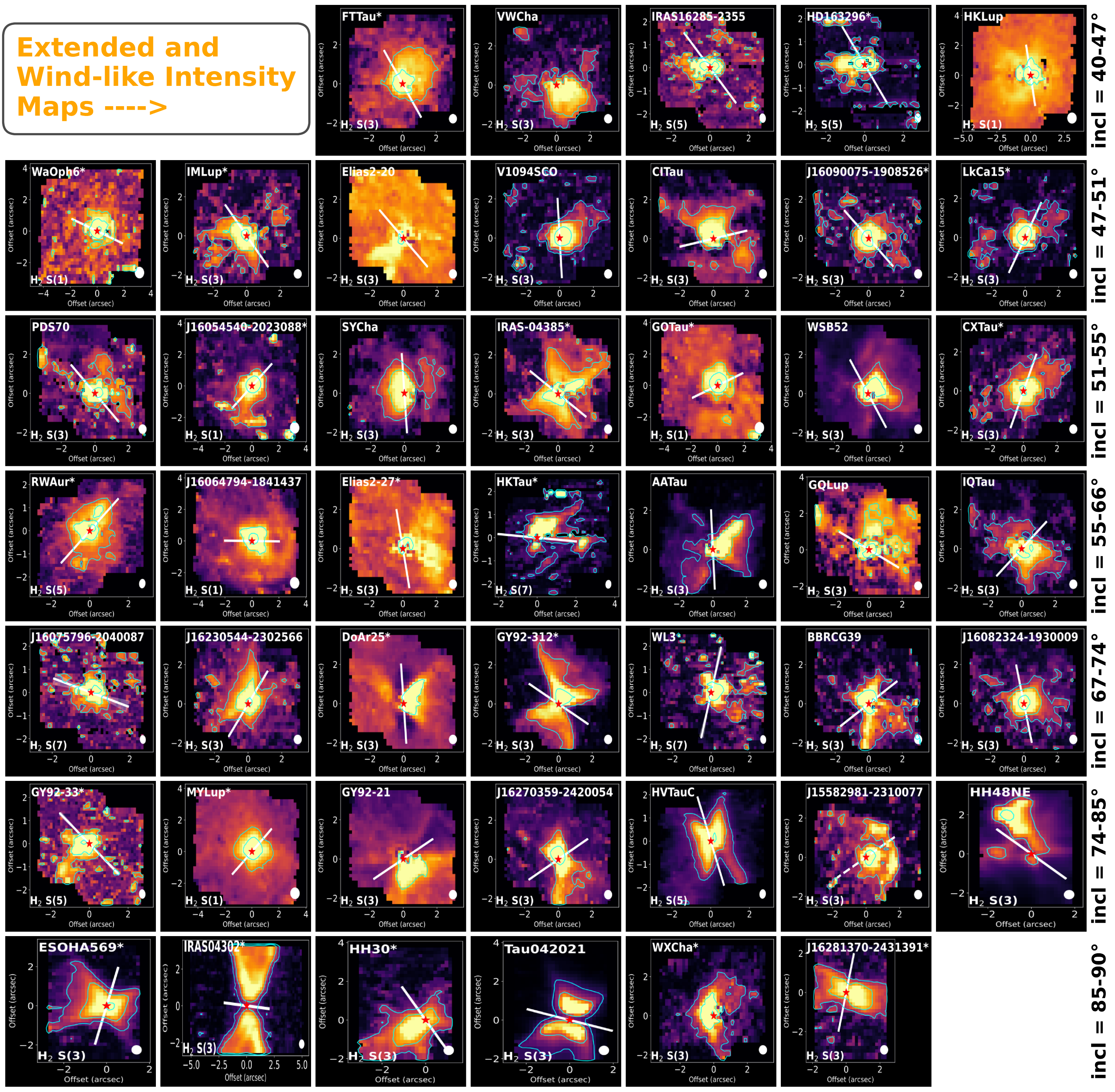}
    \caption{Intensity maps for sources with H$_2$ winds with the red star highlighting the corresponding continuum centroid location (for incl$>$80$^{\circ}$ disks, the centroid is shifted to roughly the center of the dark lane visible in continuum or line map), and white line showing the disk PA where available (Table \ref{tab:source_properties}). Primarily, S(3) maps are shown; however, when the wind is not detected in S(3), S(1), S(5), or S(7) maps are presented instead. Multiple star systems without clear H$_2$ winds are not included. highly-inclined disks are included. The maps are ordered by increasing disk inclination. Sources with asterisks (*) are further discussed in Section \ref{sec:individual_sources}. The direction of North in each map can be inferred from Figure \ref{fig:multi}.}
    \label{fig:h2_wind_maps}
\end{figure*}

The extended H$_2$ emission identified in the previous section may not all trace winds, as extended H$_2$ emission can also arise from other physical origins, such as extended disk surfaces \citep[e.g.,][]{Beck2019}, infalling material \citep[e.g.,][]{Bitner2008}, or shocked interstellar medium (ISM) filaments \citep[e.g.,][]{Neufeld2006}. We therefore next assess which sources exhibit H$_2$ S(1), S(3), S(5), and S(7) line emission consistent with a conical wind morphology.

At JWST's resolution, published observations of Class~I/II objects have revealed cone-like wind emission with the vertex located near the star (or at the centroid of the unresolved continuum), as expected. In projection, such emission appears as an isosceles triangle with the vertex near the stellar position and a vertex angle set by the wind opening angle. Typical wind semi-opening angles observed are $\lesssim$ 50$^{\circ}$ \citep{Arulanantham2024, Delabrosse2024, Nisini2024b, Tychoniec2024, Pascucci2025,Narang2026b}. Motivated by these observational findings, we developed a method to classify wind emission in H$_2$ integrated intensity maps, as described below and illustrated in Figure~\ref{fig:wind_classification_flowchart}.

For each H$_2$ emission map, we first determine the centroid of the unresolved (or slightly resolved) continuum and overplot the disk position angle derived from ALMA observations reported in the literature and collected in Table \ref{tab:source_properties}. Using this information, we define two opposing triangular regions, each with its vertex at the continuum centroid and a vertex angle of 100$^{\circ}$ (twice the largest wind semi-opening angle in \citealt{Pascucci2025}, Figure~\ref{fig:wind_classification_flowchart}). If the H$_2$ emission traces a wind-like morphology, we expect enhanced line intensity within these triangular regions relative to the regions outside. We therefore compute the ratio of the average line intensity inside the triangular regions to that outside them, which we refer to as the line-map ratio (LR). In systems with a binary companion within the field of view, the H$_2$ emission can be complex. In such cases, we classify wind emission only when a clear conical structure is present (e.g., VWCha); otherwise, we do not attempt to classify the H$_2$ emission (e.g., HTLup).

To account for the effects of the PSF shape, we compute an analogous ratio using the continuum image near the line wavelength and refer to this as the continuum-map ratio (CR). If LR~$\leq$~CR, we conclude that the line emission is not wind-like. In cases where LR~$>$~CR, we further examine whether the line emission extends beyond the continuum. This additional check is necessary because, particularly for S(5), we observe an artificial arm of enhanced emission that coincides with one of the six petal-shaped wings of the MIRI PSF, always towards the left in the image plane. Details of this artifact are provided in Section \ref{sec:chan1_psf_artifact}, and an example can be seen in Figure \ref{fig:wind_classification_flowchart} in the H$_2$ S(5) emission map of DoAr33. To identify such false-positive cases, we overlay continuum contours (3$\sigma$, 15$\sigma$, 90$\sigma$, and 300$\sigma$, where $\sigma$ is the standard deviation of the background) on the line intensity map. While this LR--CR comparison inherently assumes the continuum itself is not morphologically wind-like, this is consistent with theoretical predictions that micron-sized grains decouple from the wind at high altitudes \citep[e.g.,][]{Giacalone2019} and with the mostly PSF-like continuum maps observed toward these sources (Figure \ref{fig:multi}).

If the line emission: (a) extends beyond the continuum, (b) is unattributable to a known artifact, and (c) the ratio of emission within the 100$^{\circ}$ triangle to that outside is greater in the line map than the continuum map, then we classify the emission as wind-like. This method performs well when the continuum centroid can be determined with confidence. In disks that are close to edge-on, the continuum is generally spatially resolved, with the star located in the `dark lane’ between the two disk surfaces. This configuration causes the 2D Gaussian fit to produce an incorrect continuum centroid, so the automated method is less reliable for edge-on disks. Nevertheless, we note that conical wind morphologies in these systems are the easiest to identify visually. Hence, we use visual inspection for 10 disks that have inclinations larger than 80$^{\circ}$. We also note that in cases where infall is morphologically similar to a wind, i.e., the emission is spatially extended and dominates the region perpendicular to the disk's PA, our method is unable to distinguish between infall and winds. However, such a configuration for the infall is expected to be rare since we have selected predominantly Class~II sources that have lost the natal envelope. Finally, to account for any wind emission having a semi-opening angle greater than 50$^{\circ}$, we repeat the entire analysis adopting a semi-opening angle of 60$^{\circ}$. With this wider angle, we detect previously unidentified winds toward LkCa15, CXTau, WXCha, and J16230544-2302566 in the S(3) line, and toward HVTauC in the S(5) line. Among these, the wind-like morphology toward WXCha in the MIRI data had previously been noted by \cite{Kurtovic2026}, while winds toward HVTauC were detected earlier through ground-based IFU observations of the H$_2$ 2.12~\micron{} line \citep{Beck2008,Beck2010}.

This method successfully classifies all sources except two--SYCha (in all H$_2$ lines) and FTTau (in S(3) and S(5))--which are visually consistent with wind emission \citep[see also][who analyzed the H$_2$ wind emission from SYCha]{Schwarz2025} but are initially misclassified (false negatives) at the LR-CR comparison stage. Both sources exhibit a bowl-like wind morphology rather than the more commonly observed cone-like structure (Figure~\ref{fig:sycha_fttau_outliers}). In addition to the differing morphology, we find that the continuum centroid for these sources lies inside the `bowl' rather than at the apex (Figure~\ref{fig:sycha_fttau_outliers}), which leads to an underestimated LR and, consequently, misclassification. 

This difference is driven by inclination effects, as both FTTau (40.5$^{\circ}$) and SYCha (51.1$^{\circ}$) have relatively low disk inclinations. Such a bowl-like morphology has been observed previously toward DG\,TauA \citep{Beck2008,Agra-amboage2014}, which has a similar inclination of 45$^{\circ}$. Using 3D toy models, \cite{Agra-amboage2014} demonstrated that such a bowl-like shape can result from viewing directly down the interior of the conical cavity, i.e., when the disk inclination $\lesssim$ the wind semi-opening angle. In this geometry, the stellar position is not at the apex but slightly `inside the bowl'. Following this interpretation, we conclude that both FTTau and SYCha have H$_2$ wind cavities with true semi-opening angles $\gtrsim$40$^{\circ}$ and $\gtrsim$50$^{\circ}$, respectively. For SYCha, this is consistent with the estimates of 50-70$^{\circ}$ derived using psf-deconvolved intensity maps by \cite{Schwarz2025}. In summary, we find that 58\% (28/48) of targets with extended emission in H$_2$~S(1) show conical wind morphology, while the same morphology is seen in 58\% (32/55) sources in S(3), 51\% (30/59) in S(5), and 82\% (27/33) in S(7). Interestingly, although S(7) emission is less frequently found to be spatially extended, whenever it is detected as extended, it appears to be a reliable tracer of disk winds, in contrast to earlier reports that found no S(7) emission associated with disk winds \citep{Narang2026b}.

Approximately 35\% of sources with extended H$_2$ emission that do not trace conical winds display diverse morphologies. The most striking among these is extended emission oriented along the disk position angle, with the clearest cases being WXCha, HVTauC, and HH30 in S(1); RYLup, Sz65, and GKTau in S(1) and S(3), and J1607-2040 in S(1), S(3), and S(5) lines (Figure~\ref{fig:multi}). The other prevalent morphology is distinctive to binary systems. We observed that all binaries with spatially resolved companions in the MIRI continuum present continuous H$_2$ emission bridging the two components, suggesting gas interaction between them. The most prominent example is J1615-2242 in the S(5) line (see Figure~\ref{fig:multi}). However, in many cases, determining whether the observed bridge is genuine or stems from overlap of individual H$_2$ emissions remains uncertain, given MIRI's relatively modest spatial resolution (pixel sizes of $\sim$0.13--0.27~arcsec) relative to disk scales (disk extent of 50~au at 140~pc corresponding to 0.35~arcsec).

\subsection{Jet Classification in Extended \neii{} Emission}
\label{sec:jet_classification}

\begin{figure*}
    \centering
    \includegraphics[width=\linewidth]{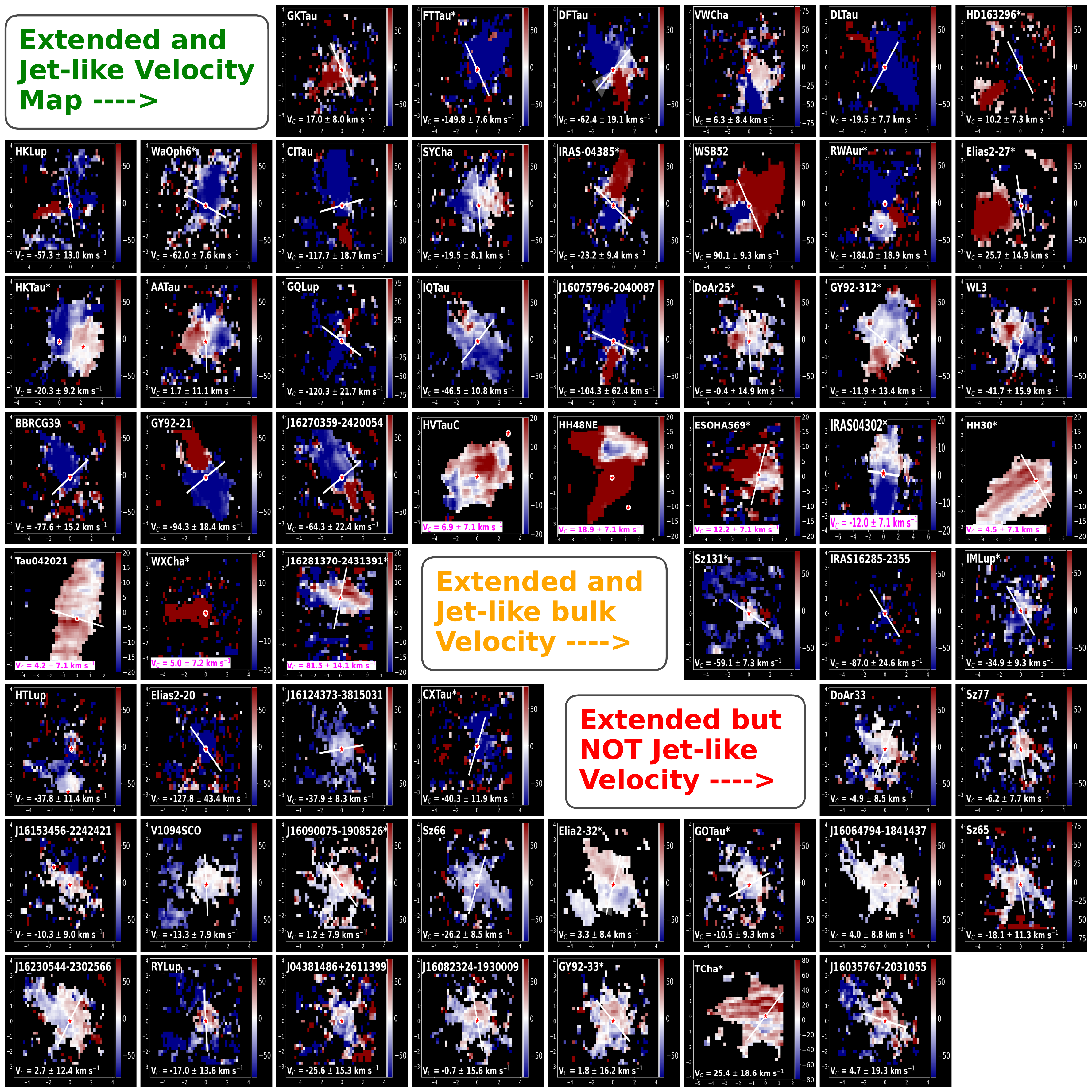}
    \caption{\neii{} velocity maps for sources with spatially extended \neii{} emission. In each of the three categories, the sources are ordered by increasing disk inclination (read from left to right and top to bottom). The velocities (in each pixel and v$_C$) are corrected for stellar radial velocity and disk inclination. Only pixels with flux greater than 2$\sigma$ are shown, where $\sigma$ is the standard deviation of the background calculated iteratively. The red star in each map marks the centroid of the corresponding continuum, and the white horizontal passing through it highlights the known disk PA (Table \ref{tab:source_properties}). Sources with asterisks are further discussed in Section \ref{sec:individual_sources}. The highly-inclined disks (incl$>$80$^{\circ}$) are indicated by highlighting their v$_C$ in magenta, and their maps (as well as v$_C$) are not deprojected for inclination. The \neii{} jet-like velocity maps show broader emission than expected for jets; however, this is a PSF effect, and deconvolution reveals \neii{} jets that are narrower than H$_2$ winds \citep[e.g., SYCha,][]{Schwarz2025}. The \neii{} spectra towards each source in this figure and the corresponding Gaussian fit to estimate v$_C$ are shown in Section \ref{sec:neii_spectra}.}
    \label{fig:velocity_maps}
\end{figure*}

High-resolution ground-based spectroscopy has shown that the \neii{} 12.81 \micron{} line can trace both fast jets \citep[e.g.,][]{vanBoekel2009} and slow ($\lesssim$10~km~s$^{-1}$) winds \citep[e.g.,][]{Pascucci2009} at the Class~II stage, contrary to H$_2$ which has been seen to trace jets only at the Class~0 stage \citep[e.g.,][or \cite{Bally2016}, \cite{Lee2020} for reviews]{Caratti_2024,Francis2026}. Consequently, we adopt a different approach to classify whether extended \neii{} emission is jet-like or not and outline our methodology here. In addition to morphology, we use the fact that jets flow at higher velocities than winds \citep[peak centroid velocities with respect to stellar velocities $\gtrsim$ 30~km~s$^{-1}$,][]{Simon2016}, and at the MIRI spectral resolution ($\sim$2700 or $\sim$100~km~s$^{-1}$), combined with the achieved wavelength calibration uncertainty of $\sim$5~km~s$^{-1}$ at 12.8~\micron{} \citep{Patapis2024}\footnote{The latest estimates can be found at \url{https://jwst-docs.stsci.edu/jwst-calibration-status/miri-calibration-status/miri-mrs-calibration-status}}, velocity shifts as small as 15~km~s$^{-1}$ can be measured, allowing us to estimate jet velocities. This approach has already been demonstrated to work for a few sources in our sample \citep[e.g.,][]{Kurtovic2026}. Accordingly, we leverage the combination of MIRI’s superior spatial and moderate spectral resolution to classify jets traced by \neii{}; the classification strategy is described below.

We first construct pixel-by-pixel velocity maps in \neii{}, following the procedure described in Section~\ref{sec:cont_sub}. These velocities are measured in the barycentric reference frame, that is, relative to the center of mass of the solar system. We then correct the velocity in each pixel for the known stellar heliocentric (relative to the center of Sun) radial velocity, collected from the literature and listed in Table \ref{tab:source_properties}, as well as for the known MIRI systematic offset in band~3A (-3~km~s$^{-1}$ for wavelength range 11.55-13.47 \micron{})\footnote{Retrieved from the JWST User Documentation at \url{https://jwst-docs.stsci.edu/jwst-mid-infrared-instrument/miri-observing-modes/miri-medium-resolution-spectroscopy}}. When the stellar radial velocity is unavailable, we use the average heliocentric radial velocity of the star-forming region in which the source lies. For Taurus, we calculate an average radial velocity of 18~km~s$^{-1}$ and a standard deviation of 3~km~s$^{-1}$ based on 36 stars reported by \cite{Nisini2024a}, excluding any multiple star systems. Similarly, for Ophiuchus it is found to be -6.85 $\pm$ 1.8 km~s$^{-1}$ \citep{Rigliaco+2016} and for Chamaeleon, we use -11.4 $\pm$ 2 km~s$^{-1}$ \citep{Biazzo2012}. Since the difference between barycentric and heliocentric reference frames is at most a few tens $\sim$m~s$^{-1}$ \citep{Wright2014} -- significantly smaller than the velocity uncertainties in this work ($\sim$km~s$^{-1}$), we do not convert between the two frames while applying the stellar radial velocity correction. We particularly lack radial velocity measurements for several highly-inclined disks ($>$80$^{\circ}$); however, these sources also exhibit the most convincing jet morphologies in \neii{} (e.g., HH30 in Figure \ref{fig:velocity_maps}). 

After applying the radial-velocity and offset corrections, we de-project the velocities using the known disk inclination angles to derive the true flow velocities (inclinations are listed in Table \ref{tab:source_properties}). The deprojection is performed for all sources except the ones that are highly inclined ($>$80$^{\circ}$) as it can lead to spurious values. The resulting pixel-by-pixel velocity maps are shown in Figure~\ref{fig:velocity_maps}. Overlaid on these maps, we also report the centroid velocity (v$_c$) of the bulk flow (or bulk velocity), derived from fitting the \neii{} line from the spectrum integrated over the entire IFU. The v$_c$ is shown in white when the velocities are deprojected using the known disk inclination and in magenta when the maps are not deprojected. When listing the bulk velocity, we also report the associated uncertainties, which account for the 1$\sigma$ statistical uncertainty, 5~km~s$^{-1}$ wavelength calibration uncertainty, and disk inclination uncertainty. The 1$\sigma$ statistical uncertainty is estimated by fitting 1000 realizations of the \neii{} line following \cite{Bajaj2024}. Based on these maps (Figures \ref{fig:velocity_maps}, and \ref{fig:multi}), we classify the extended \neii{} emission into three categories:

(i) \textbf{Extended emission with a jet-like velocity map}: These cases exhibit emission aligned with the stellar position and oriented perpendicular to the disk position angle, with velocities exceeding $\pm$30~km~s$^{-1}$ and a clear velocity gradient across the disk mid-plane (i.e., blue on one side and red on the other). For highly asymmetric jets, e.g., FTTau, we also consider a map to be jet-like if one side of the disk PA is dominated by blue (or red) and the other side does not show any significant emission. In many sources that exhibit both red- and blue-shifted emission in the velocity maps, the reported bulk velocity is dominated by the stronger component (or the mix of both). Recovering both the red and blue components would require a two-component fit; however, we perform only a single-component fit, as the scope of this work is limited to classification. 

\begin{figure*}
    \centering
    \includegraphics[width=0.62\linewidth]{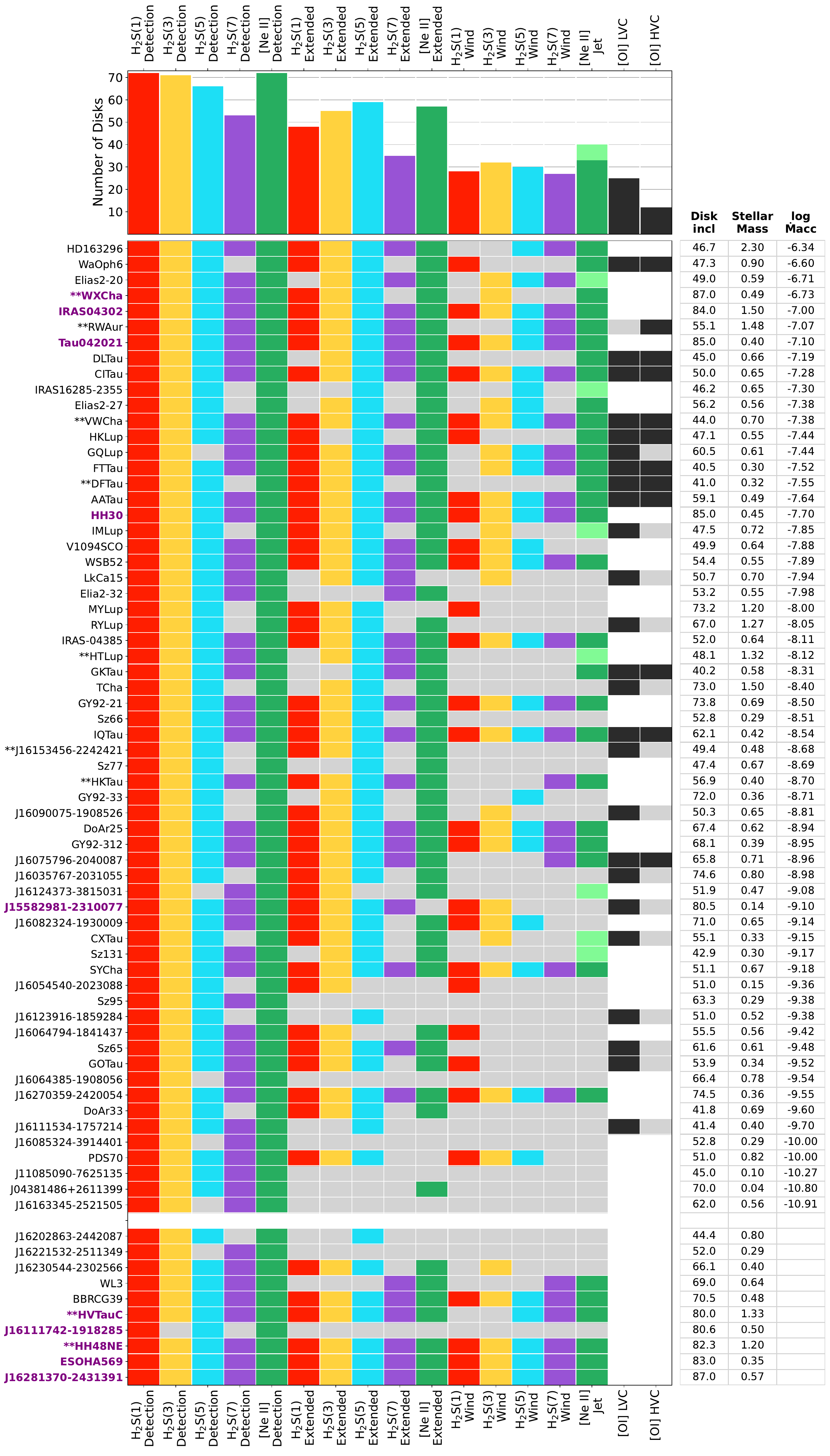}
    \caption{Figure summarizing which of the five analyzed spectral lines are detected, spatially extended, and exhibit outflow-like signatures. For example, when H$_2$ S(1) is detected toward a given source, the corresponding block is highlighted in red; when it is not detected, the block is shown in gray. For \neii{} jet classification, the additional lighter shade of green highlights the cases showing `Extended emission with a jet-like bulk velocity', whereas the darker shade highlights `Extended emission with a jet-like velocity map' (see Section \ref{sec:jet_classification} for the classification criteria). The \oi{} low-velocity component (LVC) and high-velocity component (HVC) detections are compiled from the literature: white block indicates that no published data is available for that source, while gray block indicates that data exists but the corresponding component is absent in the spectrum. The total number of sources with published \oi{} detections is equivalent to the number of sources showing \oi{} LVC, as this component was detected towards all but one source (RWAurA). The sources are ordered from top to bottom by decreasing mass accretion rate and, when not available, by increasing disk inclination. Sources whose names are prefixed with `**' denote multiple systems with companion(s) within the same FOV, while sources whose names are displayed in purple are highly-inclined. To the right side, key stellar and disk properties of each source are provided, references to which can be found in Table \ref{tab:source_properties_references}.}
    \label{fig:det_ext_out_stat} 
\end{figure*}

The \neii{} spectrum and a single-component Gaussian fit for each source in Figure \ref{fig:velocity_maps} are shown in Figure \ref{fig:neii_spectra_fit_extended_sources} in the Appendix. CITau is an example of spectrally resolved jet components. We find that about 58\% (33/57) of the targets with extended \neii{} emission fall into this `Extended emission with a jet-like velocity map' category, showing a clear jet-like velocity map. Of these, only VWCha had a prior ground-based \neii{} HVC detection, while spatially resolved \neii{} jets toward CITau, SYCha, WSB52, Elias2-27, Tau042021, and Elias2-20 have been recently reported with MIRI \citep{Arulanantham2024,Schwarz2025,Devaraj2026,Narang2026b}. The remaining 26 jet-driving-sources are new detections in \neii{}. In addition, 10 unique systems show \oi{} HVC detections (Figure \ref{fig:det_ext_out_stat}) and in each case, the \oi{} HVC velocity shift (blue or red) matches the dominant \neii{} component identified here.

(ii) \textbf{Extended emission with a jet-like bulk velocity}: In these cases, one or more of the criteria for a jet-like velocity map are not satisfied; however, the bulk velocity exceeds 30~km~s$^{-1}$. The \neii{} line towards these targets is well detected as demonstrated in Section \ref{sec:neii_spectra}. Physically, these sources may represent jets that are spatially unclear (not well resolved morphologically, or too faint to be well detected in individual pixels) but have sufficient brightness asymmetry between the blue and red lobes to create a bulk velocity shift $|v_c|$ $>$ 30~km~s$^{-1}$  in the integrated spectrum, not explainable by a low-velocity PE wind (where we expect $|v_w|$ $<$ 30~km~s$^{-1}$). We find that approximately 12\% (7/57) of the targets showing extended emission fall into this category, exhibiting a highly red/blue-shifted bulk velocity.

(iii) \textbf{Extended but not jet-like emission}: These cases fail to meet one or more of the criteria for a jet-like velocity map, and the bulk velocity is also $<$30~km~s$^{-1}$. In these sources, \neii{} likely traces the disk atmosphere and/or atomic winds. We find that most of the \neii{} intensity maps in these cases are consistent with only marginally resolved emission. For example, TCha, classified here as not jet-like, but found to show marginally extended emission, has been shown to likely trace a photoevaporative wind with JWST MIRI \citep{Bajaj2024,Sellek2024}, in agreement with the small blueshift in \neii{} measured through high-resolution ground-based spectroscopy \citep{Pascucci2009,Sacco2012}. In many of these maps, a low-amplitude ($<$20~km~s$^{-1}$ after deprojection) blue–red contrast is observed in velocities around the centroid. This could arise from uneven slice illumination \citep[e.g., see][for a more detailed explanation and its effect]{Agra-amboage2014}, which was not corrected for in the calibration pipeline version used in this study. About 30\% (17/57) of the targets with extended \neii{} emission fall into this `Extended but not jet-like emission' category.

Finally, 20\% (15/72) sources do not show any spatially resolved \neii{} emission. This may arise for several reasons, which we outline below. (a) We identify two sources, Sz95 and J16085324-3914401, that exhibit jet like bulk velocities of $-45 \pm 11$ km~s$^{-1}$ and $-69 \pm 10$ km~s$^{-1}$, respectively. These likely represent cases in which the jet contribution to the \neii{} line intensity has decreased to the point that the jet can no longer be spatially resolved. Although they don't have \oi{} data, this interpretation is consistent with their relatively small jet velocities, given the observed correlation between \oi{}-HVC jet velocity and \oi{}-HVC flux \citep{Banzatti2019}. (b) In other sources, the velocities are consistent with a \neii{} LVC, suggesting emission from the disk atmosphere and/or weak winds. For example, three such sources with available \oi{} data, J16111534-1757214, LkCa15, and J16123916-1859284, show only \oi{} LVC emission \citep{Banzatti2019,Fang2023,Nisini2024a}. The compact \neii{} emission in these systems is therefore consistent with one of our aforementioned explanations, a weak wind origin. (c) Finally, for sources in the Upper~Sco region, we detect weak \neii{} background emission that becomes significant with SNR $>$ 3 when integrated over the full field of view. This may indicate the presence of substantial EUV and/or X-ray radiation in Upper~Sco and will be explored further in a forthcoming publication (N. Bajaj et al. 2026, in prep).

\section{Empirical results} \label{sec:results}

A schematic summary of which of the five lines (H$_2$ S(1), S(3), S(5), S(7), and \neii{}) are detected, spatially extended, and trace an extended wind/jet is presented in Figure \ref{fig:det_ext_out_stat}. The sources are arranged in order of decreasing mass accretion rate, and when not available, by increasing disk inclination. It is immediately apparent that the jet and wind fractional detection rate is highest for the highest accretors. It is also very high for the highly-inclined disks (\textit{i} $>$ 80$^{\circ}$), with H$_2$ winds detected towards all but one. We note that the line detections in this table are based on spectra integrated over the entire FOV, and in some cases, such as H$_2$ S(1), the detections may be attributable to diffuse emission in the background. Based on this summary, we look at statistics of extended H$_2$ wind and \neii{} jet detections with MIRI in Section \ref{sec:basic_stats}, followed by an identification of sources that exhibit asymmetric jet and/or wind morphologies in Section \ref{sec:asymmetric_morphology_discussion}. We then investigate how the outflow detection rates depend on disk inclination (Section \ref{sec:trend_with_inclination}), stellar mass (Section \ref{sec:trend_with_mass}), and accretion rate (Section \ref{sec:trend_with_accretion}). Finally, we compare the outflows detected in \neii{} (extended emission with jet-like velocity map + extended emission with jet-like bulk velocity) and H$_2$ with those detected through ground-based high-resolution \oi{} spectroscopy in Section \ref{sec:OI_NeII_H2_comparison}. 

We note that throughout this section, we exclude 9 known multiple star systems (i.e., stellar companions with masses $>$0.1~M$_{\odot}$, indicated in Figure \ref{fig:det_ext_out_stat} with `**' before their name) in which both components fall within the FOV, as their H$_2$ emission is often difficult to interpret. However, we retain the wide separation binary Sz65 and Sz66, since they were observed separately with no overlap in the FOV. Sources with separations larger than half the MIRI FOV are wide binaries, in which disk evolution has been shown to proceed similarly to that around single stars \citep[e.g.,][]{Harris2012,ZhangY2023}. In total, 9 observations are excluded from the analysis, but their emission morphologies and \neii{} velocity maps are discussed individually in Section \ref{sec:individual_sources}. In Sections \ref{sec:trend_with_inclination}, \ref{sec:trend_with_mass}, and \ref{sec:trend_with_accretion}, we show the detection fractions of outflows and their associated uncertainties, calculated as described in Section \ref{sec:uncertainty_estimation}. We also plot the total integrated line SNR, calculated following the procedure outlined in Section \ref{sec:SNR_calculation}.

\subsection{Outflow detection rates: Comparison between \neii{} jets and H$_2$ winds}
\label{sec:basic_stats}

\begin{figure}
    \centering
    \includegraphics[width=\linewidth]{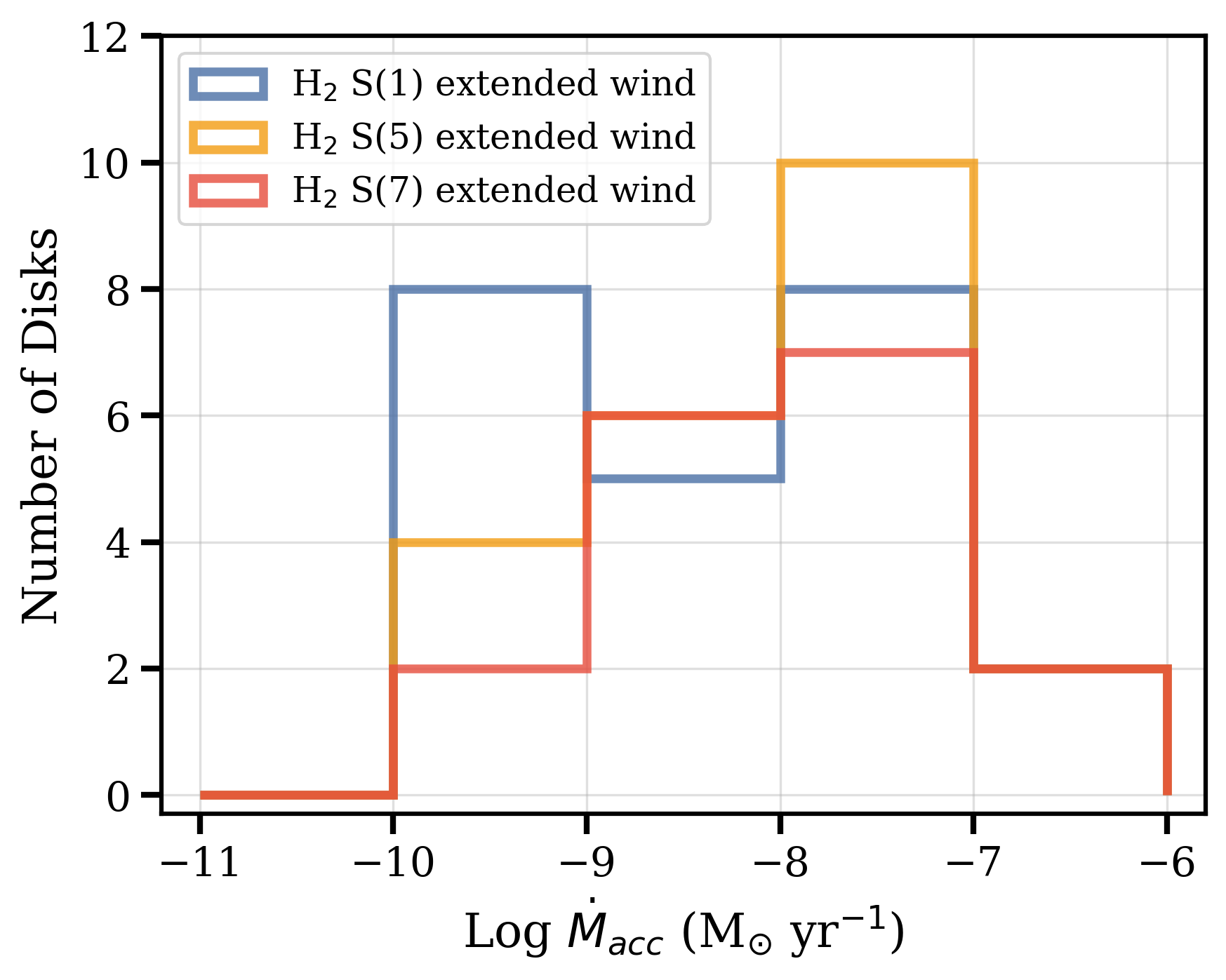}
    \caption{Histogram of mass accretion rates for sources with extended H$_2$ winds detected in S(1) (blue), S(5) (orange), and S(7) (red). Multiple-star systems are excluded, and so are several highly-inclined disks ($i > 80^\circ$) lacking reliable accretion rate measurements (Table \ref{tab:source_properties}).}
    \label{fig:s1_s5_hist_macc}
\end{figure}

Out of the 72 targets in our sample, more than half of them (46) exhibit a wind morphology in at least one of the four H$_2$ lines analyzed in this work. In comparison, \cite{Narang2026b} detected conical H$_2$ wind morphology towards 10 disks, in line with the number of inclined disks in their sample (18) being four times lower than this work. Interestingly, we find that only 18 out of 46 sources show extended wind morphology simultaneously in all four H$_2$ lines, while the remaining display winds in only one, two, or three of the investigated H$_2$ lines. The number of sources showing extended wind signatures is comparable across the individual lines, with 28 detections in H$_2$~S(1), 32 detections in H$_2$~S(3), 30 in H$_2$~S(5), and 27 in H$_2$~S(7). Interestingly, when comparing the number of disks exhibiting extended H$_2$ winds in S(1), S(5), and/or S(7) to $\dot{M}_{\rm acc}$ (see Figure \ref{fig:s1_s5_hist_macc}), we find that the incidence of extended S(5) and S(7) winds decreases with decreasing accretion rate, whereas S(1) winds show no significant trend. Because H$_2$~S(5) and S(7) have higher excitation temperatures and trace hotter gas than S(1), this points to an evolution in wind properties, which we discuss in Section~\ref{sec:atomic_molecular_evolution_discussion}. Similar to H$_2$, we find that a little more than half of our sample (40 sources) shows evidence for a spatially resolved jet in \neii{}, based on their morphology and kinematics (Section~\ref{sec:jet_classification}). 

The \neii{} jet detections and the H$_2$ wind detections do not fully overlap. 34 sources exhibit both a jet in \neii{} and a wind in at least one H$_2$ line. The remaining sources are distributed in two distinct non-overlapping groups. First, 12 sources display a wind morphology in at least one H$_2$ line (out of a total of 46 wind-sources) without a corresponding \neii{} jet (however, see objects such as J1623-2302 in Section \ref{sec:individual_sources}), and all 12 are single-star systems. Second, 6 sources show \neii{} jet detections (out of a total of 40 jet-sources) without clear H$_2$ winds, however, all of them show extended emission in multiple H$_2$ lines. We note that two of these are binaries where the wind identification may be complicated by H$_2$ emission from the two systems (DFTau, HTLup). Of the four single sources, two have \oi{} data available, and both show LVC emission (DLTau and GKTau, Figure \ref{fig:det_ext_out_stat}), indicating the presence of an atomic wind despite the absence of detectable extended molecular wind emission.

Interestingly, the 34 sources exhibiting both \neii{} jets and extended H$_2$ winds are preferentially higher accretors compared to the sources exhibiting only extended H$_2$ winds or only \neii{} jets. We discuss the implications of this result in Section \ref{sec:atomic_molecular_evolution_discussion}.

\subsection{Intrinsically asymmetric jet and wind launching} \label{sec:asymmetric_morphology_discussion}

Based on Figure \ref{fig:velocity_maps}, many sources display asymmetry in the spatial extent of the \neii{} jet between the two sides of the disk. For a few edge-on Class~II disks, extinction along both jet lobes is demonstrated to be comparable, suggesting that any observed asymmetry likely reflects the intrinsic properties of the jet launching mechanism \citep{Bajaj2025}. However, most targets in our sample with reliable \neii{} velocity measurements are moderately inclined (40–80$^{\circ}$), where projection effects become substantial. In these systems, the blue-shifted lobe is expected to appear both extended and brighter even when launched symmetrically, producing an apparent asymmetry that is challenging to distinguish from intrinsic effects without independent constraints on extinction (e.g., FTTau in Figure \ref{fig:velocity_maps}). 

In contrast, sources displaying brighter and more extended red-shifted lobes cannot be attributed to projection effects and therefore require intrinsically asymmetric jet launching. We identify five clear cases among the non-edge-on disks where velocities can be confidently measured: GKTau, IRAS-04385, WSB52, Elias2-27, and WXCha\footnote{We do not classify WXCha as edge-on \citep[as reported in literature with a large uncertainty,][]{Banzatti2015} given its \neii{} velocities and discuss in greater detail in Section \ref{sec:individual_sources}.}. The detection of intrinsically asymmetric jets supports a magnetic origin for these flows (Section \ref{sec:asymmetric_morphology_discussion}); however, the mechanism underlying such asymmetries remains unclear, and these observations provide important constraints for models of jet launching.

Among these five sources with red-shifted-lobe-dominated \neii{} jet emission, GKTau does not display clear evidence for an extended H$_2$ wind \citep[though it has an \oi{} wind,][]{Banzatti2019}, whereas two sources, IRAS-04385 (S(3) and S(5)) and WSB52 (S(1) and S(3)), show dominant molecular wind emission coinciding with the red-shifted jet lobe. In contrast, the remaining two sources, Elias2-27 and WXCha, display dominant H$_2$ wind emission on the opposite side of the jet, with blue-shifted winds and red-shifted jets (see S(3) maps in Figure \ref{fig:multi}). The implications of these results are discussed in Section \ref{sec:H2_MHD_driven}.

\subsection{Trend with disk inclination: outflow detection rate constant from 40$^{\circ}$ to 80$^{\circ}$}
\label{sec:trend_with_inclination}

\begin{figure}
    \centering
    \includegraphics[width=\linewidth]{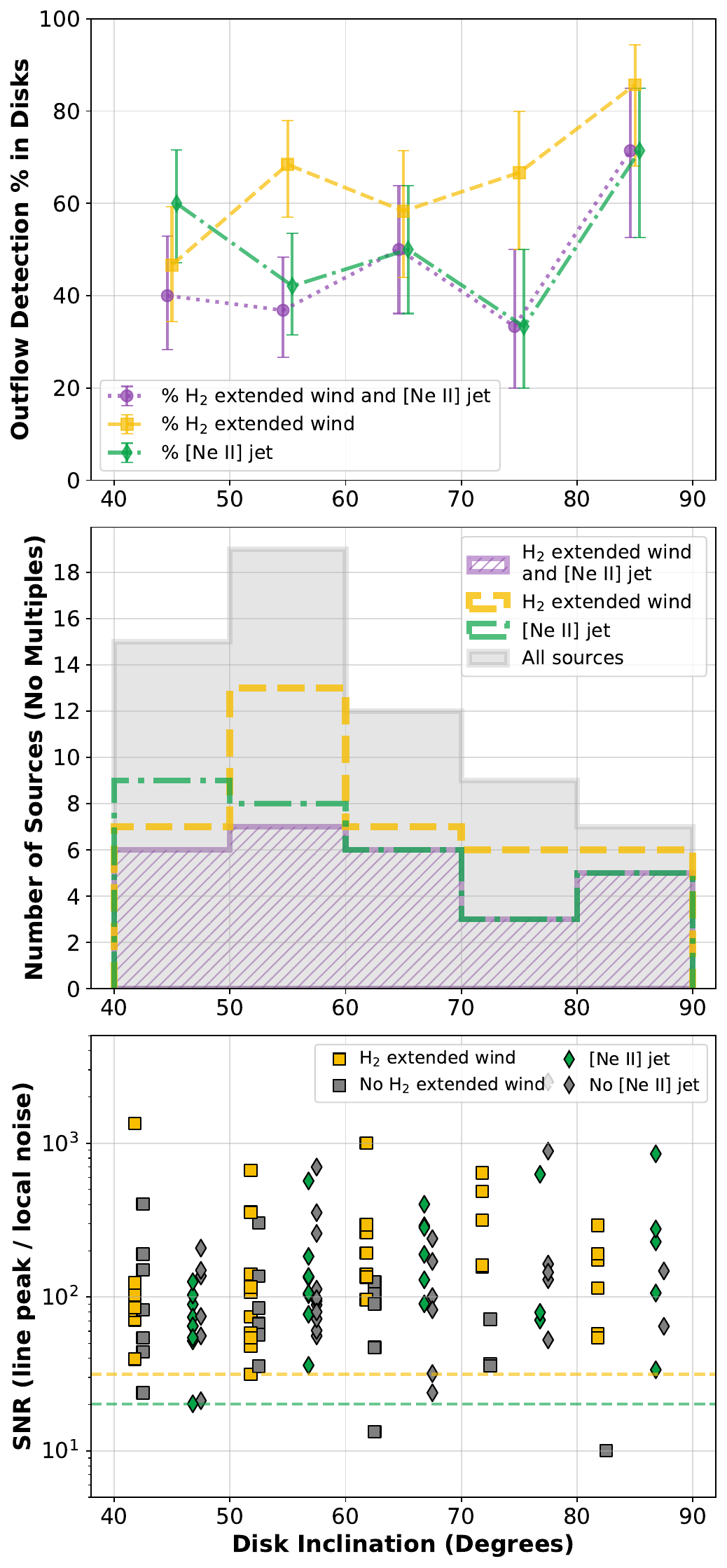}
    \caption{The top panel shows the histogram of sources exhibiting the detections listed in the legend, with the full sample of sources with literature disk inclination measurements shown in gray. The middle panel shows the corresponding outflow detection fractions in each inclination bin, with uncertainties estimated using the method described in Section~\ref{sec:uncertainty_estimation}. The points are plotted with a small offset in x-axis from the bin center to ensure that each error bar is visible. The lower panel shows the line SNR (see Section \ref{sec:SNR_calculation} for details on how it's estimated) for the individual sources in each bin, with boxes highlighting the average H$_2$ SNR and diamonds displaying \neii{} SNR. We have excluded the known multiple star systems from this plot. The dashed lines in the lower panel highlight the minimum line SNR for which an extended jet (green) or an extended wind (yellow) is detected.
    }
    \label{fig:fraction_disks_outflows_inclination}
\end{figure}

To check for any effects of disk inclination on outflow detections, we plot the disk fractions exhibiting outflows (winds, jets, or both) as a function of inclination, separated in bins of 10$^{\circ}$ from 40$^{\circ}$ to 90$^{\circ}$ (see Figure \ref{fig:fraction_disks_outflows_inclination}). Interestingly, we find that the fraction of disks showing (a) jets, (b) winds, and as a consequence, (c) both jets and winds, remains roughly constant from $\sim$40$^{\circ}$ to $\sim$80$^{\circ}$ before increasing toward $\sim$90$^{\circ}$ as expected. 

For jets, we find that the outflow detection fraction is slightly higher in the 40-50$^{\circ}$ bin than the average of 50-80$^{\circ}$ bins, in addition to the 80-90$^{\circ}$ bin. This trend can be explained by how velocity and morphology contribute differently to our classification. At lower inclinations ($\lesssim$60$^{\circ}$), the line-of-sight velocity component is relatively strong, making jets easier to identify through their kinematic signatures. At higher inclinations ($\gtrsim$70$^{\circ}$), the velocity information becomes less informative, but the jet structure itself is more clearly resolved because contamination from disk emission is reduced. Together, these effects could lead to small increases in the jet detection fraction at lower (40–50$^{\circ}$) and higher (80–90$^{\circ}$) inclinations. In contrast, H$_2$ extended winds show increased detections only in the highest inclination bin, which is consistent with the fact that we can only use the morphology of H$_2$ for wind classification, which is best resolved at the highest disk inclinations (80–90$^{\circ}$).

\subsection{Trend with stellar mass: outflow detection rate similar from 0.1 to 1 solar mass}
\label{sec:trend_with_mass}

\begin{figure}
    \centering
    \includegraphics[width=\linewidth]{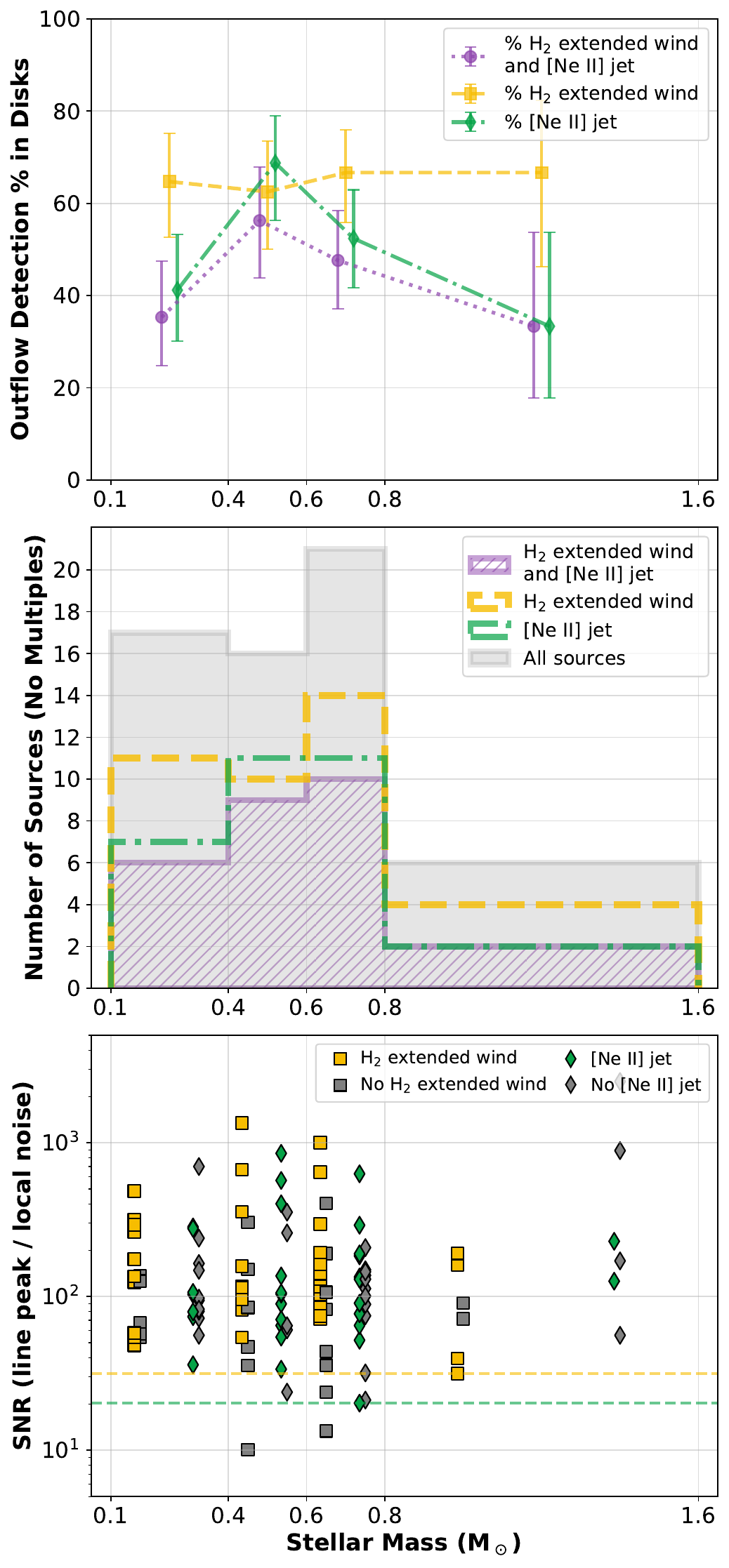}
    \caption{Similar to Figure \ref{fig:fraction_disks_outflows_inclination} but plotted against stellar mass. The bin edges are selected to have a similar number of sources in each bin. Since stellar mass estimates are available for all edge-on disks, we include them in the analysis.}
    \label{fig:fraction_disks_outflows_mass}
\end{figure}

Before discussing the outflow detection rates as a function of mass accretion rate, which is key to wind-driven accretion theory, we first discuss its dependence on stellar mass. It is important to disentangle the effects of stellar mass and mass accretion rate, as a relatively young disk around a low-mass star can exhibit a mass accretion rate comparable to that of a more evolved disk around a relatively higher-mass star \citep{Alcala2017,Fang2023}. As a result, accretion rate alone is not a unique tracer of disk evolution without accounting for stellar mass. To address this, we examine the outflow detection rates in four bins of stellar masses with bin edges selected such that there are similar number of sources in each bin, except the largest bin where the lack of MIRI observations of single sources with stellar masses $>$ 0.8~M$_{\odot}$ is apparent (Figure~\ref{fig:fraction_disks_outflows_mass}). We find that the H$_2$ extended wind detection fraction is remarkably constant across all stellar mass bins, whereas the jet detection fraction slightly increases between the 0.1-0.4 and 0.4-0.6~M$_{\odot}$ bins and then somewhat decreases towards higher masses up to 1.6~M$_{\odot}$. The small variations in \neii{} jet detection fractions possibly arise from (a) small number statistics in the highest mass bin, which is reflected in its error bar, and (b) uncertainties in stellar mass, which based on the table of stellar and disk properties produced by \cite{Manara2023}, we find an average 3$\sigma$ (where $\sigma$ is the standard deviation) of $\sim$0.1~M$_{\odot}$ between the four values of stellar masses listed in the table. Within the uncertainties, the outflow detection fraction is roughly independent of the stellar mass in the range $\sim$0.1-1~M$_{\odot}$ (where most of our single sources lie), with further investigation needed above this range.

The approximately steady outflow detection rate suggests that jet and wind launching is a common feature of the protostellar phase, governed by disk-mediated accretion and/or magnetic processes that operate similarly across a wide range of stellar properties. Although the strength and kinematics of outflows may still vary with stellar mass, their occurrence appears largely mass independent in the 0.1-1~M$_{\odot}$ range, pointing to a broadly self-similar launching mechanism.

\subsection{Trend with accretion rate: Outflow detection rate increases with accretion}
\label{sec:trend_with_accretion}

\begin{figure}
    \centering
    \includegraphics[width=\linewidth]{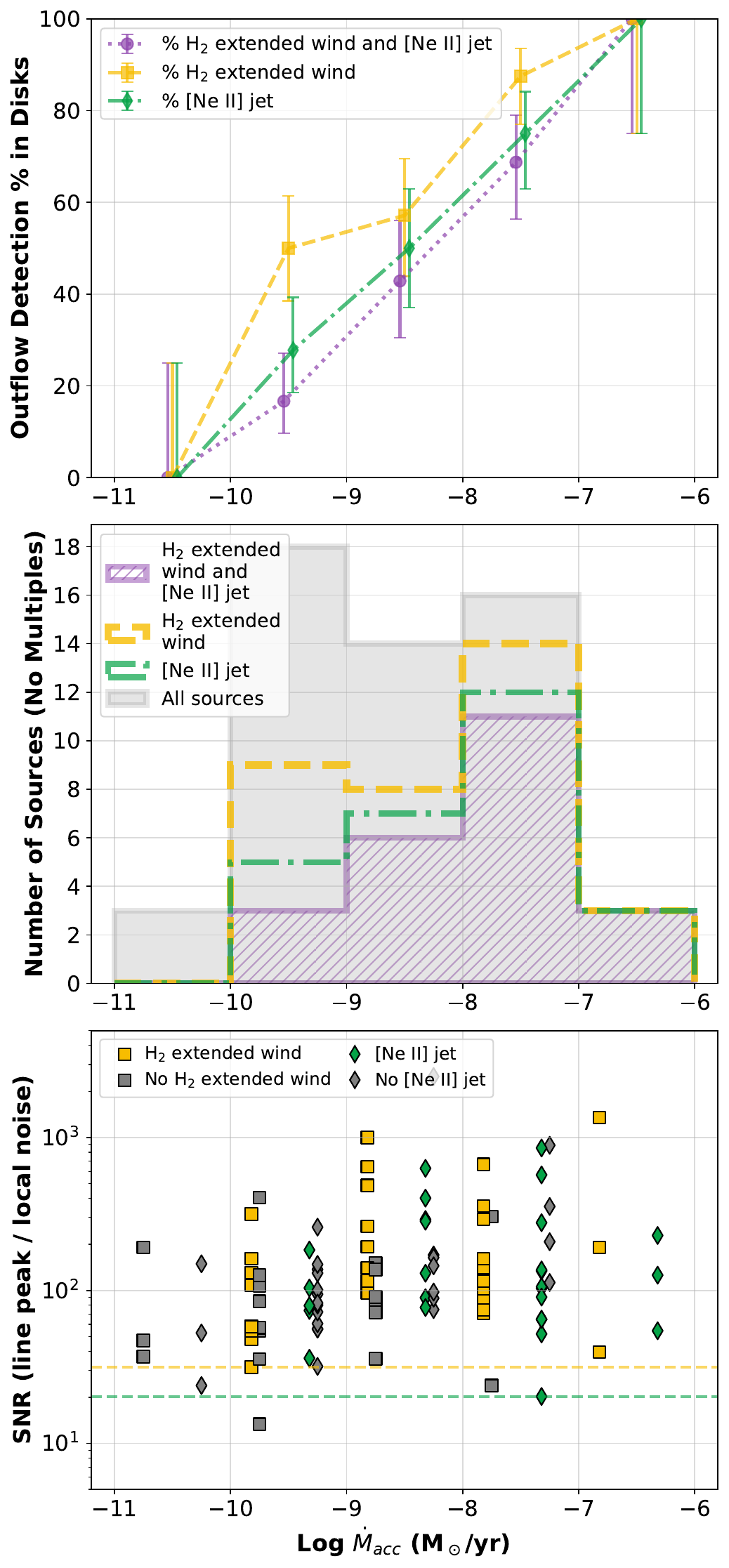}
    \caption{Similar to Figure \ref{fig:fraction_disks_outflows_mass} but plotted against mass accretion rate. The highly-inclined disks that lack mass accretion rate estimates are not included in this analysis.}
    \label{fig:fraction_disks_outflows_macc}
\end{figure}

Following the stellar mass analysis, we now examine how the fraction of sources exhibiting extended jets, winds, or both varies with stellar mass accretion rate (see Figure~\ref{fig:fraction_disks_outflows_macc}). This enables us to examine the relationship between outflows and accretion activity. We find that the fraction of disks exhibiting either jets, extended H$_2$ winds, or both increases steadily with increasing accretion rate. In addition, the detection fractions of extended jets and winds in most individual bins are similar to each other, except the [-10,-9] bin, where the wind detection fraction is higher, which is consistent with the population of low accretors identified in Section \ref{sec:basic_stats} that shows extended H$_2$ wind without a corresponding \neii{} jet. 

This trend of lower outflow detection rates at lower accretion rates is not due to sensitivity limits. The lower panel of Figure \ref{fig:fraction_disks_outflows_macc} shows the line SNR for all the sources in each accretion bin and highlights the lowest SNR for which an extended wind in H$_2$ and/or a jet in \neii{} is detected. Considering this as our outflow detection limit, we find that almost all the sources have line SNR above this detection limit. 


\begin{figure*}
    \centering
    \includegraphics[width=1\textwidth]{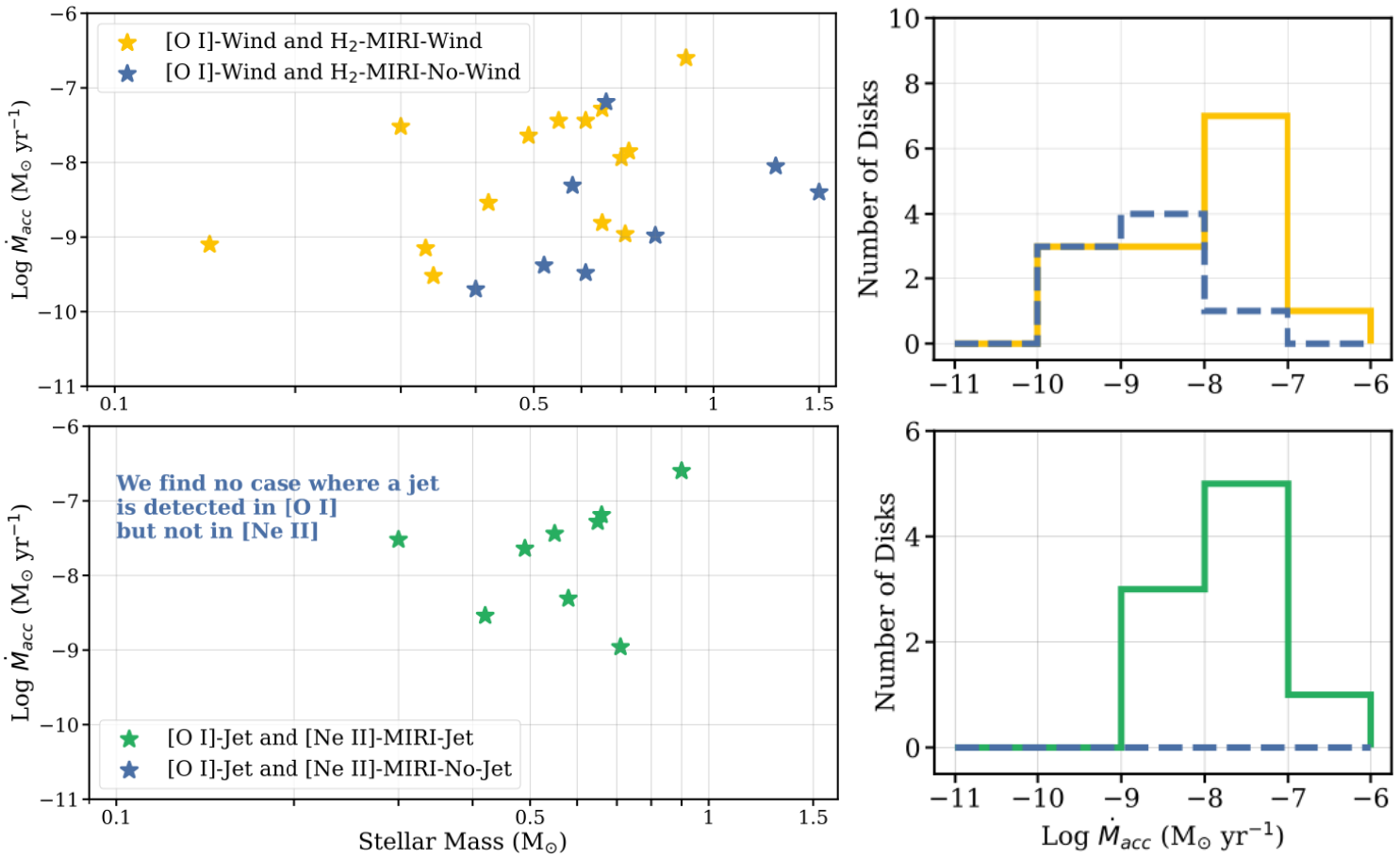}
    \caption{Left panels: Stellar mass versus mass accretion rate for sources with wind detections in both H$_2$ and \oi{}, or just \oi{} and jet detections in both \neii{} and \oi{}, or just \oi{}. The color of each point, along with the adjacent histograms, corresponds to the categories listed in the legend. Right panels: Histograms of the mass accretion rates for the same source categories shown in the left panels. Known multiple star systems exhibiting these properties are excluded from the plot. We also note that although many highly-inclined disks show jet and wind detections in the MIRI data, they are not included here due to the lack of accretion rate estimates and/or \oi{} data.}
    \label{fig:H2_NeII_OI_comparison}
\end{figure*}

\subsection{Comparison of \oi{}, \neii{}, and H$_2$ outflow detections}
\label{sec:OI_NeII_H2_comparison}

Much of our pre-JWST understanding of jets and winds from T~Tauri star-disk systems comes from ground-based high-resolution spectroscopy (8000 $\lesssim$ R $\lesssim$ 115,000), primarily using the \oi{}~6300\AA{} line \citep[e.g.,][see also \cite{Pascucci2023} for a review]{Simon2016,Fang2018,McGinnis2018,Nisini2018,Banzatti2019,Fang2023,Nisini2024a}. To assess whether our JWST observations recover the full population of high-velocity components (HVCs) and low-velocity components (LVCs) previously identified from the ground, we compare these literature detections with jets traced by \neii{} and extended molecular winds traced by H$_2$ using MIRI (Figure \ref{fig:H2_NeII_OI_comparison}). Of the 72 disks in our sample, we found published \oi{} high-resolution spectroscopy for 26 of them (highest inclination $\sim$ 80.5$^{\circ}$), leading to a sample overlap of about 36\%. 

In the lower-left panel of Figure \ref{fig:H2_NeII_OI_comparison}, we plot stellar mass versus mass accretion rate for sources where a jet is detected in both \oi{} (from the literature) and \neii{} (this work). We do not find any case where a jet is detected in \oi{} but not in \neii{} (supposed to be navy-blue points in Figure \ref{fig:H2_NeII_OI_comparison}), meaning that all previously reported \oi{} jets are recovered in \neii{} MIRI, indicating comparable sensitivity of these tracers to jet emission. This is especially important because, while ground-based spectroscopy provides very precise jet velocities \citep[e.g.,][]{Sacco2012}, it often lacks the superior sensitivity and spatial resolution of MIRI aboard JWST in the mid-IR. With MIRI, we can spatially resolve these same jets while simultaneously measuring their velocities in \neii{}. This opens up the possibility of understanding their excitation, mass loss rates, variability along the jet, and the interaction of atomic jets with the broader molecular winds \citep[e.g.,][]{Nisini2024b, Bajaj2025}. 

A similar comparison is shown in the top panels of Figure \ref{fig:H2_NeII_OI_comparison}, where we highlight sources exhibiting (i) winds in both H$_2$ (extended) and \oi{} and (ii) wind in \oi{} but not in H$_2$. We find that a little less than half of the sources with \oi{} LVC detections do not show an extended wind in any of the H$_2$ lines. A closer inspection reveals that these sources predominantly lie at lower accretion rates (top-right panel of Figure \ref{fig:H2_NeII_OI_comparison}) compared to the cases where an extended wind is seen in H$_2$ and an LVC is detected in \oi{}. Moreover, this trend is even stronger when interpreted in different stellar bins; i.e., the only low accretors that show extended H$_2$ winds are the lowest-mass stars. This absence of extended H$_2$ wind emission at lower accretion rates is not driven by sensitivity limitations, as we detect extended H$_2$ winds in systems with accretion rates as low as $\sim$10$^{-10}$~M$_{\odot}$~yr$^{-1}$ (Figure \ref{fig:H2_NeII_OI_comparison}, also see Section \ref{sec:trend_with_accretion} where only one source with $log\dot M_{acc} > -10$ has SNR below the wind detection threshold). Similarly, \cite{Gangi2020}, in a high-resolution spectroscopic survey of 37 T~Tauri star--disk systems, found \oi{} LVC towards nearly all disks but H$_2$~$\nu$1-0~S(1) LVC only towards $\sim$50\% of the sample, consistent with the detection fraction of extended winds seen here in the lower excitation lines of H$_2$~$\nu$=0-0. 

\section{Discussion} \label{sec:Discussion}

\begin{table*}
    \centering
    \caption{A summary of key statistics and results}
    \renewcommand{\arraystretch}{1.2}
    \begin{tabular}{|p{0.5\linewidth}|p{0.5\linewidth}|}
        \hline
        \multicolumn{1}{|c|}{\textbf{H$_2$ Statistics}} & \multicolumn{1}{c|}{\textbf{\neii{} Statistics}} \\ \hline
        64/72 sources show extended emission in at least one H$_2$ line. &  57/72 sources show extended emission in the \neii{} line. \\ \hline
        46/64 extended H$_2$ sources show a wind morphology in at least one H$_2$ line. & 33/57 extended \neii{} sources show jet-like velocity maps, and 7 others show jet-like bulk velocities. \\ \hline
        18/64 extended H$_2$ sources show other morphologies (e.g., along the disk plane or binaries) & 17/57 show mostly PSF-like morphology consistent with LVC (or zero velocity). \\ \hline
        \multicolumn{2}{|c|}{\textbf{Combined Results}} \\ \hline
        \multicolumn{2}{|p{\linewidth}|}{34 disks host both H$_2$ winds and \neii{} jets and are preferentially higher accretors than the 12 hosting only H$_2$ winds (Section~\ref{sec:basic_stats}).}\\ \hline
        \multicolumn{2}{|p{\linewidth}|}{Of the 4 single disks showing \neii{} jets without clear H$_2$ winds, 2 have \oi{} data and both show LVC (wind) (Section \ref{sec:basic_stats}).} \\ \hline
        \multicolumn{2}{|p{\linewidth}|}{The outflow detection rate positively correlates with accretion without any significant variations with stellar mass or disk inclination (Sections \ref{sec:trend_with_inclination}, \ref{sec:trend_with_mass} and \ref{sec:trend_with_accretion}).} \\ \hline
        \multicolumn{2}{|p{\linewidth}|}{Higher accretors show an extended H$_2$ wind along with \oi{} LVC whereas the lower accretors only show \oi{} LVC (Section \ref{sec:OI_NeII_H2_comparison}).} \\
        \hline
    \end{tabular}
    \label{tab:summary_table}
\end{table*}

Table~\ref{tab:summary_table} summarizes the empirical results presented in Section~\ref{sec:results}. Here, we integrate them to evaluate the origin of H$_2$ winds and their evolution. In particular, we evaluate whether photoevaporation or MHD disk winds are more likely to drive the spatially extended H$_2$ emission observed with MIRI (Section \ref{sec:H2_MHD_driven}). By combining the results of this paper with those from ground-based high-resolution spectroscopy, we expand upon the empirical evolutionary sequence for outflows launched from the disk (Section \ref{sec:atomic_molecular_evolution_discussion}), first developed by \cite{Pascucci2020}.

\subsection{Most Spatially resolved H$_2$ winds are likely MHD-driven} \label{sec:H2_MHD_driven}

Growing evidence suggests a physical connection between fast, collimated jets and slower disk winds. High-resolution optical spectroscopy shows that the \oi{} LVC (winds) centroid velocities and line widths correlate with both the HVC (jets) equivalent width (EW) and the accretion luminosity \citep{Banzatti2019}. Spatially resolved studies further support this link; NIRSpec IFU observations of four moderate-to-high accretors with edge-on disks \citep{Bajaj2025} consistently found a nested morphology with the faster jet encased in the slower H$_2$ wind, as expected from MHD disk wind models \citep{Pascucci2025}. With a sample more than an order of magnitude larger than that of \citet{Pascucci2025} and spanning a broader range of accretion rates, we find that the majority of \neii{} jet sources ($\sim$85\%, 34/40) have a corresponding extended H$_2$ wind (Section~\ref{sec:basic_stats}). For the remaining systems, those that are single and lack an H$_2$ wind detection (four total) exhibit an \oi{} wind when high-resolution optical spectra are available (two, Figure~\ref{fig:det_ext_out_stat}). Together, these results establish a one-to-one correspondence between atomic jets and winds (atomic or molecular tracers).

Moreover, Figure~\ref{fig:fraction_disks_outflows_macc} shows a strong positive correlation between $\dot{M}_{\mathrm{acc}}$ and H$_2$ wind detection rates. Such a correlation can be explained by an origin of extended H$_2$ emission in MHD disk winds, where denser winds that drive higher accretion rates also produce higher line fluxes in the extended wind at distances resolvable by MIRI ($\gtrsim$20~au). Particularly, the majority of our wind-sample (34/46) that shows both extended H$_2$ winds and \neii{} jets and consists of preferentially higher accretors ($\dot{M}_{acc} \gtrsim 10^{-9}~M_{\odot}~yr^{-1}$), are more consistent with MHD-disk wind interpretation than PE as the presence of jets/inner-disk-winds (which cannot be PE-driven) toward higher accretors will attenuate stellar X-rays and UV more readily than toward lower accretors, where the jet weakens (Figure~\ref{fig:fraction_disks_outflows_macc}). However, in the 12 lower-accretion systems ($\dot{M}_{acc} \lesssim 10^{-8.5}~M_{\odot}~yr^{-1}$) where extended H$_2$ emission is present without a detected jet, the origin remains ambiguous and may be consistent with either MHD disk winds or PE winds. Modeling of the extended H$_2$ wind emission detected in this work under both frameworks is therefore required to assess their viability \citep[e.g.,][S.~Clark et al. 2026, in prep.]{Nakatani2026}.

Finally, disk wind asymmetries not attributable to extinction provide direct evidence for non-ideal MHD processes, such as the Hall effect, in wind launching. Although our sample spans a wide range of inclinations and is not optimally suited for a systematic asymmetry analysis, two exceptional sources, IRAS-04385 and WSB52, exhibit predominantly redshifted jet and wind emission that cannot be explained by projection or extinction effects and therefore must be intrinsically asymmetric (see Section \ref{sec:asymmetric_morphology_discussion} for more details). These systems are therefore only consistent with MHD disk winds \citep[e.g., see][for MHD launching models that incorporate non-ideal effects, capable of producing asymmetric jets and winds]{Mori2025,Tu2025}.

\subsection{Evolution of Winds from Molecular to Atomic} \label{sec:atomic_molecular_evolution_discussion}

From a combined analysis of high-resolution optical and infrared spectra, \citet{Pascucci2020} found that \neii{} HVCs (jets) are detected toward high accretors, while LVCs (winds) appear in sources with low $\dot{M}_{\rm acc}$, low \oi{} luminosity, and large infrared spectral indices (n$_{13-31}$) indicative of inner dust-disk depletion. To explain these trends, they proposed a scenario \citep[see Fig.~9 in][]{Pascucci2020} where dense molecular inner winds in high accretors shield the disk surface from hard stellar X-rays \citep[e.g.,][]{Hollenbach2009}. Only when accretion subsides and, along with it, the jet, the \oi{} LVC and the inner molecular wind, X-rays can reach the disk surface, driving increased \neii{} LVC emission and an outer wind. By spatially identifying molecular winds in H$_2$ and jets in \neii{}, this work can directly test and expand upon this scenario.

(i) We find that the predicted molecular winds are closely associated with the jets and both are preferentially detected towards moderate-to-high accretors, i.e., up to 85\% of the sources with $\dot{M}_{acc} \gtrsim 10^{-8}~M_{\odot}~yr^{-1}$ (Figure \ref{fig:fraction_disks_outflows_macc} top panel). These atomic jets and molecular winds could indeed shield high-energy stellar photons, preferentially reducing photoevaporation towards less evolved sources (higher accretors).

(ii) We directly observe a weakening of the hotter, likely inner, molecular wind with decreasing accretion rate, evidenced by the declining detection rate of extended H$_2$ S(5) and S(7) winds (see Figure \ref{fig:s1_s5_hist_macc}), which trace ($E_{\rm up} \sim 4500$ K and $7200$~K) hotter inner wind layers compared to S(1) ($E_{\rm up} \sim 1000$ K), which is more similar in excitation to \neii{} ($E_{\rm up} \sim 1100$ K). This reduced inner molecular wind will lead to less shielding of high-energy stellar photons as disks evolve, allowing more photons to reach the outer disk.

(iii) Our data indicate that the jets subside earlier than the extended H$_2$ winds because we identify 12 preferentially lower accretors ($\dot{M}_{acc} \lesssim 10^{-8.5}~M_{\odot}~yr^{-1}$) that exhibit an extended H$_2$ wind without a corresponding \oi{} or \neii{} jet (Section \ref{sec:basic_stats} and \ref{sec:OI_NeII_H2_comparison}). In these 12 sources, we find spatially compact or slightly resolved \neii{} emission, as opposed to extended wind-like morphologies in H$_2$, suggesting an origin of \neii{} emission closer to the star compared to H$_2$. This may point to a change in the origin of the wind.

(iv) Finally, by leveraging the subset of MIRI targets with literature high-resolution \oi{} spectroscopy data (Section \ref{sec:OI_NeII_H2_comparison} and Figure \ref{fig:H2_NeII_OI_comparison}), we find that the \oi{} winds (LVC) persist at lower $\dot{M}_{acc}$ than the H$_2$ molecular winds for any stellar mass bin. Together with the low detection fraction of inner molecular winds (traced by H$_2$~S(5) and S(7)) and preferential tracing of winds in \neii{} (opposed to jets) towards more evolved disks, this suggests a transition from molecular winds to predominantly atomic winds.

These results provide empirical support for the disk wind evolution sketched in \citet{Pascucci2020}. Crucially, this work characterizes the detection frequency of molecular winds in mostly Class~II disks, traces the evolution of the inner wind, and defines the $\dot{M}_{acc}$ regimes where the jet and the molecular and atomic winds dominate.

\subsection{Connections to Outflows from Class 0 and I Disks} \label{sec:connections_to_class0_discussion}

In the past, spatially resolved outflows have been more commonly detected and hence, studied, towards Class~0/I protostars than Class~II \citep[e.g.,][]{Bally1983,Reipurth2004,Frank2014,Bally2016}. High-velocity collimated molecular jets in CO, SiO, and SO (at mm wavelengths) are detected towards Class~0 sources \citep[e.g.,][]{Lee2020,Podio2021}, whereas, atomic jets are more frequently found towards Class~I sources \citep{Hartigan1995,Nisini2005} with a molecular component also found in ro-vibrational H$_2$ lines \citep{Davis2001,Davis2011}. Class~0 and I sources are also known to have wind-like morphologies on small scales ($<$ 500~au) detected primarily at mm wavelengths, e.g., in SO and CO \citep[e.g.,][]{Bjerkeli2016,Tabone2017,Lee2020}. While a possible interpretation of some of these wind-like morphologies is ambient material swept up by a fast wide-angle X-wind \citep[e.g.,][]{Shang2020,Shang2023}, the comparison of nested morphologies in a handful of Class~0, I, and II disks points to radially extended MHD winds \citep[][, in PPVII]{Pascucci2023}. JWST now enables us to extend these comparisons at IR wavelengths.

Recently, \cite{Francis2026} analyzed JWST MIRI data of 16 Class~0 and 7 Class~I sources, all of which have mass accretion rates $>$10$^{-7}$~M$_{\odot}$~yr$^{-1}$, greatly complementing our sample of mostly Class~II disks with accretion rates $<$10$^{-7}$~M$_{\odot}$~yr$^{-1}$. They found spatially resolved atomic/molecular jets and molecular winds towards all but one Class~I source, leading to an outflow detection rate of $>$95\%. This extends the increasing trend of outflow detection rates with increasing mass accretion rate, seen here at lower $\dot{M}_{acc}$, up to $>$10$^{-5}$~M$_{\odot}$~yr$^{-1}$ (Section \ref{sec:trend_with_accretion} and Figure \ref{fig:fraction_disks_outflows_macc}).

At Class~0, \cite{Francis2026} found that high-J H$_2$ lines (e.g., S(7)) trace narrow jets with shocks and knots, whereas at Class~I, jets become atomic \citep[][Ressler et al. 2026, in prep]{Tychoniec2024,vanDishoeck2025} and high-J H$_2$ lines trace wider winds alongside low-J lines. At Class~II, this trend continues: high-J and low-J H$_2$ lines trace broader winds (see Figure~\ref{fig:multi}), while jets are detected in atomic and ionized lines such as \oi{} and \neii{} (see Figures~\ref{fig:velocity_maps} and \ref{fig:H2_NeII_OI_comparison}). Additionally, towards lower $\dot{M}_{acc}$ ($\lesssim$10$^{-8.5}$~M$_{\odot}$~yr$^{-1}$), we identify the transition of winds from being dominantly molecular to atomic (Section~\ref{sec:atomic_molecular_evolution_discussion}). Along with the previously identified transition from atomic to ionized \citep{Pascucci2020}, this mirrors the evolution of jets from molecular to atomic to ionized proposed by \cite{Nisini2015}, but takes place at a relatively later stage.

At the Class~0 stage, it is argued that jet mass-loss rates must be high enough for molecules to form \citep{Glassgold1991} and remain sufficiently shielded from FUV photodissociation \citep{Tabone2020} for jets to stay molecular. A similar argument might apply to the molecular winds seen at the Class~II stage: following the empirical evolution described in Section~\ref{sec:atomic_molecular_evolution_discussion}, these winds are likely dense enough to block the majority of EUV and X-ray photons from penetrating and ionizing species such as argon and neon further out. Attenuation of X-rays by the winds will also protect H$_2$ against photoionization \citep{Nakatani2026}. We will verify this by estimating the H$_2$ mass-loss rates for all 46 molecular winds in a future publication.



Finally, \cite{Francis2026} found the H$_2$ wind semi-opening angle to increase from 10$^o$ to 45$^o$ from Class~0 to Class~I in the S(1) line, and \cite{Narang2026b} found most wind semi-opening angles for 10 Class~II sources to fall between 40$^o$ and 50$^o$, with a median of 45$^o$, albeit in the higher excitation line of S(5) \citep[hence, likely narrower than S(1), e.g.,][]{Nisini2024b,Schwarz2025,Narang2026a}. While we focus on statistics here and defer opening angle calculations to a future study, some sources clearly exhibit even wider flows. For example, SY~Cha \citep[50--70$^o$;][]{Schwarz2025}, J1623-2302, WX~Cha, HV~TauC, and CX~Tau (and possibly GY92-21) required a triangular extraction region with a semi-opening angle of 60$^o$ rather than 50$^o$ to capture the wind emission (see Section~\ref{sec:wind_classification}). Conversely, few outliers such as HH~30 \citep[with NIRSpec,][]{Pascucci2025}, Elias~2-20 \citep{Narang2026b}, and J1628-2431 are narrower than 40$^o$ (see also Figure \ref{fig:h2_wind_maps} where they are visibly narrower than others).

\section{Summary and Conclusions} \label{sec:conclusion}

In this work, we have presented a systematic analysis of \neii{} and H$_2$ S(1), S(3), S(5), and S(7) emission in a sample of 72 inclined (\textit{i}$>$40$^{\circ}$) mostly Class~II protoplanetary disks observed with JWST MIRI MRS. We developed a new framework for identifying extended emission and classifying it as wind-like or jet-like in H$_2$ and \neii{}, respectively. Other than winds, we found H$_2$ tracing extended emission aligned along the disk PA and all binary systems with spatially well-resolved companions in the continuum exhibiting a bridge in H$_2$, suggesting binary gas interactions (Figure~\ref{fig:multi}). Our main findings on jets and winds are as follows:

\begin{enumerate}

    \item \textbf{Extended emission is common and predominantly traces outflows.} The fraction of sources exhibiting spatially extended emission in our sample is as high as $\sim$89\% in the H$_2$ lines (64/72) and $\sim$79\% in the \neii{} line (57/72) (Figure \ref{fig:det_ext_out_stat} and Table \ref{tab:summary_table}). Cone-like morphologies consistent with disk winds are detected in 46 out of 64 sources with extended H$_2$ emission. Collimated high-velocity jets, with \neii{} emission extended perpendicular to the disk plane, are identified in 40 out of 57 sources. The other 17 objects with extended \neii{} emission show bulk velocities $<$30~km~s$^{-1}$ and PSF-like morphologies, consistent with marginally resolved winds \citep[e.g.,][]{Bajaj2024}.

    \item \textbf{Co-spatial jet and wind asymmetries as evidence for magnetic launching.} In IRAS~04385 and WSB~52, we found dominantly redshifted jet and molecular wind emissions, indicating intrinsic asymmetry, as the brighter redshifted side cannot be attributed to projection or extinction effects. For other systems (40$^{\circ}$$<$\textit{i}$<$80$^{\circ}$), however, such effects cannot be excluded. Since thermal winds are not expected to produce such asymmetries, the molecular winds in these two systems are likely magnetically driven, consistent with MHD disk-wind launching.

    \item \textbf{For every jet, there is a wind, and their detection rate increases with accretion.} 85\% (34/40) of the sources with \neii{} jet detections exhibit a corresponding extended H$_2$ wind detection. Of the remaining sources, 2 single-star systems have \oi{} observations reported in the literature \citep{Banzatti2019,Nisini2024a}, and a wind (LVC) is detected towards both of them, suggesting that whenever there is a jet, there is a wind (molecular or atomic). In addition, we found that only a small fraction of sources with low accretion rates exhibit extended winds in H$_2$ and jets in \neii{}. This fraction increases gradually with accretion rate, reaching as high as 80\% for accretion rates $>$10$^{-8}$~M$_{\odot}$~yr$^{-1}$ (Figure \ref{fig:fraction_disks_outflows_macc}). We verified that this trend is not driven or affected by variations in disk inclination or stellar mass across our sample.

    \item \textbf{Winds evolve from being predominantly molecular to predominantly atomic as disks evolve.} Comparing with literature \oi{}~6300\text{\AA} high-resolution spectroscopy results, available for 36\% of our sample, we found that winds towards higher accretors ($>$10$^{-8.5}$~M$_{\odot}$~yr$^{-1}$) are preferentially detected in both atomic and molecular tracers, whereas those towards lower accretors lack extended H$_2$ wind detections. In addition, we found upto 75\% fewer detections of hot, inner molecular winds traced by H$_2$~S(5) and S(7) compared to colder molecular winds traced by H$_2$~S(1), towards lower accretors.
    
\end{enumerate}

Combining all results, we identify an evolutionary sequence in molecular disk wind tracers that, together with the results (and proposed sequence) of \citet{Pascucci2020} on atomic tracers, can be summarized as follows. At earlier stages of Class~I/II disk evolution, when accretion rates are high, an atomic jet (\oi{}, \neii{} HVC) is driven alongside a hot, inner atomic (\oi{} LVC) and molecular (H$_2$, predominantly hot, S(5) and S(7)) wind. We argue that these flows are MHD-disk-winds (Section~\ref{sec:H2_MHD_driven}) and are likely dense enough to shield the outer disk from high-energy stellar photons, thereby suppressing photoevaporation at larger radii as supported by the lack of \neii{} LVC \citep{Pascucci2020}. As disks evolve toward lower accretion rates, the atomic jets fade and the hot inner molecular winds weaken below detection thresholds. The resulting tenuous inner flows could allow high-energy stellar photons to penetrate further into the disk, heating its surface. As a result, a predominantly atomic wind is seen (\neii{} LVC) along with a molecular component further out (H$_2$, predominantly cooler, S(1)), given that the H$_2$ S(1) winds are spatially more extended than \neii{} LVC. With evolution, this molecular contribution diminishes further, leaving behind a mostly atomic wind (\oi{}, \neii{} LVC). 

As the largest systematic study of molecular winds from Class~II disks to date, we conclude that extended molecular disk winds are more prevalent across the Class~II stage of disk evolution than previously recognized. This could have far-reaching implications for disk evolution and planet formation, including but not limited to: slowing inward planet migration through altered disk surface density profiles \citep[e.g.,][]{Bai2016,Ogihara2018,Kimmig2020}, and promoting the streaming instability by enabling accretion to proceed with reduced midplane turbulence \citep[via MHD wind-driven accretion, e.g.,][]{Tabone2022} while simultaneously enhancing the local dust-to-gas ratio through the vertical removal of dust-depleted gas by disk winds \citep[MHD or photoevaporative;][]{Gorti2015,Bai2016,Carrera2017}. Furthermore, JWST MIRI observations of spatially resolved outflows are capable of providing crucial constraints on models of jet and wind launching.

\begin{acknowledgments}

This work is based on observations made with the NASA/ESA/CSA James Webb Space Telescope. The JWST data presented in this article were obtained from the Mikulski Archive for Space Telescopes (MAST) at the Space Telescope Science Institute. The specific observations analyzed can be accessed via \dataset[doi: 10.17909/efwr-c167]{https://doi.org/10.17909/efwr-c167}. This material is based upon work supported by the National Aeronautics and Space Administration under Grant No. 80NSSC25K0301 issued through the NNH24ZDA001N-FINESST program. Any opinions, findings, and conclusions or recommendations expressed in this article are those of the author(s) and do not necessarily reflect the views of the National Aeronautics and Space Administration. N.S.B. and I.P. acknowledge partial support from NASA/STScI GO grants JWST-GO-01621.001 and JWST-GO-02260.001. This work has been carried out within the framework of the NCCR PlanetS supported by the Swiss National Science Foundation under grant 51NF40\_205606. A.D.S. acknowledges support from the ERC grant 101019751 MOLDISK. R.A.~acknowledges funding from the Science \& Technology Facilities Council (STFC) through Consolidated Grant ST/W000857/1. The authors acknowledge the use of GitHub Copilot through the Visual Studio Code extension to assist with debugging and optimizing Python data analysis codes, and Claude 4.6 to assist with grammar refinement. All AI-assisted outputs were reviewed, verified, and edited by the authors.

\end{acknowledgments}

\facilities{James Webb Space Telescope}

\software{Astropy \citep{astropy:2013,astropy:2018,astropy:2022}, Claude Opus 4.6 \href{https://www.anthropic.com/system-cards}{(Anthropic)}, Copilot 0.41 \href{https://github.com/features/copilot}{(GitHub)}, JWST \citep{Bushouse2026}, LMFIT \citep{lmfit}, Matplotlib \citep{Hunter:2007}, NumPy \citep{harris2020array}, Pandas \citep{pandas}, Photutils \citep{photutils}, Scikit-learn \citep{scikit-learn}, SciPy \citep{2020SciPy-NMeth}, VSCode \href{https://code.visualstudio.com/}{(Microsoft Corporation)}}

\appendix

\section{H2 S(1), S(3), S(5), S(7) and \neii{} line maps with nearby continuum images}
\label{sec:line_cont_maps_all}

In this section, we show pixel-by-pixel continuum-subtracted and intensity-integrated line maps (see Figures \ref{fig:multi}) for each of the H$_2$ S(1), S(3), S(5), S(7) and \neii{} lines. On the line maps, the centroid of the corresponding continuum is highlighted with a red star, and the known disk position angle (along with the known error) is shown as a white line passing through the continuum centroid. The line maps are overlaid with cyan contours of the same map encircling 3$\sigma$, 15$\sigma$, 90$\sigma$, and 300$\sigma$ emission, where $\sigma$ is the standard deviation of the background calculated iteratively. Next to each line map is a map of the continuum at the same wavelength. The same line emission contours as those overlaid on the line maps are overlaid on these continuum maps to make a direct comparison of the shape of the line emission with the continuum. The sources marked with an asterisk are further discussed in Section \ref{sec:individual_sources}.

\begin{figure}[p]
    \centering
    \includegraphics[width=0.95\textwidth]{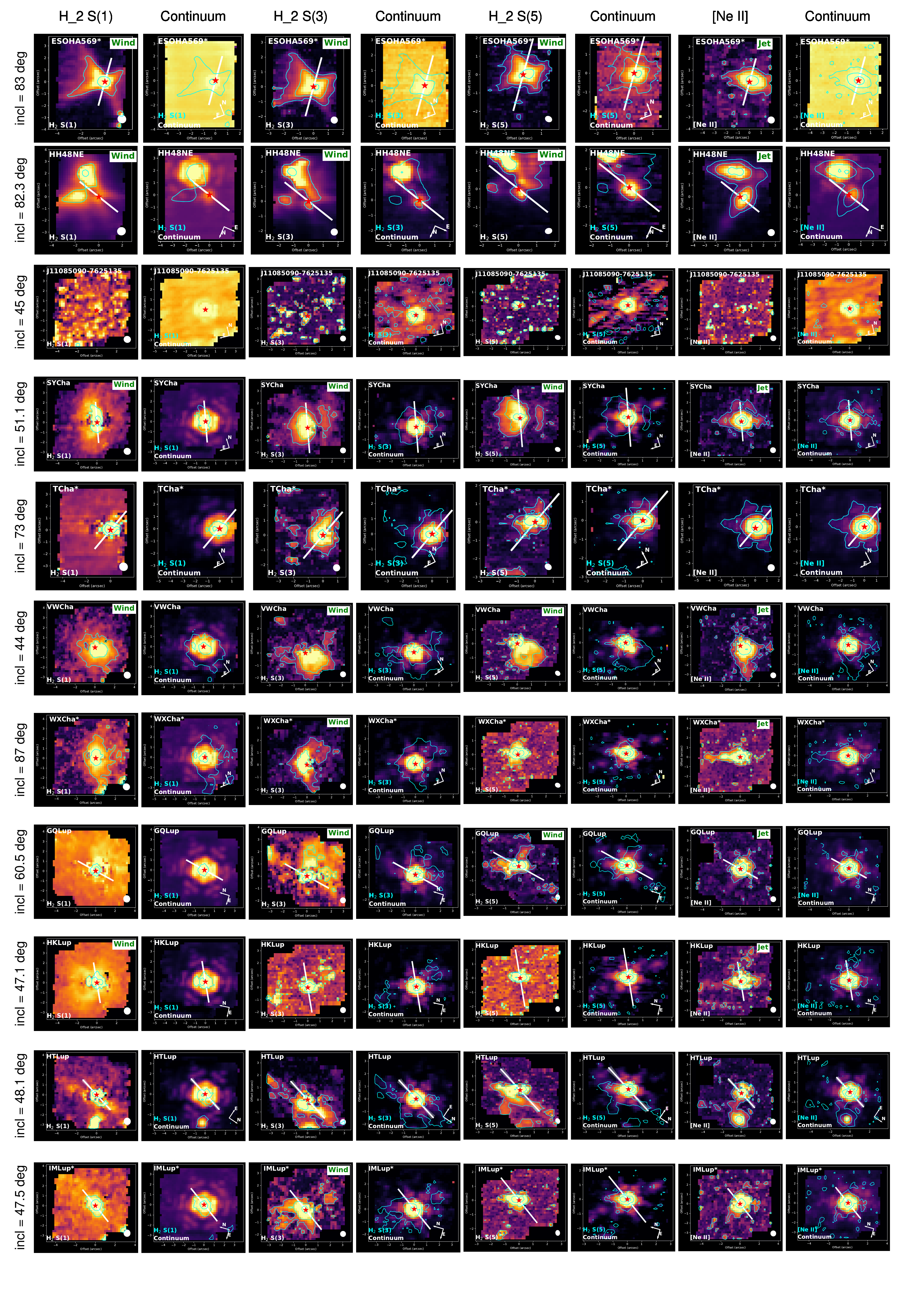}
    \caption{Caption provided at the end.}
    \label{fig:multi}
\end{figure}

\begin{figure}[p]
    \ContinuedFloat
    \centering
    \includegraphics[width=\textwidth]{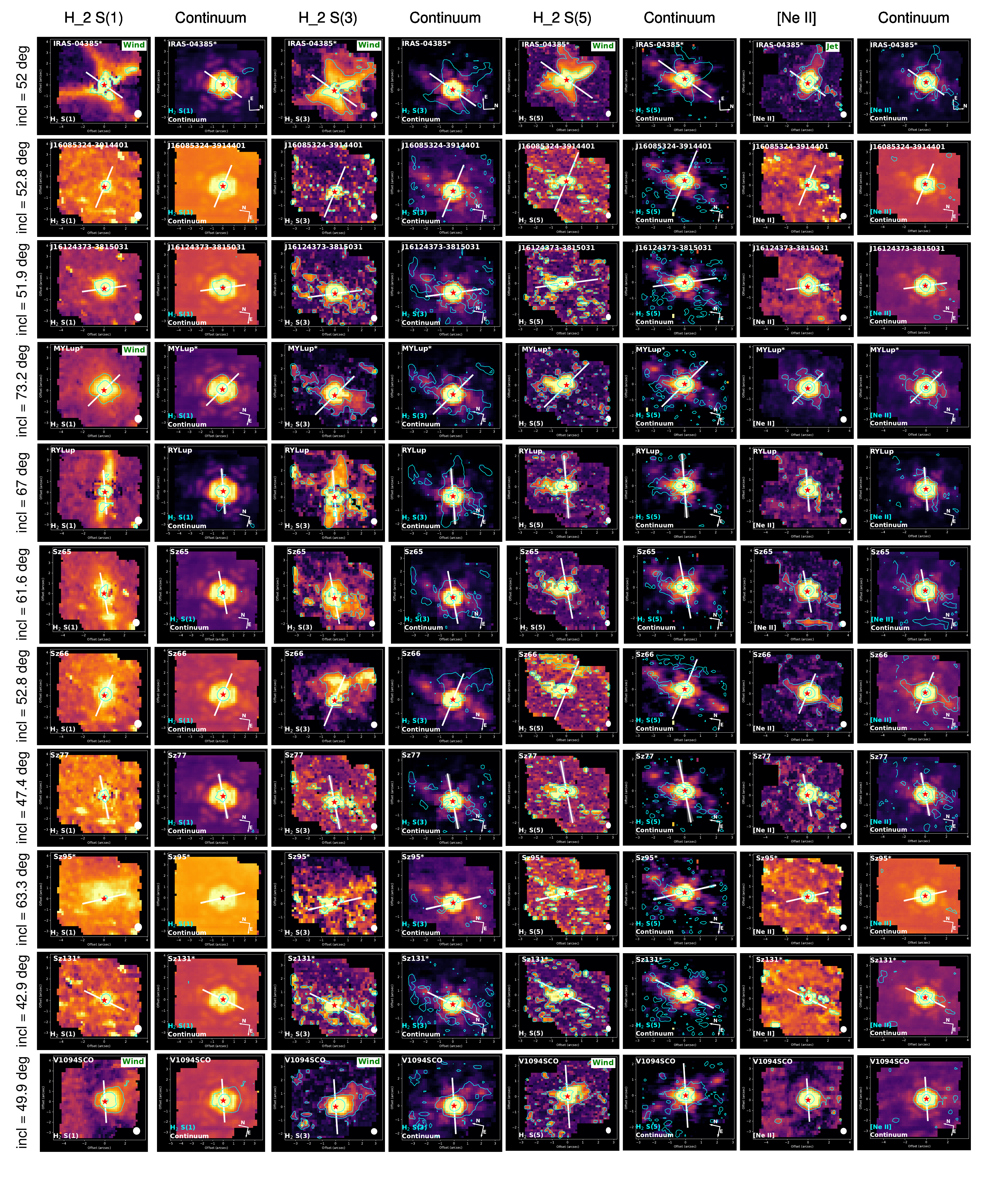}
    \caption{(continued)}
\end{figure}

\begin{figure}[p]
    \ContinuedFloat
    \centering
    \includegraphics[width=\textwidth]{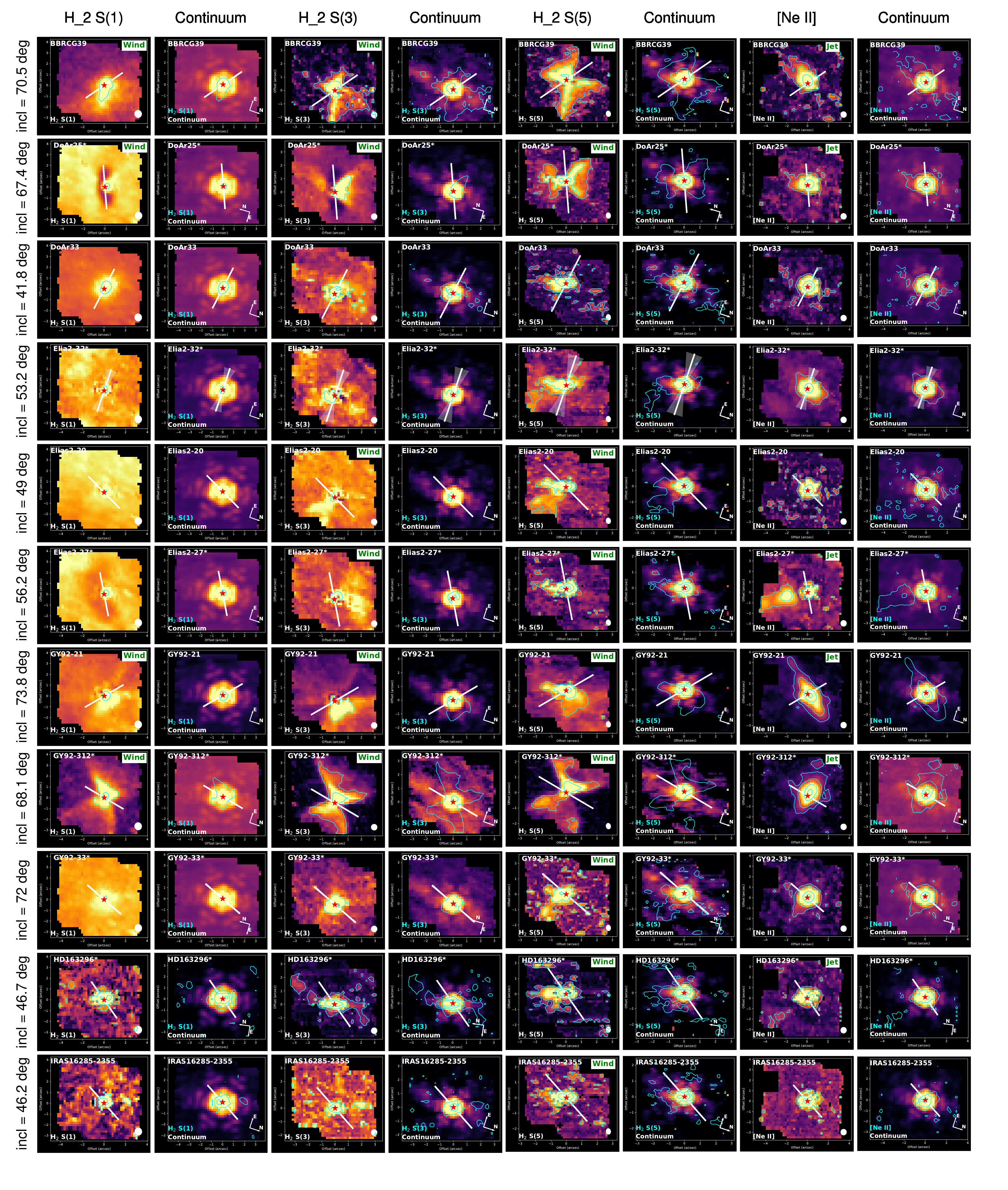}
    \caption{(continued)}
\end{figure}

\begin{figure}[p]
    \ContinuedFloat
    \centering
    \includegraphics[width=\textwidth]{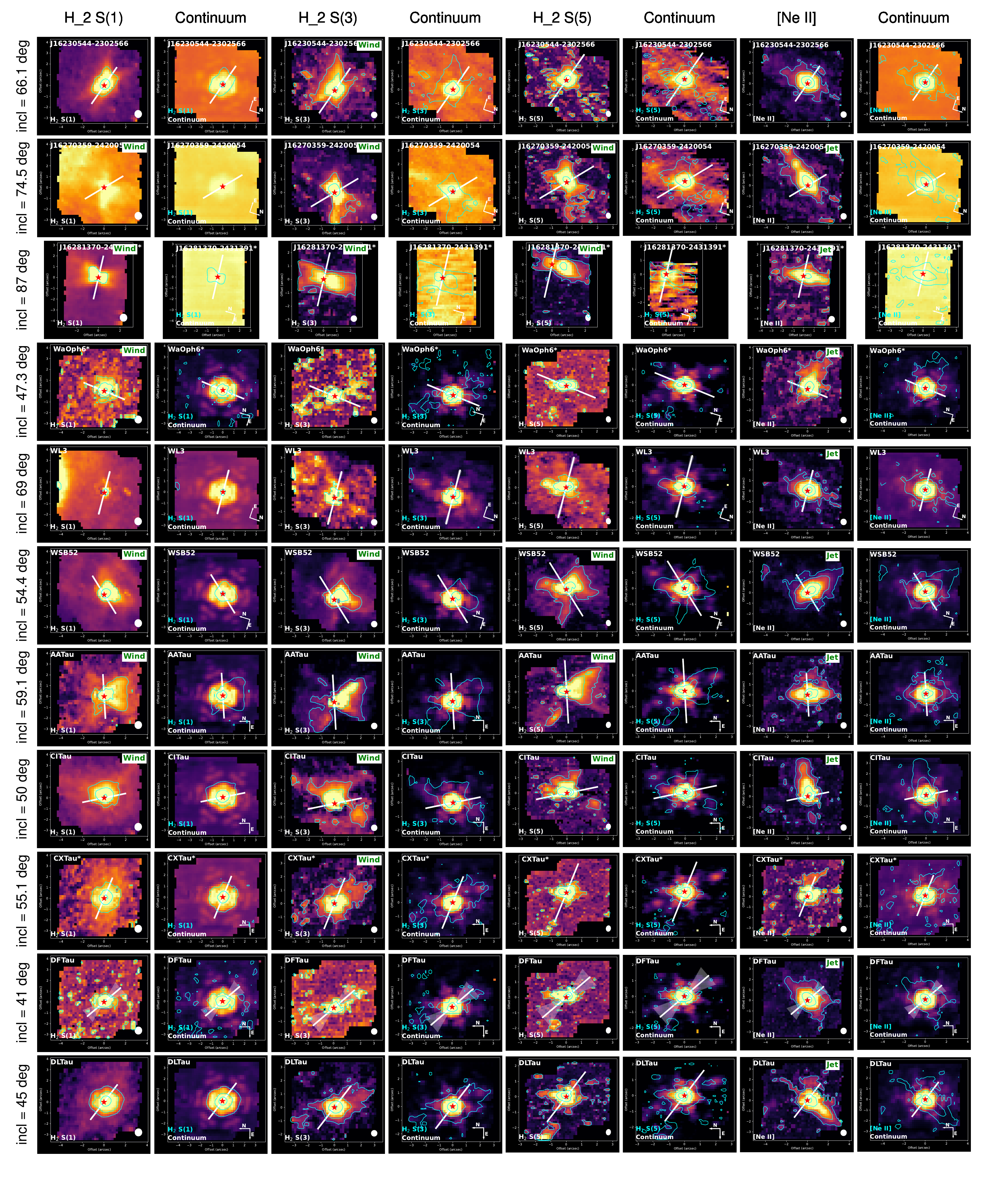}
    \caption{(continued)}
\end{figure}

\begin{figure}[p]
    \ContinuedFloat
    \centering
    \includegraphics[width=\textwidth]{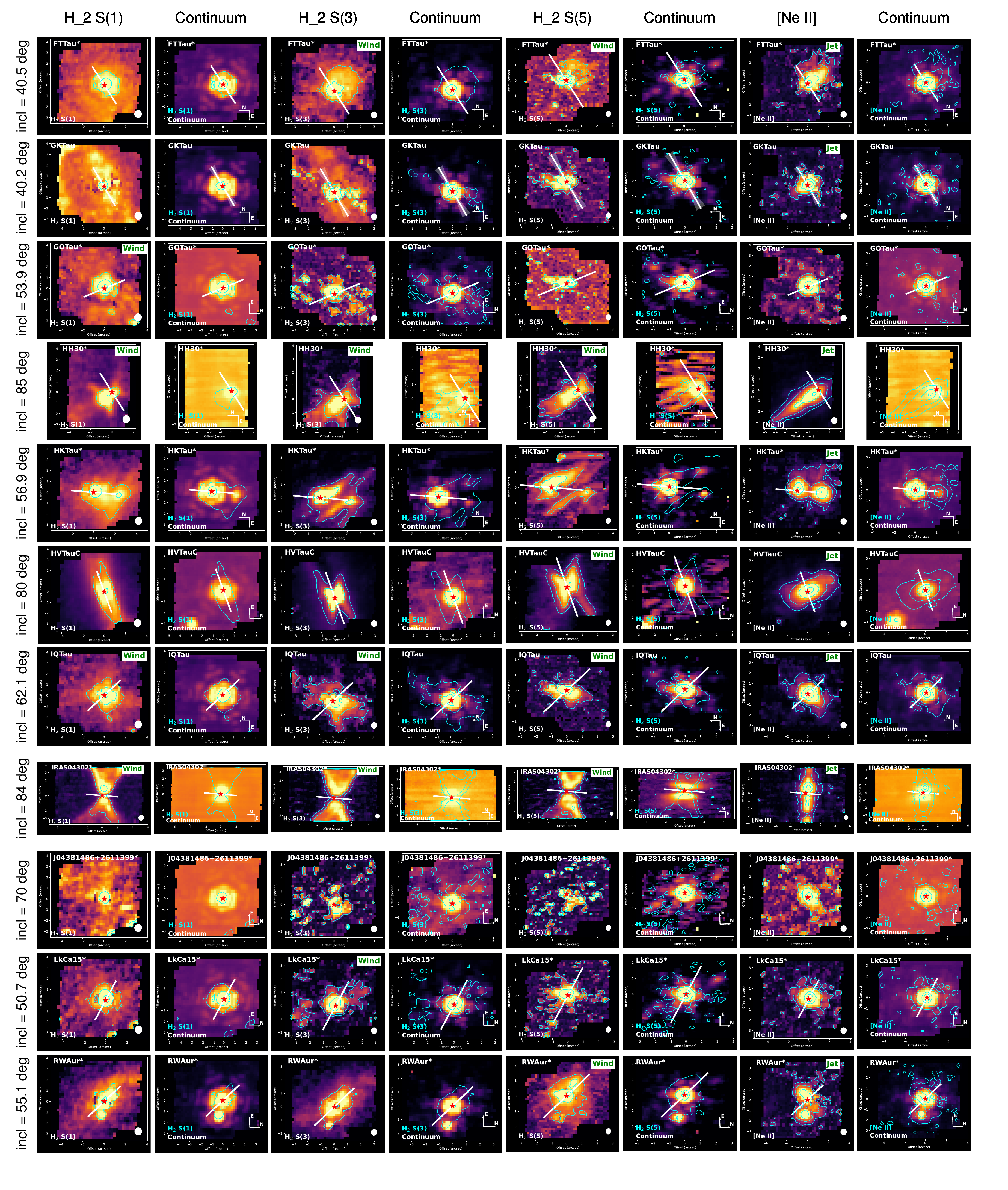}
    \caption{(continued)}
\end{figure}

\begin{figure}[p]
    \ContinuedFloat
    \centering
    \includegraphics[width=\textwidth]{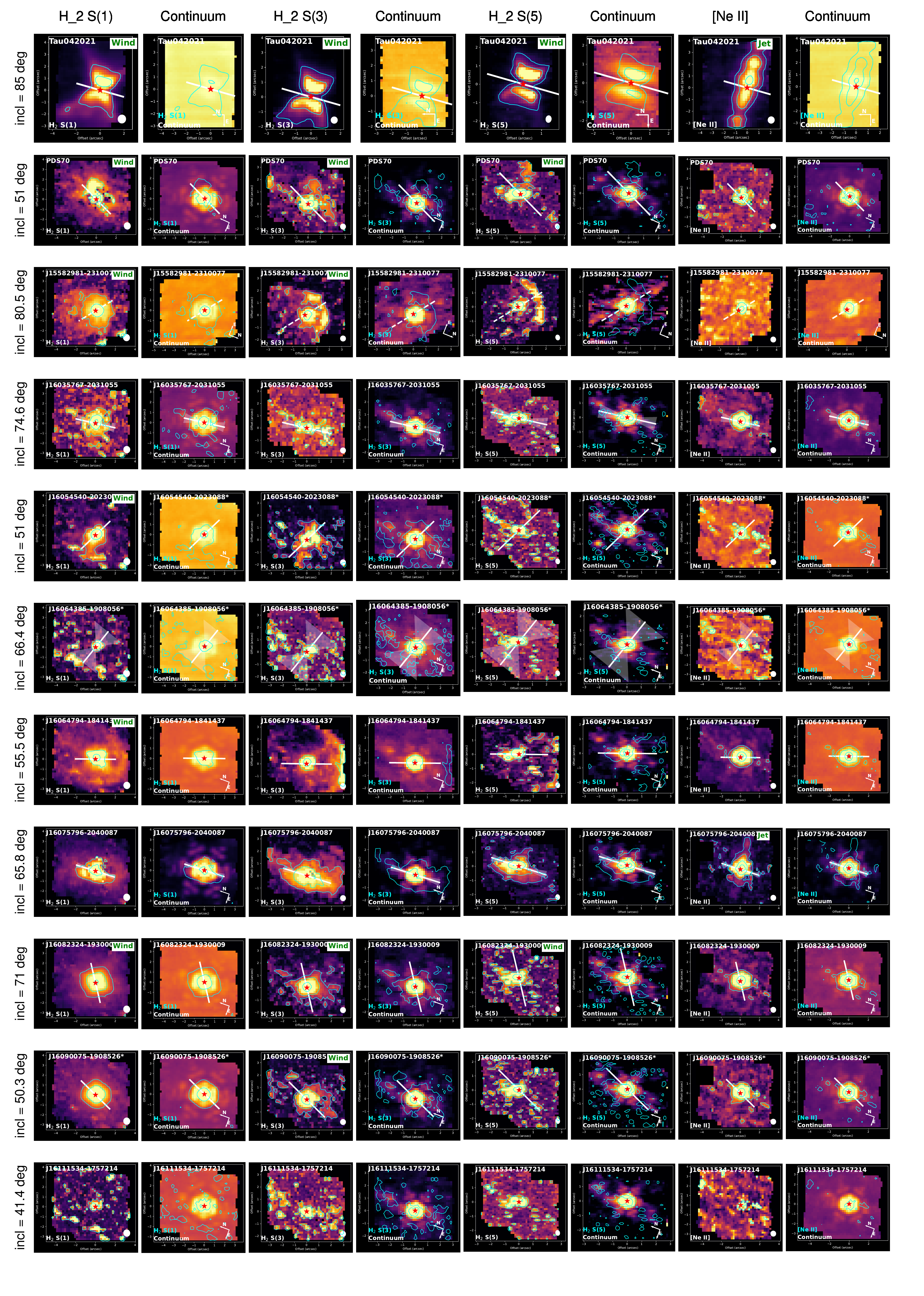}
    \caption{(continued)}
\end{figure}

\begin{figure}[p]
    \ContinuedFloat
    \centering
    \includegraphics[width=\textwidth]{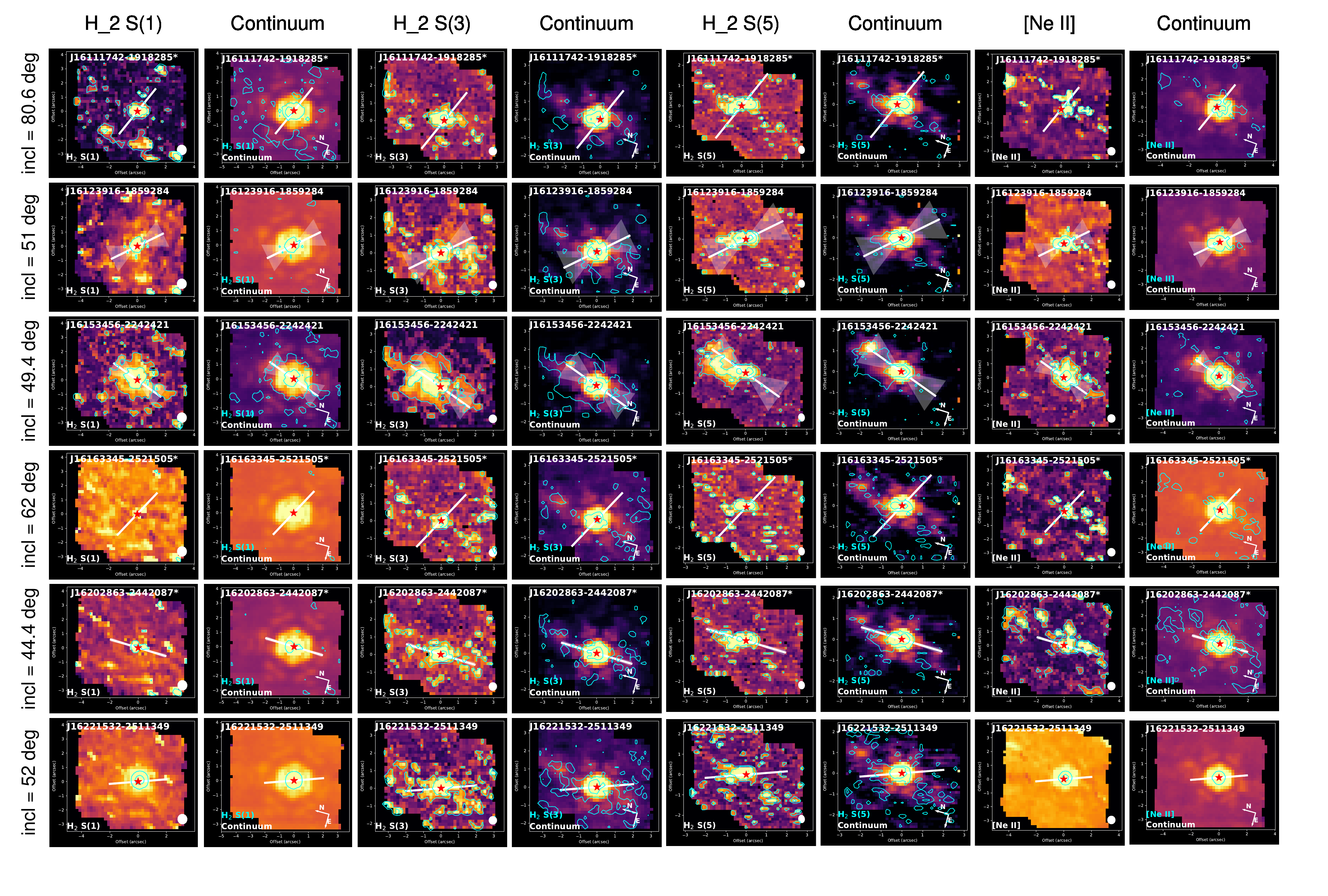}
    \caption{Continuum-subtracted and line-integrated intensity maps of H$_2$~S(1), S(3), S(5), and \neii{} lines are shown along with nearby continuum slices overlaid with line contours. All maps are displayed using an asinh normalization with a maximum percentile cut of 97.5. For maps with peak values below 35$\times \sigma$, a linear normalization with a maximum percentile cut of 97 is used instead. An example of this latter case is the continuum map of J11085090$-$7625135 at the H$_2$~S(1) wavelength. The cyan contours overlaid are those of the line emission in both the line maps and the corresponding continuum maps. The contour levels are 3$\sigma$, 15$\sigma$, 90$\sigma$, and 300$\sigma$, where $\sigma$ is the standard deviation of the background calculated iteratively using the \texttt{astropy} tool \texttt{sigma\_clipped\_stats}. The inclination of each disk is provided at the left end of the row. For many of these nearly edge-on disks, the continuum (or line map) is resolved into two disk surfaces (or wind lobes) separated by a dark lane, whose midpoint corresponds to the stellar position. In such cases, the stellar location is not reliably captured by a 2D Gaussian centroid. For these sources (incl $\geq$ 80$^{\circ}$), we adjust the stellar position to approximately the center of the dark lane. The sources are arranged in the same order as in Tables \ref{tab:source_properties} and \ref{tab:source_properties_references}. The source names in this figure are searchable. The sources with an asterisk are further discussed in Section \ref{sec:individual_sources}.}
\end{figure}

\begin{figure*}
    \ContinuedFloat
    \centering
    \includegraphics[width=\linewidth]{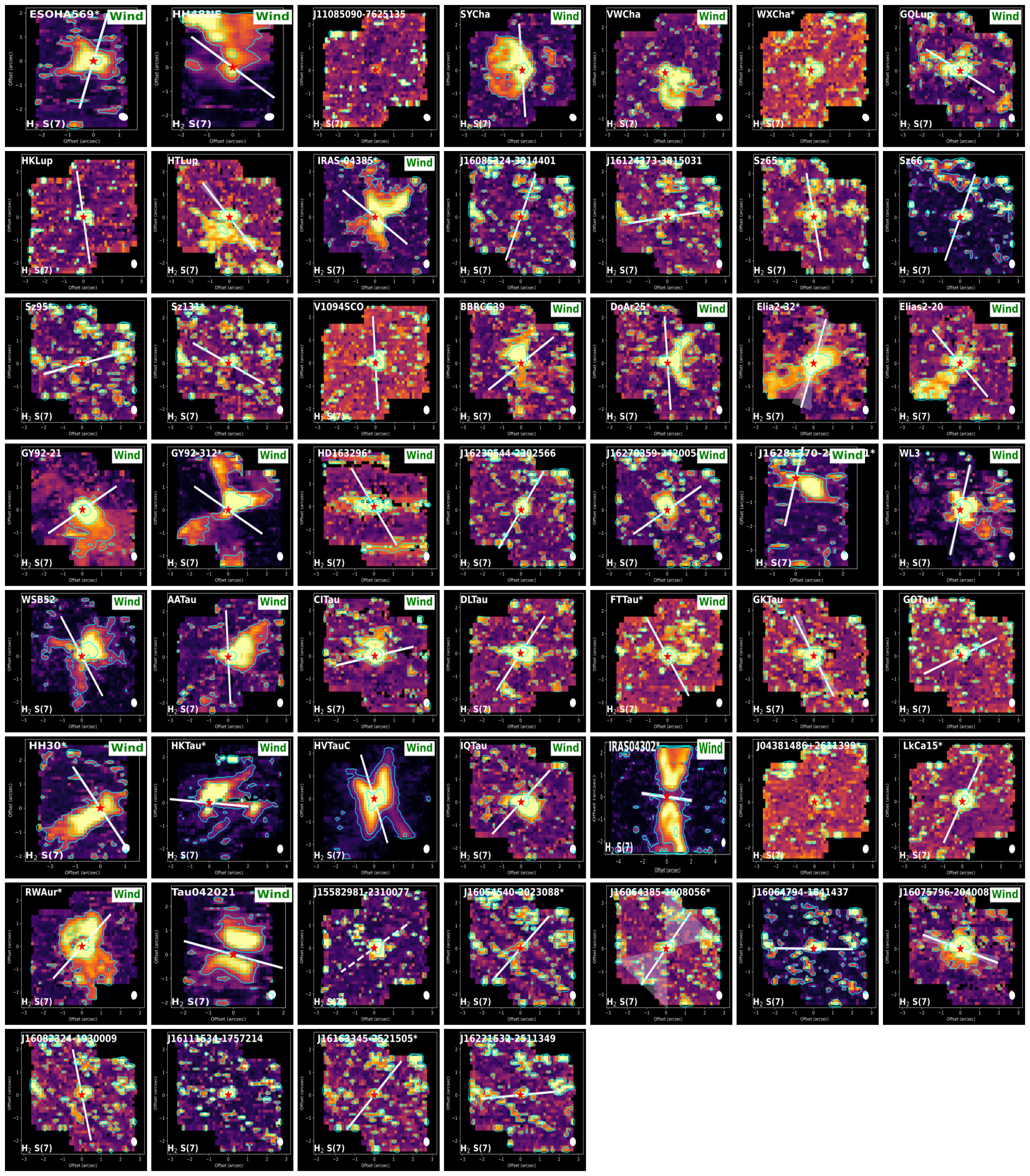}
    \caption{Integrated intensity maps of the H$_2$ S(7) line for sources with detected S(7) emission. The maps are displayed in the same manner as above.}
\end{figure*}

\section{\neii{} spectra and Gaussian fits for sources with extended \neii{} emission}
\label{sec:neii_spectra}

\begin{figure*}
    \centering
    \includegraphics[width=\linewidth]{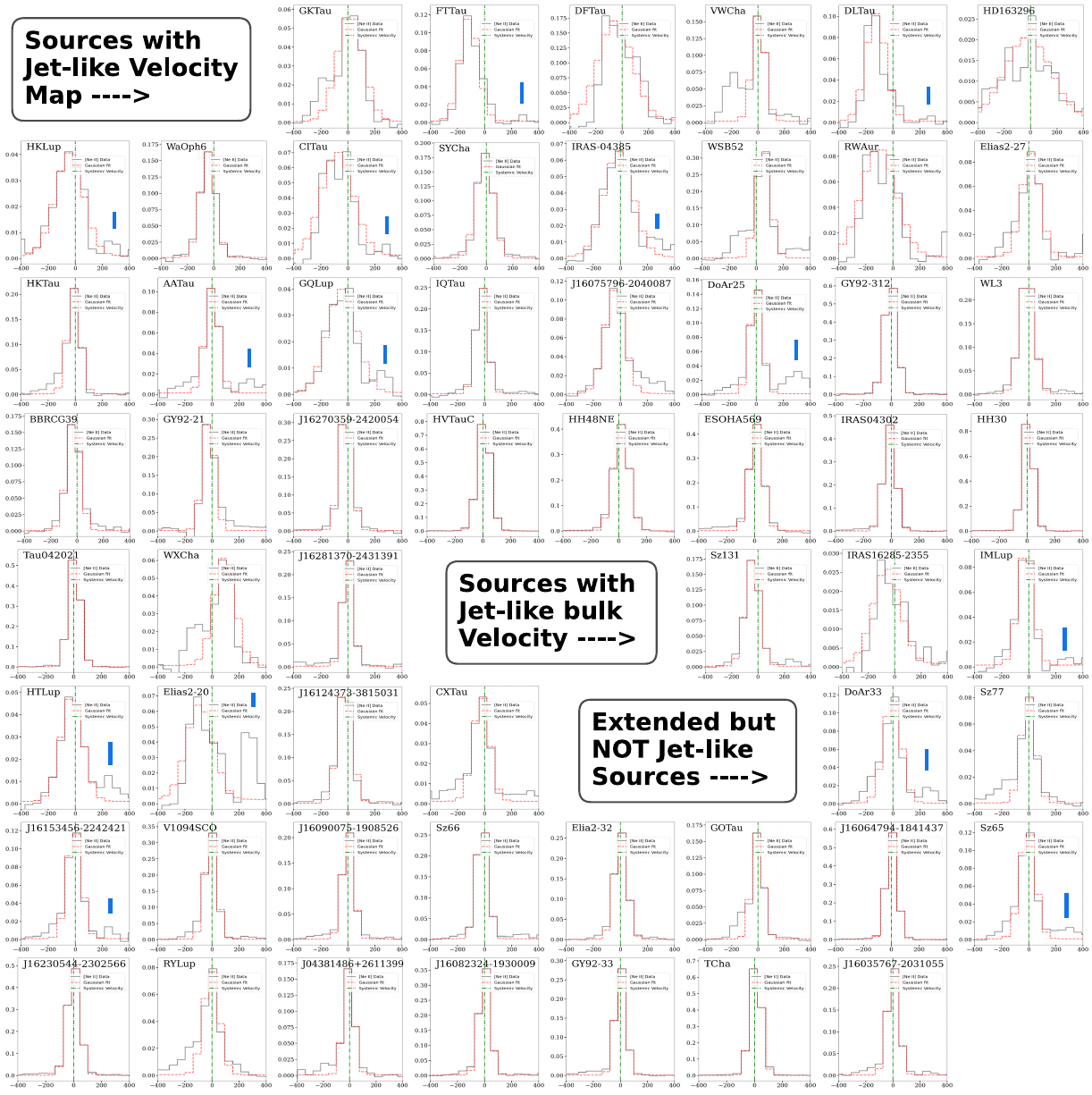}
    \caption{\neii{} spectrum for sources with extended \neii{} emission plotted in the units of velocity (km~s$^{-1}$). The sources are arranged the same way as in Figure \ref{fig:velocity_maps}. For each source, the continuum-subtracted line emission is shown in gray, and the best-fit 1D Gaussian is shown in red with the systemic velocity (shifted to 0) highlighted in green. Emission beyond 200~km~s$^{-1}$ from the systemic velocity arises from H$_2$O and is highlighted with a blue vertical line in cases where it is particularly strong.}
    \label{fig:neii_spectra_fit_extended_sources}
\end{figure*}

Here, we present the \neii{} spectra (Figure \ref{fig:neii_spectra_fit_extended_sources}) for all targets for which \neii{} velocity maps are shown in Figure \ref{fig:velocity_maps}. We first convert the wavelength axis to velocity using the \neii{} rest wavelength of 12.813548~\micron{}, retrieved from the latest version of the NIST Atomic Spectra Database \citep{NIST_ASD_2024}. We then correct the spectra for the systematic offset ($-3~\mathrm{km~s^{-1}}$) and for the stellar radial velocity. For stars without published radial velocities, we instead adopt the average radial velocity of the host star-forming region (see Section \ref{sec:jet_classification} and Table \ref{tab:source_properties} for details). Finally, we iteratively fit a single-component 1D Gaussian to the data using a least-squares method and overplot the best-fit model on the spectra in Figure \ref{fig:neii_spectra_fit_extended_sources}.

\section{Channel 1 PSF artifact}
\label{sec:chan1_psf_artifact}

\begin{figure*}
    \centering
    \includegraphics[width=0.8\linewidth]{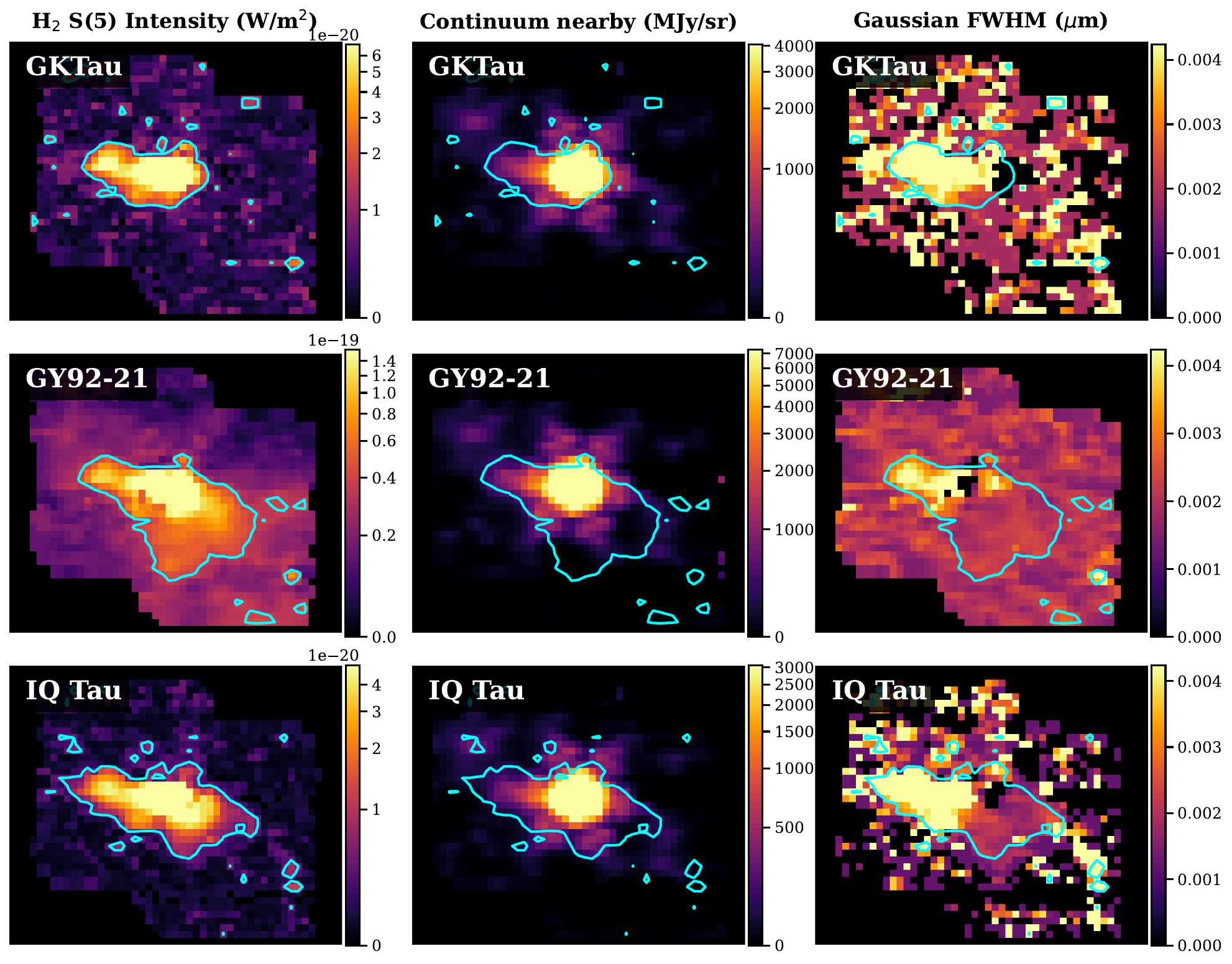}
    \caption{The left column shows the pixel-by-pixel line-integrated H$_2$~S(5) intensity map, with the 5$\sigma$ contour overlaid in cyan, where $\sigma$ is the background standard deviation calculated iteratively. The middle column presents a continuum slice near the S(5) line wavelength for the respective targets. The right column displays the full width at half maximum of the 1D Gaussian fitted at each pixel during the construction of the intensity maps. The FWHM is only shown for pixels with Gaussian fits stronger than 3$\sigma$. The cyan contours remain identical across each row.}
    \label{fig:artifact}
\end{figure*}

We have identified that, at low MIRI MRS wavelengths, particularly in Channel~1, an unaccounted PSF artifact significantly affects the interpretation of extended emission in lines that fall within this wavelength range (e.g., H$_2$~S(5), \arii{}, \hi{}~6-5). The MIRI PSF consists of a central core surrounded by six petal-shaped wings. In the spaxels corresponding to the leftward pointing wing in the image plane, we detect bright emission that is substantially broader than the nominal MIRI MRS spectral resolution at these wavelengths. This effect is illustrated in Figure~\ref{fig:artifact}. The first column presents the H$_2$~S(5) integrated intensity maps for three targets in our sample. At first glance, the S(5) map of GY92-21 appears to show an azimuthal asymmetry in the wind emission on one side of the disk plane, with only the left arm in emission. However, such a wind morphology has not been reported previously and is difficult to interpret physically. The second column shows the continuum emission near the line, which traces the PSF structure. It is clear that the asymmetric feature spatially coincides with the leftward-pointing PSF wing. This behavior is seen consistently in all three targets shown in Figure~\ref{fig:artifact}. The third column shows the FWHM of the Gaussian fitted in each spaxel. We find enhanced FWHM values in exactly the same spaxels that display the asymmetric bright emission. This indicates that the emission in those spaxels is unlikely to have a physical origin in the disks and is instead caused by broad Gaussian-like systematic noise introduced by the artifact. The strength of the FWHM enhancement varies among the targets, with GKTau showing the largest contrast, followed by GY92-21 and IQTau. Although not shown here, we observe the same behavior in the \arii{} 6.985~\micron{} line and in the \hi{}~6-5 line.

\section{Discussion of individual sources}
\label{sec:individual_sources}

\subsection{**HH48NE}

HH48 is a binary star system, with both stars well resolved in the MIRI continuum at all wavelengths. HH48NE is a known edge-on disk, consistent with the morphology of the H$_2$ emission associated with it. Both components drive jets in \neii{} (Figure \ref{fig:velocity_maps}) and winds in H$_2$ (Figure \ref{fig:h2_wind_maps}). Interestingly, all the H$_2$ maps show prominent emission seemingly from a third object that is not detected in the continuum (Figure \ref{fig:multi}). No continuum emission has been observed at this location in the ALMA mm data either \citep{Sturm2023}. Hence, this H$_2$ emission is likely a result of the interaction of winds from the two HH48 sources.

\subsection{**VWCha}

VWCha is a complicated quadruple system with VWChaA being a spectroscopic binary and VWChaBC being a 0.1~arcsec binary. The separation between VWChaA and BC systems is 0.7~arcsec \citep[or 130~au in sky-plane, e.g.,][]{Melo2003,Brandeker2001,Zsidi2022,Kurtovic2026}. In the continuum, the two binary systems are resolved in Channel 1 (see the H$_2$ S(5) continuum map in Figure \ref{fig:multi}). The velocity maps (Figure \ref{fig:velocity_maps}) clearly show two distinct flows with different velocities. The VWChaA system appears to be driving a high-velocity jet, while the BC system is likely driving a low-velocity wind. The direction of the jet and wind is similar to that seen in the H$_2$ maps; however, it remains unclear whether the H$_2$ wind is driven by VWChaA, VWChaBC, or both.

\subsection{**WXCha}

The bulk \neii{} velocity centroid of $\sim$80~km~s$^{-1}$ toward WXCha, combined with the reported disk inclination of 87$^{\circ}$ \citep{Banzatti2015}, implies a deprojected jet velocity of $\sim$1600~km~s$^{-1}$. Such a high velocity is inconsistent with observed T~Tauri jets and difficult to reconcile with current jet-launching models. However, \cite{Banzatti2015} also reported an uncertainty of 31$^{\circ}$ on the inclination. Adopting a lower inclination of 56$^{\circ}$ instead yields a deprojected jet velocity of $\sim$150~km~s$^{-1}$, consistent with typical T~Tauri jet velocities. This interpretation is supported by \cite{Fiorellino2022}, who reported observational characteristics inconsistent with an edge-on geometry. A lower inclination is also favored by the absence of extended scattered light in the SPHERE image \citep{Ginski2024}. Together, these results suggest that WXCha is likely significantly less inclined than 87$^{\circ}$.

\subsection{**HTLup}

HTLup is another complex system, with HTLup A and B separated by 0.16 arcsec and HTLupC located 2.8 arcsec away. With MIRI, we cannot resolve component B from A; however, HTLupC is clearly resolved from the AB system and lies within the field of view (Figure \ref{fig:multi}). The H$_2$ map toward HTLupC is incomplete because it is truncated by the field of view, while the map toward HTLupAB is difficult to interpret, showing emission both along the disk direction and perpendicular to it, likely suggesting contributions from both the A and B components. Interestingly, the \neii{} velocity map shows significantly higher velocities closer to HTLupAB than toward HTLupC, suggesting a possible jet launched from the AB system and a wind from the C system. If confirmed, this would extend the trend first reported by \cite{Kurtovic2026}, in which one component of a binary drives a jet while the other drives a wind in \neii{}.

\subsection{IMLup}

The \neii{} velocity map of IMLup shows highly blueshifted velocities on one side (lower left) of the disk midplane and very low velocities on the other, producing a clear gradient. The emission is oriented perpendicular to the disk midplane, suggesting a jet origin. However, the central axis of the blueshifted emission is not aligned with the continuum centroid, and the emission in the upper right relative to the centroid is too broad to be consistent with a jet. Therefore, we do not classify this map as jet-like. Nevertheless, a blueshifted jet was detected towards this source using VLT/MUSE in \oi{} \citep{Flores-Rivera2023}, consistent with the \neii{} bulk velocity centroid indicating a jet.

\subsection{IRAS-04385}

IRAS-04385 shows a clear bipolar wind in all the H$_2$ lines. However, in the H$_2$ S(1) line, the wind curves outward—away from the central axis—rather than bending inward toward it, and IRAS-04385 is the only source in our sample that shows this morphology. A similar behavior has been observed in cold CO outflows \citep[e.g.,][]{Mitchell1997}. It can likely be attributed to a sharp drop in external pressure, leading to the lateral expansion of the flow.

\subsection{RYLup}

RYLup exhibits significant extended emission along the outer disk position angle in the H$_2$ S(1) and S(3) lines. This emission extends beyond the gas disk size (1.75~arcsec; measured as 2.9 times the millimeter dust disk size), making it an especially interesting object for detailed study. We note that a modest inner disk misalignment has been identified in this object \citep[$\sim$10$^{\circ}$,][]{Bohn2022}, with potential inner disk inclination variability leading to misalignments of up to $\sim$30$^{\circ}$ \citep{Manset2009}. If the observed H$_2$ emission is tracing winds from the inner disk, then the wind emission may align with the outer disk, constraining the wind semi-opening angle to be $\sim$30$^{\circ}$, which is low but in the observed range \citep[e.g.,][]{Pascucci2025}. However, it could also just be tracing the irradiated disk surface.

\subsection{Sz66}

Sz66 is part of a wide-separation binary with Sz65 (also included in this work), with a projected sky-plane separation of $\sim$980 au \citep{Miley2024}. In H$_2$ S(3), it shows two blobs of emission: one aligned with the disk, which could be associated with extended disk emission, and another oriented at an angle away from the disk midplane. The S(5) line shows similarly extended emission on much larger scales (a structure $>$3 arcsec long at a distance of $\sim$2 arcsec from the source), which could plausibly be associated with shocked ISM gas. Interestingly, in \neii{}, the emission appears morphologically jet-like, with narrow, extended emission perpendicular to the disk position angle (Figure \ref{fig:multi}). However, the \neii{} velocity map shows low velocities (Figure \ref{fig:velocity_maps}), with a bulk velocity centroid of $-20.8 \pm 6.7$ km s$^{-1}$ and no velocity gradient above and below the disk midplane, arguing against a jet origin for the extended emission. Upon closer inspection, we find constant continuum emission in that direction (lower right of the centroid) that brightens at the \neii{} wavelength and is likely responsible for the apparent extended \neii{} emission.

\subsection{DoAr25}

DoAr25 shows a clear bipolar wind in all H$_2$ lines (Figure \ref{fig:multi}). The \neii{} map exhibits a slight extension perpendicular to the disk position angle, and the velocity map (Figure \ref{fig:velocity_maps}) shows this extended emission to be redshifted with velocities upto $>$80~km~s$^{-1}$ per pixel. While it is not the clearest case of its category, the emission in \neii{} is consistent with a jet origin. 

\subsection{Elia2-32}

The velocity map of Elia2-32 (Figure \ref{fig:velocity_maps}) reveal a \neii{} velocity gradient, with redshifted emission toward the upper left and blueshifted emission toward the lower right, giving the appearance of a jet. The axis of this potential jet is inclined by $<70^{\circ}$ from the disk PA, which itself is uncertain by 15$^{\circ}$. This suggests that Elia2-32 could host a jet, and the misalignment could be a result of uncertainty in disk PA. However, the observed velocities are lower than those typically expected for a jet and are instead consistent with the uneven slice illumination effect seen towards several other sources in this category. 

\subsection{Elias2-27}

Elias 2-27 shows a highly unusual jet/wind morphology: the H$_2$ S(3) emission traces a unipolar wind toward the blueshifted side, while the \neii{} emission reveals a unipolar jet toward the red-shifted side. ALMA studies of this disk have revealed spiral structures and a strongly perturbed vertical gas structure, consistent with infall-triggered gravitational instabilities \citep{Paneque-Carreno2021}. It is therefore possible that a combination of infall and outflow is contributing to the observed H$_2$ morphology. At the same time, the jet, launched from the very inner regions of the disk, could be asymmetric due to the asymmetric magnetospheric star-disc interaction \citep{Takasao2022}, whereas the blueshifted molecular wind, launched from further out, could be so due to the non-ideal MHD effects in the dead zone \citep{Bethune2017}.

\subsection{GY92-21}

In H$_2$ S(5), GY92-21 appears to show an asymmetric, one-armed wind in the red-shifted direction (Figures \ref{fig:multi} and \ref{fig:velocity_maps}). However, this one-armed feature aligns closely with the PSF artifact discussed in Section \ref{sec:chan1_psf_artifact} and likely does not have a physical origin. This interpretation is supported by the H$_2$ S(1) and S(3) lines, which display a bipolar wind without any indication of one red-shifted arm being brighter than the other.

\subsection{GY92-33}

The \neii{} emission toward this source is extended perpendicular to the disk PA, suggesting a jet (Figure \ref{fig:multi}). However, the velocity map shows neither a gradient across the disk midplane nor velocities high enough to be consistent with a jet (Figure \ref{fig:velocity_maps}). Therefore, we classify the \neii{} emission as not jet-like.

\subsection{HD163296}

HD163296 is the most massive star included in our sample. Since it is massive and bright, it was observed with short exposure times in an attempt to minimize saturation and its effect on the data, which has led to different artifacts in the data. E.g., the horizontal streak in the H$_2$~S(5) intensity map at the bottom is likely an artifact. Similarly, it's \neii{} line profile has also been affected (Figure \ref{fig:neii_spectra_fit_extended_sources}). We identify a \neii{} jet and an extended H$_2$ wind in this source, which is consistent with previous detections of an atomic jet \citep{Wassell2006,Ellerbroek2014,Xie2021,Kirwan2022} and a slow, wide molecular cold CO wind \citep{Klaassen2013} that is rotating \citep{Booth2021}. While this source is part of our sample, we excluded it from the empirical results derived in this work.

 \subsection{J16230544-2302566}

The H$_2$ S(1) and S(3) maps of 2MASS~J16230544–2302566 show broad, conical emission on both sides of the disk (Figure \ref{fig:multi}). The S(3) emission is classified as a wind only when adopting a wider cone semi-opening angle of 60$^{\circ}$. However, this morphology could also arise from an irradiated flared disk surface viewed edge-on. High-resolution ALMA CO imaging with kinematic information could help distinguish between these scenarios. The \neii{} emission towards this source morphologically looks jet-like (Figure \ref{fig:multi}); however, none of the pixels in its velocity map (Figure \ref{fig:velocity_maps}) show velocities $>$50~km~s$^{-1}$. The low velocities seen are instead consistent with the slice illumination effect seen towards many of the other disks in the same category.

\subsection{WaOph6}

The \neii{} emission toward WaOph6 appears morphologically broader than expected for a jet (see Figure \ref{fig:multi}). Its velocity map reveals two components: a narrow, high-velocity emission clearly tracing a jet and a low-velocity emission around it (Figure \ref{fig:velocity_maps}). A comparison with the continuum in Figure \ref{fig:multi} shows that this additional low-velocity emission around the jet closely coincides with the PSF “petals”, suggesting that it originates from point-like \neii{} emission, likely arising from the disk atmosphere.

\subsection{CXTau}

We tentatively detect a wind in this source only in the H$_2$ S(3) line and only when adopting a wider opening angle of 60$^{\circ}$ for the wind classification.. \cite{Anderson2024}, using the same MIRI dataset, reported a very wide and extended wind in the S(1) line; however, the opening angle is too wide to classify as a wind according to our method (LR/CR ratio), even when considering 60$^{\circ}$ instead of 50$^{\circ}$.

\subsection{**DFTau}

DFTau is a $\sim$0.1~arcsec equal mass binary \citep{Kutra2025} system, with the size of individual disks $\leq$3~au \citep{Grant2024}. Similar to \cite{Kurtovic2026}, we also find a high-velocity jet component associated with the primary and a low-velocity wind/disk atmosphere emission associated with the secondary. However, we do not find any extended wind-like emission in any of the H$_2$ lines, despite it showing a relatively high mass accretion rate (log~$\dot{M}_{acc}$ = -7.55~M$_{\odot}$~yr$^{-1}$) and showing an inner atomic wind traced in \oi{}, making it an interesting outlier to the overall trend found in this work where only disks with low $\dot{M}_{acc}$ with an inner \oi{} wind detection lack an extended H$_2$ MIRI wind.

\subsection{HH30}

HH30 shows an intriguing progression of H$_2$ emission morphology: the S(1) line primarily traces the disk surface with faint wind emission; S(3) traces both the disk surface and the wind; and S(5) predominantly traces the wind (Figure \ref{fig:multi}). The morphology of the higher excitation lines of H$_2$ covered by NIRSpec is also consistent with wind emission \cite{Pascucci2025}. To the authors' knowledge, this is the first time such a sequence has been spatially resolved. In \neii{}, because the disk around HH30 is nearly edge-on and its exact radial velocity is unknown, the velocity gradient between the redshifted and blueshifted lobes of the jet is not apparent. Nevertheless, the jet is readily identifiable based on its morphology. Interestingly, the \neii{} velocity map also exhibits a lateral velocity gradient, which could indicate either jet rotation or an instrumental effect, such as IFU distortion or uneven slice illumination, and therefore warrants further investigation (Figure \ref{fig:velocity_maps}).

\subsection{**HKTau}

HKTau is another binary system with a projected sky-plane separation of $\sim$2.3 arcsec \citep[e.g.,][]{McCabe2011}. The binary is well resolved with MIRI at all wavelengths, with HKTauA appearing brighter in the continuum maps (see Figure \ref{fig:multi}). Interestingly, following the trend observed in binaries so far, high \neii{} velocities are detected toward emission from HKTauA, suggesting a jet, whereas much lower velocities ($\lesssim$30 km s$^{-1}$) are seen toward HKTauB. However, HKTauB is known to be close to edge-on, and after proper deprojection, the intrinsic velocities would likely be higher. The H$_2$ emission toward HKTau is complicated by its binary nature, making it difficult to interpret, except S(7) where a clear wind morphology is seen.

\subsection{**HVTauC}

Similar to 2MASS~J16230544–2302566, the H$_2$ emission toward HVTauC is also very wide-angled, suggesting an origin in the disk surface, a wind, or a combination of both (Figure \ref{fig:multi}). The opening angle of the emission clearly decreases with increasing $J_{up}$, from S(1) to S(7). Interestingly, \cite{Beck2008,Beck2010} argued in favor of a wind based on the combination of high excitation temperature in v=1, T$\sim$2000~K and large transverse extent, which appeared difficult to jointly reproduce by pure UV/X-ray illumination of a static disk, increasing the likelihood that higher rotational transitions (e.g., S(7)) also trace a wind.

\subsection{J04381486+2611399}

2MASS~J04381486+2611399 is an M7.25 brown dwarf with a mass of $\sim$0.05~M$_{\odot}$ \citep{Luhman2004,Manara2023}. The integrated intensity maps in Figure \ref{fig:multi} show faint extended emission in both H$_2$ and \neii{}; however, the H$_2$ emission is too weak to satisfy our criteria for extended emission identification (Section \ref{sec:extended_emission_id}). In all maps, the emission is elongated along the upper-left--lower-right direction, suggestive of an outflow origin. However, the absence of constraints on the disk position angle, combined with the low signal-to-noise of the emission, prevents a robust confirmation of this interpretation. Deeper observations may better constrain the nature of this faint emission \citep[see also][for a detailed spectral analysis of this source]{Perotti2025}.



\subsection{LkCa15}

We detect a wind in this source in the S(3) line only when adopting a wider opening angle (60$^{\circ}$) for the wind classification. The S(1) emission also appears consistent with a bipolar wind morphology; however, the extended component is sufficiently faint that it is not classified as a wind by our method.

\subsection{**RWAur}

RWAur is a binary star system with a separation of 1.5 arcsec \citep[corresponding to $\sim$240 au in the sky plane;][]{Cabrit2006}. MIRI clearly resolves the two stars in the continuum at nearly all wavelengths (see the continuum map near H$_2$ S(1) at 17.035~\micron{} in Figure \ref{fig:multi}). In all H$_2$ lines, the emission appears to be dominated near RWAurA, with S(1) and S(3) showing extended emission along the disk position angle, and S(5) showing significant vertical (or lifted) emission perpendicular to the disk midplane. The extended curved H$_2$ filament in S(1) traces the same tidal arm as seen in CO by \cite{Cabrit2006}. The \neii{} velocity maps (Figure \ref{fig:velocity_maps}) indicate that RWAurA is driving a high-velocity jet, whereas RWAurB shows much lower velocities consistent with emission from the disk atmosphere or atomic winds \citep[as shown first by][]{Kurtovic2026}. The absence of a jet in RWAurB is consistent with its inner disk hole of 4~au \citep{Rodriguez2018} and low $\dot{M}_{\rm acc}$. We note that in our sample, RWAurA is the only system where \oi{} high-velocity component was detected without a low velocity component (see Figure \ref{fig:det_ext_out_stat}). Such systems are extremely rare, and out of $>$300 T Tauri stars with published \oi{} high-res spectra, only 3 such cases are found, including RWAurA \citep[others are IP~Tau and V409~Tau,][]{Banzatti2019,Fang2023,Nisini2024a}.

\subsection{J15582981-2310077 and J16064794-1841437}

The H$_2$ emission toward these two sources is particularly interesting, as it exhibits a morphology typically seen in externally photoevaporating disks \citep[e.g.,][]{Champion2017}, with a bright front on one side of the star (lower right in case of J1606-1841) and a flow on the opposite side (upper left in case of J1606-1841). This structure is most clearly seen in the S(1) line in J1606-1841 and the S(3) line in J1558-2310. These sources are discussed in detail in a forthcoming publication (N. S. Bajaj et al. 2026, in prep).

\subsection{J16035767-2031055}

The red-shifted velocities seen towards this source in \neii{} velocity maps (Figure \ref{fig:velocity_maps}) are higher than those seen towards other sources in the same category, suggesting a possible jet emanating from the source. However, due to the lack of jet-like morphology, we don't classify it as jet-like.

\subsection{J16075796-2040087}

The H$_2$ S(1), S(3) and S(5) emission toward 2MASS~J16075796–2040087 is similar to RY~Lup in that it extends significantly along the disk position angle. However, its exact morphology differs from that seen in RY~Lup. In this source, \neii{} clearly traces a bipolar jet (Figure \ref{fig:velocity_maps}), whereas in RY~Lup, \neii{} traces a low-velocity wind, as inferred from ground-based high-resolution spectroscopy \citep{Pascucci2020}.

\subsection{J16090075-1908526}

The \neii{} velocity map toward this source (Figure \ref{fig:velocity_maps}) is intriguing, showing extended emission perpendicular to the disk midplane and a clear velocity gradient, with redshifted emission on one side of the disk and blueshifted emission on the other. However, the low-velocities are consistent with the uneven slice illumination effect seen towards several other sources in the same category.

\subsection{**J16153456-2242421}

2MASS~J16153456–2242421 (also known as VV Sco or PDS 82) is a known binary system in the Upper Sco star-forming region \cite{Kraus2008,Barenfeld2019}. With MIRI, the companion is detected and spatially resolved in the continuum up to $\lesssim$17~\micron{} (Figure \ref{fig:multi}). Although the system shows extended \neii{} emission, the velocity map reveals no clear evidence of a jet driven by either star or disk (Figure \ref{fig:velocity_maps}).

\subsection{J16075796-2040087, and WL3} \label{sec:outlier_disks}

J1607-2040 and WL3 are the only single sources where the wind is detected only in the S(7) line. Both of these sources have higher disk inclinations of $\sim$66$^{\circ}$ and 69$^{\circ}$, and stellar masses of 0.71 and 0.64~M$_{\odot}$, respectively, neither of which is concerning in relation to outflow detection in lower \textit{J} transitions of H$_2$. J1607-2040 exhibits a highly extended H$_2$ emission structure that is elongated along the disk position angle, which could have led to an underestimation of any perpendicularly extended wind-like emission. WL3 shows a tentative biconical emission structure in H$_2$ S(3), possibly indicative of a weak wind; however, the emission is too faint to satisfy our classification criteria (see Figure \ref{fig:multi}).

\section{Calculation of SNR}
\label{sec:SNR_calculation}

For a given line, we compute the SNR as follows. We first integrate the spectrum over the full IFU field of view and subtract the local continuum near the line, following the procedure described in Section \ref{sec:cont_sub}. We then record the maximum flux near the line center, defined as the point closest to the line-center wavelength along with the two adjacent points on either side. To estimate the noise, we construct a separate spectrum integrated only over regions of the IFU field of view that contain background emission, explicitly excluding any contribution from the PSF. This avoids contamination from water lines in the noise estimate. We then calculate the standard deviation near the line wavelength using this background spectrum. The SNR is obtained by dividing the line peak flux by this standard deviation. 

Because the standard deviation is computed only from the background, our SNR estimates are systematically overestimated. However, since these values are used for relative comparisons between sources, this systematic offset does not affect our results or conclusions. In contrast, estimating the standard deviation within the PSF would lead to underestimated SNR values for sources with strong water emission lines, introducing a non-systematic bias.

\section{Uncertainty calculation on Outflow detection fraction}
\label{sec:uncertainty_estimation}

We estimate the uncertainty on the disk fraction exhibiting a given property using the Wilson score interval for a binomial proportion \citep{Wilson1927,Agresti1998}. For a bin containing $N$ sources, of which $k$ show the property of interest, the measured fraction is $f=k/N$. The Wilson interval is derived from the binomial likelihood and provides confidence limits that are centered on a slightly adjusted estimate of the fraction rather than directly on $f$. For the 68\% confidence level adopted here, the interval can be written as

\begin{equation}
f_{\rm c} = \frac{f + 1/(2N)}{1 + 1/N}, \qquad
\Delta f = \frac{\sqrt{f(1-f)/N + 1/(4N^2)}}{1 + 1/N},
\end{equation}

with the lower and upper bounds given by $f_{\rm c} \pm \Delta f$. Unlike Poisson-based uncertainties, which treat detections as simple event counts and are commonly approximated as $1/\sqrt{N}$, the Wilson score interval explicitly accounts for the fact that a fraction is bounded between 0 and 1 and depends on both the number of detections and the total sample size. As a result, it yields more intuitive and stable uncertainties, particularly in bins with small sample sizes or low detection fractions.

\bibliographystyle{aasjournal}
\bibliography{citations}

\begin{thebibliography}{}
\expandafter\ifx\csname natexlab\endcsname\relax\def\natexlab#1{#1}\fi
\providecommand{\url}[1]{\href{#1}{#1}}
\providecommand{\dodoi}[1]{doi:~\href{http://doi.org/#1}{\nolinkurl{#1}}}
\providecommand{\doeprint}[1]{\href{http://ascl.net/#1}{\nolinkurl{http://ascl.net/#1}}}
\providecommand{\doarXiv}[1]{\href{https://arxiv.org/abs/#1}{\nolinkurl{https://arxiv.org/abs/#1}}}

\bibitem[{{Abdurro'uf} {et~al.}(2022){Abdurro'uf}, {Accetta}, {Aerts}, {Silva Aguirre}, {Ahumada}, {Ajgaonkar}, {Filiz Ak}, {Alam}, {Allende Prieto}, {Almeida}, {Anders}, {Anderson}, {Andrews}, {Anguiano}, {Aquino-Ort{\'\i}z}, {Arag{\'o}n-Salamanca}, {Argudo-Fern{\'a}ndez}, {Ata}, {Aubert}, {Avila-Reese}, {Badenes}, {Barb{\'a}}, {Barger}, {Barrera-Ballesteros}, {Beaton}, {Beers}, {Belfiore}, {Bender}, {Bernardi}, {Bershady}, {Beutler}, {Bidin}, {Bird}, {Bizyaev}, {Blanc}, {Blanton}, {Boardman}, {Bolton}, {Boquien}, {Borissova}, {Bovy}, {Brandt}, {Brown}, {Brownstein}, {Brusa}, {Buchner}, {Bundy}, {Burchett}, {Bureau}, {Burgasser}, {Cabang}, {Campbell}, {Cappellari}, {Carlberg}, {Wanderley}, {Carrera}, {Cash}, {Chen}, {Chen}, {Cherinka}, {Chiappini}, {Choi}, {Chojnowski}, {Chung}, {Clerc}, {Cohen}, {Comerford}, {Comparat}, {da Costa}, {Covey}, {Crane}, {Cruz-Gonzalez}, {Culhane}, {Cunha}, {Dai}, {Damke}, {Darling}, {Davidson}, {Davies}, {Dawson}, {De Lee}, {Diamond-Stanic}, {Cano-D{\'\i}az}, {S{\'a}nchez},
  {Donor}, {Duckworth}, {Dwelly}, {Eisenstein}, {Elsworth}, {Emsellem}, {Eracleous}, {Escoffier}, {Fan}, {Farr}, {Feng}, {Fern{\'a}ndez-Trincado}, {Feuillet}, {Filipp}, {Fillingham}, {Frinchaboy}, {Fromenteau}, {Galbany}, {Garc{\'\i}a}, {Garc{\'\i}a-Hern{\'a}ndez}, {Ge}, {Geisler}, {Gelfand}, {G{\'e}ron}, {Gibson}, {Goddy}, {Godoy-Rivera}, {Grabowski}, {Green}, {Greener}, {Grier}, {Griffith}, {Guo}, {Guy}, {Hadjara}, {Harding}, {Hasselquist}, {Hayes}, {Hearty}, {Hern{\'a}ndez}, {Hill}, {Hogg}, {Holtzman}, {Horta}, {Hsieh}, {Hsu}, {Hsu}, {Huber}, {Huertas-Company}, {Hutchinson}, {Hwang}, {Ibarra-Medel}, {Chitham}, {Ilha}, {Imig}, {Jaekle}, {Jayasinghe}, {Ji}, {Johnson}, {Jones}, {J{\"o}nsson}, {Katkov}, {Khalatyan}, {Kinemuchi}, {Kisku}, {Knapen}, {Kneib}, {Kollmeier}, {Kong}, {Kounkel}, {Kreckel}, {Krishnarao}, {Lacerna}, {Lane}, {Langgin}, {Lavender}, {Law}, {Lazarz}, {Leung}, {Leung}, {Lewis}, {Li}, {Li}, {Lian}, {Liang}, {Lin}, {Lin}, {Lin}, {Lintott}, {Long}, {Longa-Pe{\~n}a}, {L{\'o}pez-Cob{\'a}}, {Lu},
  {Lundgren}, {Luo}, {Mackereth}, {de la Macorra}, {Mahadevan}, {Majewski}, {Manchado}, {Mandeville}, {Maraston}, {Margalef-Bentabol}, {Masseron}, {Masters}, {Mathur}, {McDermid}, {Mckay}, {Merloni}, {Merrifield}, {Meszaros}, {Miglio}, {Di Mille}, {Minniti}, {Minsley}, \& {Monachesi}}]{Abdurrouf2022}
{Abdurro'uf}, {Accetta}, K., {Aerts}, C., {et~al.} 2022, \apjs, 259, 35, \dodoi{10.3847/1538-4365/ac4414}

\bibitem[{{Agra-Amboage} {et~al.}(2014){Agra-Amboage}, {Cabrit}, {Dougados}, {Kristensen}, {Ibgui}, \& {Reunanen}}]{Agra-amboage2014}
{Agra-Amboage}, V., {Cabrit}, S., {Dougados}, C., {et~al.} 2014, \aap, 564, A11, \dodoi{10.1051/0004-6361/201220488}

\bibitem[{Agresti \& Coull(1998)}]{Agresti1998}
Agresti, A., \& Coull, B.~A. 1998, The American Statistician, 52, 119, \dodoi{10.1080/00031305.1998.10480550}

\bibitem[{{Agurto-Gangas} {et~al.}(2025){Agurto-Gangas}, {P{\'e}rez}, {Sierra}, {Miley}, {Zhang}, {Pascucci}, {Pinilla}, {Deng}, {Carpenter}, {Trapman}, {Vioque}, {Rosotti}, {Kurtovic}, {Cieza}, {Anania}, {Tabone}, {Schwarz}, {Hogerheijde}, {TorresVillanueva}, {Ruiz-Rodriguez}, \& {Gonz{\'a}lez-Ruilova}}]{Agurto-gangas2025}
{Agurto-Gangas}, C., {P{\'e}rez}, L.~M., {Sierra}, A., {et~al.} 2025, \apj, 989, 4, \dodoi{10.3847/1538-4357/adc7ab}

\bibitem[{{Alcalá, J. M.} {et~al.}(2017){Alcalá, J. M.}, {Manara, C. F.}, {Natta, A.}, {Frasca, A.}, {Testi, L.}, {Nisini, B.}, {Stelzer, B.}, {Williams, J. P.}, {Antoniucci, S.}, {Biazzo, K.}, {Covino, E.}, {Esposito, M.}, {Getman, F.}, \& {Rigliaco, E.}}]{Alcala2017}
{Alcalá, J. M.}, {Manara, C. F.}, {Natta, A.}, {et~al.} 2017, \aap, 600, A20, \dodoi{10.1051/0004-6361/201629929}

\bibitem[{{Alexander} {et~al.}(2014){Alexander}, {Pascucci}, {Andrews}, {Armitage}, \& {Cieza}}]{Alexander2014}
{Alexander}, R., {Pascucci}, I., {Andrews}, S., {Armitage}, P., \& {Cieza}, L. 2014, in Protostars and Planets VI, ed. H.~{Beuther}, R.~S. {Klessen}, C.~P. {Dullemond}, \& T.~{Henning}, 475--496, \dodoi{10.2458/azu_uapress_9780816531240-ch021}

\bibitem[{{Anderson} {et~al.}(2024){Anderson}, {Williams}, {Blake}, {Pontoppidan}, {Salyk}, {Boogert}, {Ross}, \& {Cleeves}}]{Anderson2024}
{Anderson}, A.~R., {Williams}, J.~P., {Blake}, G.~A., {et~al.} 2024, \apj, 977, 213, \dodoi{10.3847/1538-4357/ad7e16}

\bibitem[{{Andrews}(2020)}]{Andrews2020}
{Andrews}, S.~M. 2020, \araa, 58, 483, \dodoi{10.1146/annurev-astro-031220-010302}

\bibitem[{{Ansdell} {et~al.}(2020){Ansdell}, {Haworth}, {Williams}, {Facchini}, {Winter}, {Manara}, {Hacar}, {Chiang}, {van Terwisga}, {van der Marel}, \& {van Dishoeck}}]{Ansdell2020}
{Ansdell}, M., {Haworth}, T.~J., {Williams}, J.~P., {et~al.} 2020, \aj, 160, 248, \dodoi{10.3847/1538-3881/abb9af}

\bibitem[{{Arulanantham} {et~al.}(2024){Arulanantham}, {McClure}, {Pontoppidan}, {Beck}, {Sturm}, {Harsono}, {Boogert}, {Cordiner}, {Dartois}, {Drozdovskaya}, {Espaillat}, {Melnick}, {Noble}, {Palumbo}, {Pendleton}, {Terada}, \& {van Dishoeck}}]{Arulanantham2024}
{Arulanantham}, N., {McClure}, M.~K., {Pontoppidan}, K., {et~al.} 2024, \apjl, 965, L13, \dodoi{10.3847/2041-8213/ad35c9}

\bibitem[{{Arulanantham} {et~al.}(2025){Arulanantham}, {Salyk}, {Pontoppidan}, {Banzatti}, {Zhang}, {{\"O}berg}, {Long}, {Carr}, {Najita}, {Pascucci}, {Colmenares}, {Xie}, {Huang}, {Green}, {Andrews}, {Blake}, {Bergin}, {Pinilla}, {Vioque}, {Dahl}, {Raul}, {Krijt}, \& {The Jdiscs Collaboration}}]{Arulanantham2025}
{Arulanantham}, N., {Salyk}, C., {Pontoppidan}, K., {et~al.} 2025, \aj, 170, 67, \dodoi{10.3847/1538-3881/addd01}

\bibitem[{{Aso} {et~al.}(2017){Aso}, {Ohashi}, {Aikawa}, {Machida}, {Saigo}, {Saito}, {Takakuwa}, {Tomida}, {Tomisaka}, \& {Yen}}]{Aso2017}
{Aso}, Y., {Ohashi}, N., {Aikawa}, Y., {et~al.} 2017, \apj, 849, 56, \dodoi{10.3847/1538-4357/aa8264}

\bibitem[{{Astropy Collaboration} {et~al.}(2013){Astropy Collaboration}, {Robitaille}, {Tollerud}, {Greenfield}, {Droettboom}, {Bray}, {Aldcroft}, {Davis}, {Ginsburg}, {Price-Whelan}, {Kerzendorf}, {Conley}, {Crighton}, {Barbary}, {Muna}, {Ferguson}, {Grollier}, {Parikh}, {Nair}, {Unther}, {Deil}, {Woillez}, {Conseil}, {Kramer}, {Turner}, {Singer}, {Fox}, {Weaver}, {Zabalza}, {Edwards}, {Azalee Bostroem}, {Burke}, {Casey}, {Crawford}, {Dencheva}, {Ely}, {Jenness}, {Labrie}, {Lim}, {Pierfederici}, {Pontzen}, {Ptak}, {Refsdal}, {Servillat}, \& {Streicher}}]{astropy:2013}
{Astropy Collaboration}, {Robitaille}, T.~P., {Tollerud}, E.~J., {et~al.} 2013, \aap, 558, A33, \dodoi{10.1051/0004-6361/201322068}

\bibitem[{{Astropy Collaboration} {et~al.}(2018){Astropy Collaboration}, {Price-Whelan}, {Sip{\H{o}}cz}, {G{\"u}nther}, {Lim}, {Crawford}, {Conseil}, {Shupe}, {Craig}, {Dencheva}, {Ginsburg}, {Vand erPlas}, {Bradley}, {P{\'e}rez-Su{\'a}rez}, {de Val-Borro}, {Aldcroft}, {Cruz}, {Robitaille}, {Tollerud}, {Ardelean}, {Babej}, {Bach}, {Bachetti}, {Bakanov}, {Bamford}, {Barentsen}, {Barmby}, {Baumbach}, {Berry}, {Biscani}, {Boquien}, {Bostroem}, {Bouma}, {Brammer}, {Bray}, {Breytenbach}, {Buddelmeijer}, {Burke}, {Calderone}, {Cano Rodr{\'\i}guez}, {Cara}, {Cardoso}, {Cheedella}, {Copin}, {Corrales}, {Crichton}, {D'Avella}, {Deil}, {Depagne}, {Dietrich}, {Donath}, {Droettboom}, {Earl}, {Erben}, {Fabbro}, {Ferreira}, {Finethy}, {Fox}, {Garrison}, {Gibbons}, {Goldstein}, {Gommers}, {Greco}, {Greenfield}, {Groener}, {Grollier}, {Hagen}, {Hirst}, {Homeier}, {Horton}, {Hosseinzadeh}, {Hu}, {Hunkeler}, {Ivezi{\'c}}, {Jain}, {Jenness}, {Kanarek}, {Kendrew}, {Kern}, {Kerzendorf}, {Khvalko}, {King}, {Kirkby}, {Kulkarni},
  {Kumar}, {Lee}, {Lenz}, {Littlefair}, {Ma}, {Macleod}, {Mastropietro}, {McCully}, {Montagnac}, {Morris}, {Mueller}, {Mumford}, {Muna}, {Murphy}, {Nelson}, {Nguyen}, {Ninan}, {N{\"o}the}, {Ogaz}, {Oh}, {Parejko}, {Parley}, {Pascual}, {Patil}, {Patil}, {Plunkett}, {Prochaska}, {Rastogi}, {Reddy Janga}, {Sabater}, {Sakurikar}, {Seifert}, {Sherbert}, {Sherwood-Taylor}, {Shih}, {Sick}, {Silbiger}, {Singanamalla}, {Singer}, {Sladen}, {Sooley}, {Sornarajah}, {Streicher}, {Teuben}, {Thomas}, {Tremblay}, {Turner}, {Terr{\'o}n}, {van Kerkwijk}, {de la Vega}, {Watkins}, {Weaver}, {Whitmore}, {Woillez}, {Zabalza}, \& {Astropy Contributors}}]{astropy:2018}
{Astropy Collaboration}, {Price-Whelan}, A.~M., {Sip{\H{o}}cz}, B.~M., {et~al.} 2018, \aj, 156, 123, \dodoi{10.3847/1538-3881/aabc4f}

\bibitem[{{Astropy Collaboration} {et~al.}(2022){Astropy Collaboration}, {Price-Whelan}, {Lim}, {Earl}, {Starkman}, {Bradley}, {Shupe}, {Patil}, {Corrales}, {Brasseur}, {N{"o}the}, {Donath}, {Tollerud}, {Morris}, {Ginsburg}, {Vaher}, {Weaver}, {Tocknell}, {Jamieson}, {van Kerkwijk}, {Robitaille}, {Merry}, {Bachetti}, {G{"u}nther}, {Aldcroft}, {Alvarado-Montes}, {Archibald}, {B{'o}di}, {Bapat}, {Barentsen}, {Baz{'a}n}, {Biswas}, {Boquien}, {Burke}, {Cara}, {Cara}, {Conroy}, {Conseil}, {Craig}, {Cross}, {Cruz}, {D'Eugenio}, {Dencheva}, {Devillepoix}, {Dietrich}, {Eigenbrot}, {Erben}, {Ferreira}, {Foreman-Mackey}, {Fox}, {Freij}, {Garg}, {Geda}, {Glattly}, {Gondhalekar}, {Gordon}, {Grant}, {Greenfield}, {Groener}, {Guest}, {Gurovich}, {Handberg}, {Hart}, {Hatfield-Dodds}, {Homeier}, {Hosseinzadeh}, {Jenness}, {Jones}, {Joseph}, {Kalmbach}, {Karamehmetoglu}, {Ka{l}uszy{'n}ski}, {Kelley}, {Kern}, {Kerzendorf}, {Koch}, {Kulumani}, {Lee}, {Ly}, {Ma}, {MacBride}, {Maljaars}, {Muna}, {Murphy}, {Norman}, {O'Steen},
  {Oman}, {Pacifici}, {Pascual}, {Pascual-Granado}, {Patil}, {Perren}, {Pickering}, {Rastogi}, {Roulston}, {Ryan}, {Rykoff}, {Sabater}, {Sakurikar}, {Salgado}, {Sanghi}, {Saunders}, {Savchenko}, {Schwardt}, {Seifert-Eckert}, {Shih}, {Jain}, {Shukla}, {Sick}, {Simpson}, {Singanamalla}, {Singer}, {Singhal}, {Sinha}, {Sip{H{o}}cz}, {Spitler}, {Stansby}, {Streicher}, {{{S}}umak}, {Swinbank}, {Taranu}, {Tewary}, {Tremblay}, {Val-Borro}, {Van Kooten}, {Vasovi{'c}}, {Verma}, {de Miranda Cardoso}, {Williams}, {Wilson}, {Winkel}, {Wood-Vasey}, {Xue}, {Yoachim}, {Zhang}, {Zonca}, \& {Astropy Project Contributors}}]{astropy:2022}
{Astropy Collaboration}, {Price-Whelan}, A.~M., {Lim}, P.~L., {et~al.} 2022, \apj, 935, 167, \dodoi{10.3847/1538-4357/ac7c74}

\bibitem[{{Bacciotti} {et~al.}(1999){Bacciotti}, {Eisl{\"o}ffel}, \& {Ray}}]{Bacciotti1999}
{Bacciotti}, F., {Eisl{\"o}ffel}, J., \& {Ray}, T.~P. 1999, \aap, 350, 917

\bibitem[{{Bai} \& {Stone}(2013)}]{Bai2013}
{Bai}, X.-N., \& {Stone}, J.~M. 2013, \apj, 769, 76, \dodoi{10.1088/0004-637X/769/1/76}

\bibitem[{{Bai} {et~al.}(2016){Bai}, {Ye}, {Goodman}, \& {Yuan}}]{Bai2016}
{Bai}, X.-N., {Ye}, J., {Goodman}, J., \& {Yuan}, F. 2016, \apj, 818, 152, \dodoi{10.3847/0004-637X/818/2/152}

\bibitem[{{Bajaj} {et~al.}(2024){Bajaj}, {Pascucci}, {Gorti}, {Alexander}, {Sellek}, {Morrison}, {Gaspar}, {Clarke}, {Xie}, {Ballabio}, \& {Deng}}]{Bajaj2024}
{Bajaj}, N.~S., {Pascucci}, I., {Gorti}, U., {et~al.} 2024, \aj, 167, 127, \dodoi{10.3847/1538-3881/ad22e1}

\bibitem[{{Bajaj} {et~al.}(2025){Bajaj}, {Pascucci}, {Beck}, {Edwards}, {Cabrit}, {Najita}, {Schwarz}, {Semenov}, {Salyk}, {Gorti}, {Brittain}, {Krijt}, {Ruaud}, \& {Page}}]{Bajaj2025}
{Bajaj}, N.~S., {Pascucci}, I., {Beck}, T.~L., {et~al.} 2025, \aj, 169, 296, \dodoi{10.3847/1538-3881/adc73c}

\bibitem[{{Balbus} \& {Hawley}(1991)}]{Balbus1991}
{Balbus}, S.~A., \& {Hawley}, J.~F. 1991, \apj, 376, 214, \dodoi{10.1086/170270}

\bibitem[{{Bally}(2016)}]{Bally2016}
{Bally}, J. 2016, \araa, 54, 491, \dodoi{10.1146/annurev-astro-081915-023341}

\bibitem[{{Bally} \& {Lada}(1983)}]{Bally1983}
{Bally}, J., \& {Lada}, C.~J. 1983, \apj, 265, 824, \dodoi{10.1086/160729}

\bibitem[{{Banzatti} {et~al.}(2019){Banzatti}, {Pascucci}, {Edwards}, {Fang}, {Gorti}, \& {Flock}}]{Banzatti2019}
{Banzatti}, A., {Pascucci}, I., {Edwards}, S., {et~al.} 2019, \apj, 870, 76, \dodoi{10.3847/1538-4357/aaf1aa}

\bibitem[{{Banzatti} \& {Pontoppidan}(2015)}]{Banzatti2015}
{Banzatti}, A., \& {Pontoppidan}, K.~M. 2015, \apj, 809, 167, \dodoi{10.1088/0004-637X/809/2/167}

\bibitem[{{Banzatti} {et~al.}(2017){Banzatti}, {Pontoppidan}, {Salyk}, {Herczeg}, {van Dishoeck}, \& {Blake}}]{Banzatti2017}
{Banzatti}, A., {Pontoppidan}, K.~M., {Salyk}, C., {et~al.} 2017, \apj, 834, 152, \dodoi{10.3847/1538-4357/834/2/152}

\bibitem[{{Banzatti} {et~al.}(2021){Banzatti}, {Ballering}, {Bosman}, {Herczeg}, {Kalyaan}, {Krijt}, {Lambrechts}, {Long}, {Meyer}, {Oberg}, {Pascucci}, {Pinilla}, {Pontoppidan}, {Rosotti}, {Salyk}, {Vazquez}, \& {Watkins}}]{Banzatti2021}
{Banzatti}, A., {Ballering}, N., {Bosman}, A., {et~al.} 2021, {The infrared water spectrum as a tracer of pebble delivery to rocky planets}, JWST Proposal. Cycle 1, ID. \#1640

\bibitem[{{Barenfeld} {et~al.}(2017){Barenfeld}, {Carpenter}, {Sargent}, {Isella}, \& {Ricci}}]{Barenfeld2017}
{Barenfeld}, S.~A., {Carpenter}, J.~M., {Sargent}, A.~I., {Isella}, A., \& {Ricci}, L. 2017, \apj, 851, 85, \dodoi{10.3847/1538-4357/aa989d}

\bibitem[{{Barenfeld} {et~al.}(2019){Barenfeld}, {Carpenter}, {Sargent}, {Rizzuto}, {Kraus}, {Meshkat}, {Akeson}, {Jensen}, \& {Hinkley}}]{Barenfeld2019}
{Barenfeld}, S.~A., {Carpenter}, J.~M., {Sargent}, A.~I., {et~al.} 2019, \apj, 878, 45, \dodoi{10.3847/1538-4357/ab1e50}

\bibitem[{{Bary} {et~al.}(2003){Bary}, {Weintraub}, \& {Kastner}}]{Bary2003}
{Bary}, J.~S., {Weintraub}, D.~A., \& {Kastner}, J.~H. 2003, \apj, 586, 1136, \dodoi{10.1086/367719}

\bibitem[{{Beck} \& {Bary}(2019)}]{Beck2019}
{Beck}, T.~L., \& {Bary}, J.~S. 2019, \apj, 884, 159, \dodoi{10.3847/1538-4357/ab4259}

\bibitem[{{Beck} {et~al.}(2010){Beck}, {Bary}, \& {McGregor}}]{Beck2010}
{Beck}, T.~L., {Bary}, J.~S., \& {McGregor}, P.~J. 2010, \apj, 722, 1360, \dodoi{10.1088/0004-637X/722/2/1360}

\bibitem[{{Beck} {et~al.}(2008){Beck}, {McGregor}, {Takami}, \& {Pyo}}]{Beck2008}
{Beck}, T.~L., {McGregor}, P.~J., {Takami}, M., \& {Pyo}, T.-S. 2008, \apj, 676, 472, \dodoi{10.1086/527528}

\bibitem[{{B{\'e}thune} {et~al.}(2017){B{\'e}thune}, {Lesur}, \& {Ferreira}}]{Bethune2017}
{B{\'e}thune}, W., {Lesur}, G., \& {Ferreira}, J. 2017, \aap, 600, A75, \dodoi{10.1051/0004-6361/201630056}

\bibitem[{{Biazzo} {et~al.}(2012){Biazzo}, {Alcal{\'a}}, {Covino}, {Frasca}, {Getman}, \& {Spezzi}}]{Biazzo2012}
{Biazzo}, K., {Alcal{\'a}}, J.~M., {Covino}, E., {et~al.} 2012, \aap, 547, A104, \dodoi{10.1051/0004-6361/201219680}

\bibitem[{{Bitner} {et~al.}(2008){Bitner}, {Richter}, {Lacy}, {Herczeg}, {Greathouse}, {Jaffe}, {Salyk}, {Blake}, {Hollenbach}, {Doppmann}, {Najita}, \& {Currie}}]{Bitner2008}
{Bitner}, M.~A., {Richter}, M.~J., {Lacy}, J.~H., {et~al.} 2008, \apj, 688, 1326, \dodoi{10.1086/592317}

\bibitem[{{Bjerkeli} {et~al.}(2016){Bjerkeli}, {van der Wiel}, {Harsono}, {Ramsey}, \& {J{\o}rgensen}}]{Bjerkeli2016}
{Bjerkeli}, P., {van der Wiel}, M. H.~D., {Harsono}, D., {Ramsey}, J.~P., \& {J{\o}rgensen}, J.~K. 2016, \nat, 540, 406, \dodoi{10.1038/nature20600}

\bibitem[{{Blandford} \& {Payne}(1982)}]{Blandford1982}
{Blandford}, R.~D., \& {Payne}, D.~G. 1982, \mnras, 199, 883, \dodoi{10.1093/mnras/199.4.883}

\bibitem[{{Bohn} {et~al.}(2022){Bohn}, {Benisty}, {Perraut}, {van der Marel}, {W{\"o}lfer}, {van Dishoeck}, {Facchini}, {Manara}, {Teague}, {Francis}, {Berger}, {Garcia-Lopez}, {Ginski}, {Henning}, {Kenworthy}, {Kraus}, {M{\'e}nard}, {M{\'e}rand}, \& {P{\'e}rez}}]{Bohn2022}
{Bohn}, A.~J., {Benisty}, M., {Perraut}, K., {et~al.} 2022, \aap, 658, A183, \dodoi{10.1051/0004-6361/202142070}

\bibitem[{{Booth} {et~al.}(2021){Booth}, {Tabone}, {Ilee}, {Walsh}, {Aikawa}, {Andrews}, {Bae}, {Bergin}, {Bergner}, {Bosman}, {Calahan}, {Cataldi}, {Cleeves}, {Czekala}, {Guzm{\'a}n}, {Huang}, {Law}, {Le Gal}, {Long}, {Loomis}, {M{\'e}nard}, {Nomura}, {{\"O}berg}, {Qi}, {Schwarz}, {Teague}, {Tsukagoshi}, {Wilner}, {Yamato}, \& {Zhang}}]{Booth2021}
{Booth}, A.~S., {Tabone}, B., {Ilee}, J.~D., {et~al.} 2021, \apjs, 257, 16, \dodoi{10.3847/1538-4365/ac1ad4}

\bibitem[{Bradley {et~al.}(2026)Bradley, Sipőcz, Robitaille, Tollerud, Vinícius, Deil, Barbary, Wilson, Busko, Donath, Günther, Cara, Lim, Meßlinger, Conseil, Droettboom, Bostroem, Bray, Bratholm, Burnett, Jamieson, Ginsburg, Taranu, Barentsen, Craig, Morris, Perrin, \& Rathi}]{photutils}
Bradley, L., Sipőcz, B.~M., Robitaille, T.~P., {et~al.} 2026, Photutils, 3.0.0,  Zenodo, \dodoi{10.5281/zenodo.19636730}

\bibitem[{{Brandeker} {et~al.}(2001){Brandeker}, {Liseau}, {Artymowicz}, \& {Jayawardhana}}]{Brandeker2001}
{Brandeker}, A., {Liseau}, R., {Artymowicz}, P., \& {Jayawardhana}, R. 2001, \apjl, 561, L199, \dodoi{10.1086/324676}

\bibitem[{{Bushouse} {et~al.}(2026){Bushouse}, {Eisenhamer}, {Dencheva}, {Davies}, {Greenfield}, {Morrison}, {Hodge}, {Simon}, {Grumm}, {Droettboom}, {Slavich}, {Sosey}, {Pauly}, {Miller}, {Jedrzejewski}, {Hack}, {Davis}, {Crawford}, {Law}, {Gordon}, {Regan}, {Cara}, {MacDonald}, {Bradley}, {Shanahan}, {Jamieson}, {Teodoro}, {Williams}, {Pena-Guerrero}, {Graham}, {Molter}, {Brandt}, {Hayes}, {Cooper}, {Clarke}, \& {Filippazzo}}]{Bushouse2026}
{Bushouse}, H., {Eisenhamer}, J., {Dencheva}, N., {et~al.} 2026, {JWST Calibration Pipeline}, 2.0.1,  Zenodo, \dodoi{10.5281/zenodo.20058613}

\bibitem[{{Cabrit} {et~al.}(2006){Cabrit}, {Pety}, {Pesenti}, \& {Dougados}}]{Cabrit2006}
{Cabrit}, S., {Pety}, J., {Pesenti}, N., \& {Dougados}, C. 2006, \aap, 452, 897, \dodoi{10.1051/0004-6361:20054047}

\bibitem[{{Cahill} {et~al.}(2019){Cahill}, {Whelan}, {Hu{\'e}lamo}, \& {Alcal{\'a}}}]{Cahill2019}
{Cahill}, E., {Whelan}, E.~T., {Hu{\'e}lamo}, N., \& {Alcal{\'a}}, J. 2019, \mnras, 484, 4315, \dodoi{10.1093/mnras/stz280}

\bibitem[{{Caratti o Garatti} {et~al.}(2024){Caratti o Garatti}, {Ray}, {Kavanagh}, {McCaughrean}, {Gieser}, {Giannini}, {van Dishoeck}, {Justtanont}, {van Gelder}, {Francis}, {Beuther}, {Tychoniec}, {Nisini}, {Navarro}, {Devaraj}, {Reyes}, {Nazari}, {Klaassen}, {G{\"u}del}, {Henning}, {Lagage}, {{\"O}stlin}, {Vandenbussche}, {Waelkens}, \& {Wright}}]{Caratti_2024}
{Caratti o Garatti}, A., {Ray}, T.~P., {Kavanagh}, P.~J., {et~al.} 2024, \aap, 691, A134, \dodoi{10.1051/0004-6361/202451350}

\bibitem[{{Carpenter} {et~al.}(2025){Carpenter}, {Esplin}, {Luhman}, {Mamajek}, \& {Andrews}}]{Carpenter2025}
{Carpenter}, J.~M., {Esplin}, T.~L., {Luhman}, K.~L., {Mamajek}, E.~E., \& {Andrews}, S.~M. 2025, \apj, 978, 117, \dodoi{10.3847/1538-4357/ad8ebc}

\bibitem[{{Carrera} {et~al.}(2017){Carrera}, {Gorti}, {Johansen}, \& {Davies}}]{Carrera2017}
{Carrera}, D., {Gorti}, U., {Johansen}, A., \& {Davies}, M.~B. 2017, \apj, 839, 16, \dodoi{10.3847/1538-4357/aa6932}

\bibitem[{{Champion} {et~al.}(2017){Champion}, {Bern{\'e}}, {Vicente}, {Kamp}, {Le Petit}, {Gusdorf}, {Joblin}, \& {Goicoechea}}]{Champion2017}
{Champion}, J., {Bern{\'e}}, O., {Vicente}, S., {et~al.} 2017, \aap, 604, A69, \dodoi{10.1051/0004-6361/201629404}

\bibitem[{{Cieza} {et~al.}(2010){Cieza}, {Schreiber}, {Romero}, {Mora}, {Merin}, {Swift}, {Orellana}, {Williams}, {Harvey}, \& {Evans}}]{Cieza2010}
{Cieza}, L.~A., {Schreiber}, M.~R., {Romero}, G.~A., {et~al.} 2010, \apj, 712, 925, \dodoi{10.1088/0004-637X/712/2/925}

\bibitem[{{Cieza} {et~al.}(2011){Cieza}, {Olofsson}, {Harvey}, {Pinte}, {Mer{\'\i}n}, {Augereau}, {Evans}, {Najita}, {Henning}, \& {M{\'e}nard}}]{Cieza2011}
{Cieza}, L.~A., {Olofsson}, J., {Harvey}, P.~M., {et~al.} 2011, \apjl, 741, L25, \dodoi{10.1088/2041-8205/741/2/L25}

\bibitem[{{Dahm}(2010)}]{Dahm2010}
{Dahm}, S.~E. 2010, \aj, 140, 1444, \dodoi{10.1088/0004-6256/140/5/1444}

\bibitem[{{Davis} {et~al.}(2001){Davis}, {Ray}, {Desroches}, \& {Aspin}}]{Davis2001}
{Davis}, C.~J., {Ray}, T.~P., {Desroches}, L., \& {Aspin}, C. 2001, \mnras, 326, 524, \dodoi{10.1046/j.1365-8711.2001.04560.x}

\bibitem[{{Davis} {et~al.}(2011){Davis}, {Cervantes}, {Nisini}, {Giannini}, {Takami}, {Whelan}, {Smith}, {Ray}, {Chrysostomou}, \& {Pyo}}]{Davis2011}
{Davis}, C.~J., {Cervantes}, B., {Nisini}, B., {et~al.} 2011, \aap, 528, A3, \dodoi{10.1051/0004-6361/201015897}

\bibitem[{{Delabrosse} {et~al.}(2024){Delabrosse}, {Dougados}, {Cabrit}, {Tabone}, {Tychoniec}, {Ray}, {Podio}, \& {McClure}}]{Delabrosse2024}
{Delabrosse}, V., {Dougados}, C., {Cabrit}, S., {et~al.} 2024, \aap, 688, A173, \dodoi{10.1051/0004-6361/202449176}

\bibitem[{{Deng} {et~al.}(2025){Deng}, {Vioque}, {Pascucci}, {P{\'e}rez}, {Zhang}, {Kurtovic}, {Trapman}, {TorresVillanueva}, {Agurto-Gangas}, {Carpenter}, {Pinilla}, {Gorti}, {Tabone}, {Sierra}, {Rosotti}, {Cieza}, {Anania}, {Gonz{\'a}lez-Ruilova}, {Hogerheijde}, {Miley}, {Ruiz-Rodriguez}, {Ruaud}, \& {Schwarz}}]{Deng2025}
{Deng}, D., {Vioque}, M., {Pascucci}, I., {et~al.} 2025, \apj, 989, 3, \dodoi{10.3847/1538-4357/add43a}

\bibitem[{{Devaraj} {et~al.}(2026){Devaraj}, {van Dishoeck}, {Ray}, {Tychoniec}, {Garatti}, {Francis}, {Gieser}, {van Gelder}, {Tobin}, {Beuther}, {Kavanagh}, {Justtanont}, {Drechsler}, {Navarro}, \& {Perotti}}]{Devaraj2026}
{Devaraj}, R., {van Dishoeck}, E.~F., {Ray}, T.~P., {et~al.} 2026, arXiv e-prints, arXiv:2601.17820.
\newblock \doarXiv{2601.17820}

\bibitem[{{Duch{\^e}ne} {et~al.}(2024){Duch{\^e}ne}, {M{\'e}nard}, {Stapelfeldt}, {Villenave}, {Wolff}, {Perrin}, {Pinte}, {Tazaki}, \& {Padgett}}]{Duchene2024}
{Duch{\^e}ne}, G., {M{\'e}nard}, F., {Stapelfeldt}, K.~R., {et~al.} 2024, \aj, 167, 77, \dodoi{10.3847/1538-3881/acf9a7}

\bibitem[{{Dutrey} {et~al.}(2017){Dutrey}, {Guilloteau}, {Pi{\'e}tu}, {Chapillon}, {Wakelam}, {Di Folco}, {Stoecklin}, {Denis-Alpizar}, {Gorti}, {Teague}, {Henning}, {Semenov}, \& {Grosso}}]{Dutrey2017}
{Dutrey}, A., {Guilloteau}, S., {Pi{\'e}tu}, V., {et~al.} 2017, \aap, 607, A130, \dodoi{10.1051/0004-6361/201730645}

\bibitem[{{Dzyurkevich} {et~al.}(2013){Dzyurkevich}, {Turner}, {Henning}, \& {Kley}}]{Dzyurkevich2013}
{Dzyurkevich}, N., {Turner}, N.~J., {Henning}, T., \& {Kley}, W. 2013, \apj, 765, 114, \dodoi{10.1088/0004-637X/765/2/114}

\bibitem[{{Eisner} {et~al.}(2005){Eisner}, {Hillenbrand}, {White}, {Akeson}, \& {Sargent}}]{Eisner2005}
{Eisner}, J.~A., {Hillenbrand}, L.~A., {White}, R.~J., {Akeson}, R.~L., \& {Sargent}, A.~I. 2005, \apj, 623, 952, \dodoi{10.1086/428828}

\bibitem[{{Ellerbroek} {et~al.}(2014){Ellerbroek}, {Podio}, {Dougados}, {Cabrit}, {Sitko}, {Sana}, {Kaper}, {de Koter}, {Klaassen}, {Mulders}, {Mendigut{\'\i}a}, {Grady}, {Grankin}, {van Winckel}, {Bacciotti}, {Russell}, {Lynch}, {Hammel}, {Beerman}, {Day}, {Huelsman}, {Werren}, {Henden}, \& {Grindlay}}]{Ellerbroek2014}
{Ellerbroek}, L.~E., {Podio}, L., {Dougados}, C., {et~al.} 2014, \aap, 563, A87, \dodoi{10.1051/0004-6361/201323092}

\bibitem[{{Facchini} {et~al.}(2019){Facchini}, {van Dishoeck}, {Manara}, {Tazzari}, {Maud}, {Cazzoletti}, {Rosotti}, {van der Marel}, {Pinilla}, \& {Clarke}}]{Facchini2019}
{Facchini}, S., {van Dishoeck}, E.~F., {Manara}, C.~F., {et~al.} 2019, \aap, 626, L2, \dodoi{10.1051/0004-6361/201935496}

\bibitem[{{Fang} {et~al.}(2023){Fang}, {Pascucci}, {Edwards}, {Gorti}, {Hillenbrand}, \& {Carpenter}}]{Fang2023}
{Fang}, M., {Pascucci}, I., {Edwards}, S., {et~al.} 2023, \apj, 945, 112, \dodoi{10.3847/1538-4357/acb2c9}

\bibitem[{{Fang} {et~al.}(2018){Fang}, {Pascucci}, {Edwards}, {Gorti}, {Banzatti}, {Flock}, {Hartigan}, {Herczeg}, \& {Dupree}}]{Fang2018}
---. 2018, \apj, 868, 28, \dodoi{10.3847/1538-4357/aae780}

\bibitem[{{Federman} {et~al.}(2024){Federman}, {Megeath}, {Rubinstein}, {Gutermuth}, {Narang}, {Tyagi}, {Manoj}, {Anglada}, {Atnagulov}, {Beuther}, {Bourke}, {Brunken}, {Caratti o Garatti}, {Evans}, {Fischer}, {Furlan}, {Green}, {Habel}, {Hartmann}, {Karnath}, {Klaassen}, {Linz}, {Looney}, {Osorio}, {Muzerolle Page}, {Nazari}, {Pokhrel}, {Rahatgaonkar}, {Rocha}, {Sheehan}, {Slavicinska}, {Stanke}, {Stutz}, {Tobin}, {Tychoniec}, {Van Dishoeck}, {Watson}, {Wolk}, \& {Yang}}]{Federman2024}
{Federman}, S.~A., {Megeath}, S.~T., {Rubinstein}, A.~E., {et~al.} 2024, \apj, 966, 41, \dodoi{10.3847/1538-4357/ad2fa0}

\bibitem[{{Fiorellino} {et~al.}(2022){Fiorellino}, {Zsidi}, {K{\'o}sp{\'a}l}, {{\'A}brah{\'a}m}, {B{\'o}di}, {Hussain}, {Manara}, \& {P{\'a}l}}]{Fiorellino2022}
{Fiorellino}, E., {Zsidi}, G., {K{\'o}sp{\'a}l}, {\'A}., {et~al.} 2022, \apj, 938, 93, \dodoi{10.3847/1538-4357/ac912d}

\bibitem[{{Flores} {et~al.}(2023){Flores}, {Ohashi}, {Tobin}, {J{\o}rgensen}, {Takakuwa}, {Li}, {Lin}, {van't Hoff}, {Plunkett}, {Yamato}, {Sai (Insa Choi)}, {Koch}, {Yen}, {Aikawa}, {Aso}, {de Gregorio-Monsalvo}, {Kido}, {Kwon}, {Lee}, {Lee}, {Looney}, {Santamar{\'\i}a-Miranda}, {Sharma}, {Thieme}, {Williams}, {Han}, {Narayanan}, \& {Lai}}]{Flores2023}
{Flores}, C., {Ohashi}, N., {Tobin}, J.~J., {et~al.} 2023, \apj, 958, 98, \dodoi{10.3847/1538-4357/acf7c1}

\bibitem[{{Flores-Rivera} {et~al.}(2023){Flores-Rivera}, {Flock}, {Kurtovic}, {Husemann}, {Banzatti}, {Ringqvist}, {Kamann}, {M{\"u}ller}, {Fendt}, {Garc{\'\i}a Lopez}, {Marleau}, {Henning}, {Carrasco-Gonz{\'a}lez}, {van Boekel}, {Keppler}, {Launhardt}, \& {Aoyama}}]{Flores-Rivera2023}
{Flores-Rivera}, L., {Flock}, M., {Kurtovic}, N.~T., {et~al.} 2023, \aap, 670, A126, \dodoi{10.1051/0004-6361/202141664}

\bibitem[{{Francis} \& {van der Marel}(2020)}]{Francis2020}
{Francis}, L., \& {van der Marel}, N. 2020, \apj, 892, 111, \dodoi{10.3847/1538-4357/ab7b63}

\bibitem[{{Francis} {et~al.}(2026){Francis}, {Tychoniec}, {van Dishoeck}, {Sellek}, {Garatti}, {Le Gouellec}, {Gieser}, {Beuther}, {Vorster}, {Ressler}, {Nazari}, {Tabone}, {Assani}, {Devaraj}, {Tobin}, {Navarro}, {Cort{\'e}s}, {Girart}, {G{\"u}del}, {Henning}, {{\"O}stlin}, {Wright}, \& {Ray}}]{Francis2026}
{Francis}, L., {Tychoniec}, {\L}., {van Dishoeck}, E.~F., {et~al.} 2026, arXiv e-prints, arXiv:2604.13773, \dodoi{10.48550/arXiv.2604.13773}

\bibitem[{{Frank} {et~al.}(2014){Frank}, {Ray}, {Cabrit}, {Hartigan}, {Arce}, {Bacciotti}, {Bally}, {Benisty}, {Eisl{\"o}ffel}, {G{\"u}del}, {Lebedev}, {Nisini}, \& {Raga}}]{Frank2014}
{Frank}, A., {Ray}, T.~P., {Cabrit}, S., {et~al.} 2014, in Protostars and Planets VI, ed. H.~{Beuther}, R.~S. {Klessen}, C.~P. {Dullemond}, \& T.~{Henning}, 451--474, \dodoi{10.2458/azu_uapress_9780816531240-ch020}

\bibitem[{{Frasca} {et~al.}(2017){Frasca}, {Biazzo}, {Alcal{\'a}}, {Manara}, {Stelzer}, {Covino}, \& {Antoniucci}}]{Frasca2017}
{Frasca}, A., {Biazzo}, K., {Alcal{\'a}}, J.~M., {et~al.} 2017, \aap, 602, A33, \dodoi{10.1051/0004-6361/201630108}

\bibitem[{{Gammie}(1996)}]{Gammie1996}
{Gammie}, C.~F. 1996, \apj, 457, 355, \dodoi{10.1086/176735}

\bibitem[{{Gangi} {et~al.}(2020){Gangi}, {Nisini}, {Antoniucci}, {Giannini}, {Biazzo}, {Alcal{\'a}}, {Frasca}, {Munari}, {Arkharov}, {Harutyunyan}, {Manara}, {Rigliaco}, \& {Vitali}}]{Gangi2020}
{Gangi}, M., {Nisini}, B., {Antoniucci}, S., {et~al.} 2020, \aap, 643, A32, \dodoi{10.1051/0004-6361/202038534}

\bibitem[{{Gangi} {et~al.}(2022){Gangi}, {Antoniucci}, {Biazzo}, {Frasca}, {Nisini}, {Alcal{\'a}}, {Giannini}, {Manara}, {Giunta}, {Harutyunyan}, {Munari}, \& {Vitali}}]{Gangi2022}
{Gangi}, M., {Antoniucci}, S., {Biazzo}, K., {et~al.} 2022, \aap, 667, A124, \dodoi{10.1051/0004-6361/202244042}

\bibitem[{{Gardner} {et~al.}(2025){Gardner}, {Isella}, {Li}, {Li}, {Bae}, {Barraza-Alfaro}, {Benisty}, {Cataldi}, {Curone}, {Eisner}, {Facchini}, {Fasano}, {Flock}, {Follette}, {Fukagawa}, {Galloway-Sprietsma}, {Garg}, {Hall}, {Huang}, {Ilee}, {Ireland}, {Izquierdo}, {Johns-Krull}, {Kanagawa}, {Kraus}, {Lesur}, {Liu}, {Longarini}, {Loomis}, {Menard}, {Orihara}, {Pinte}, {Price}, {Ricci}, {Rosotti}, {Sallum}, {Stadler}, {Teague}, {Wafflard-Fernandez}, {Wilner}, {Winter}, {W{\"o}lfer}, {Yen}, {Yoshida}, {Zawadzki}, \& {Zhu}}]{Gardner2025}
{Gardner}, C.~H., {Isella}, A., {Li}, H., {et~al.} 2025, \apjl, 984, L16, \dodoi{10.3847/2041-8213/adc432}

\bibitem[{{Garufi} {et~al.}(2014){Garufi}, {Podio}, {Kamp}, {M{\'e}nard}, {Brittain}, {Eiroa}, {Montesinos}, {Alonso-Mart{\'\i}nez}, {Thi}, \& {Woitke}}]{Garufi2014}
{Garufi}, A., {Podio}, L., {Kamp}, I., {et~al.} 2014, \aap, 567, A141, \dodoi{10.1051/0004-6361/201321987}

\bibitem[{{Giacalone} {et~al.}(2019){Giacalone}, {Teitler}, {K{\"o}nigl}, {Krijt}, \& {Ciesla}}]{Giacalone2019}
{Giacalone}, S., {Teitler}, S., {K{\"o}nigl}, A., {Krijt}, S., \& {Ciesla}, F.~J. 2019, \apj, 882, 33, \dodoi{10.3847/1538-4357/ab311a}

\bibitem[{{Ginski} {et~al.}(2024){Ginski}, {Garufi}, {Benisty}, {Tazaki}, {Dominik}, {Ribas}, {Engler}, {Birnstiel}, {Chauvin}, {Columba}, {Facchini}, {Goncharov}, {Hagelberg}, {Henning}, {Hogerheijde}, {van Holstein}, {Huang}, {Muto}, {Pinilla}, {Kanagawa}, {Kim}, {Kurtovic}, {Langlois}, {Manara}, {Milli}, {Momose}, {Orihara}, {Pawellek}, {Pinte}, {Rab}, {Schmidt}, {Snik}, {Wahhaj}, {Williams}, \& {Zurlo}}]{Ginski2024}
{Ginski}, C., {Garufi}, A., {Benisty}, M., {et~al.} 2024, \aap, 685, A52, \dodoi{10.1051/0004-6361/202244005}

\bibitem[{{Glassgold} {et~al.}(1991){Glassgold}, {Mamon}, \& {Huggins}}]{Glassgold1991}
{Glassgold}, A.~E., {Mamon}, G.~A., \& {Huggins}, P.~J. 1991, \apj, 373, 254, \dodoi{10.1086/170045}

\bibitem[{{Gontcharov}(2006)}]{Gontcharov2006}
{Gontcharov}, G.~A. 2006, Astronomical and Astrophysical Transactions, 25, 145, \dodoi{10.1080/10556790600916780}

\bibitem[{{Gorti} {et~al.}(2015){Gorti}, {Hollenbach}, \& {Dullemond}}]{Gorti2015}
{Gorti}, U., {Hollenbach}, D., \& {Dullemond}, C.~P. 2015, \apj, 804, 29, \dodoi{10.1088/0004-637X/804/1/29}

\bibitem[{{Grant} {et~al.}(2024){Grant}, {Kurtovic}, {van Dishoeck}, {Henning}, {Kamp}, {Nowacki}, {Perraut}, {Banzatti}, {Temmink}, {Christiaens}, {Samland}, {Gasman}, {Tabone}, {G{\"u}del}, {Lagage}, {Arabhavi}, {Barrado}, {Caratti o Garatti}, {Glauser}, {Jang}, {Kanwar}, {Lahuis}, {Morales-Calder{\'o}n}, {Olofsson}, {Perotti}, {Schwarz}, {Vlasblom}, {Garcia Lopez}, \& {Long}}]{Grant2024}
{Grant}, S.~L., {Kurtovic}, N.~T., {van Dishoeck}, E.~F., {et~al.} 2024, \aap, 689, A85, \dodoi{10.1051/0004-6361/202450768}

\bibitem[{{Greenwood} {et~al.}(2017){Greenwood}, {Kamp}, {Waters}, {Woitke}, {Thi}, {Rab}, {Aresu}, \& {Spaans}}]{Greenwood2017}
{Greenwood}, A.~J., {Kamp}, I., {Waters}, L.~B.~F.~M., {et~al.} 2017, \aap, 601, A44, \dodoi{10.1051/0004-6361/201629389}

\bibitem[{{Gressel} {et~al.}(2015){Gressel}, {Turner}, {Nelson}, \& {McNally}}]{Gressel2015}
{Gressel}, O., {Turner}, N.~J., {Nelson}, R.~P., \& {McNally}, C.~P. 2015, \apj, 801, 84, \dodoi{10.1088/0004-637X/801/2/84}

\bibitem[{{Guerra-Alvarado} {et~al.}(2025){Guerra-Alvarado}, {van der Marel}, {Williams}, {Pinilla}, {Mulders}, {Lambrechts}, \& {Sanchez}}]{Guerra-Alvarado2025}
{Guerra-Alvarado}, O.~M., {van der Marel}, N., {Williams}, J.~P., {et~al.} 2025, \aap, 696, A232, \dodoi{10.1051/0004-6361/202453338}

\bibitem[{{Guti{\'e}rrez Albarr{\'a}n} {et~al.}(2020){Guti{\'e}rrez Albarr{\'a}n}, {Montes}, {G{\'o}mez Garrido}, {Tabernero}, {Gonz{\'a}lez Hern{\'a}ndez}, {Marfil}, {Frasca}, {Lanzafame}, {Klutsch}, {Franciosini}, {Randich}, {Smiljanic}, {Korn}, {Gilmore}, {Alfaro}, {Baratella}, {Bayo}, {Bensby}, {Bonito}, {Carraro}, {Delgado Mena}, {Feltzing}, {Gonneau}, {Heiter}, {Hourihane}, {Jim{\'e}nez Esteban}, {Jofre}, {Masseron}, {Monaco}, {Morbidelli}, {Prisinzano}, {Roccatagliata}, {Sousa}, {Van der Swaelmen}, {Worley}, \& {Zaggia}}]{Gutierrezalbarran2020}
{Guti{\'e}rrez Albarr{\'a}n}, M.~L., {Montes}, D., {G{\'o}mez Garrido}, M., {et~al.} 2020, \aap, 643, A71, \dodoi{10.1051/0004-6361/202037620}

\bibitem[{Harris {et~al.}(2020)Harris, Millman, van~der Walt, Gommers, Virtanen, Cournapeau, Wieser, Taylor, Berg, Smith, Kern, Picus, Hoyer, van Kerkwijk, Brett, Haldane, del R{\'{i}}o, Wiebe, Peterson, G{\'{e}}rard-Marchant, Sheppard, Reddy, Weckesser, Abbasi, Gohlke, \& Oliphant}]{harris2020array}
Harris, C.~R., Millman, K.~J., van~der Walt, S.~J., {et~al.} 2020, Nature, 585, 357, \dodoi{10.1038/s41586-020-2649-2}

\bibitem[{{Harris} {et~al.}(2012){Harris}, {Andrews}, {Wilner}, \& {Kraus}}]{Harris2012}
{Harris}, R.~J., {Andrews}, S.~M., {Wilner}, D.~J., \& {Kraus}, A.~L. 2012, \apj, 751, 115, \dodoi{10.1088/0004-637X/751/2/115}

\bibitem[{{Hartigan} {et~al.}(1995){Hartigan}, {Edwards}, \& {Ghandour}}]{Hartigan1995}
{Hartigan}, P., {Edwards}, S., \& {Ghandour}, L. 1995, \apj, 452, 736, \dodoi{10.1086/176344}

\bibitem[{{Hartmann} {et~al.}(1998){Hartmann}, {Calvet}, {Gullbring}, \& {D'Alessio}}]{Hartmann1998}
{Hartmann}, L., {Calvet}, N., {Gullbring}, E., \& {D'Alessio}, P. 1998, \apj, 495, 385, \dodoi{10.1086/305277}

\bibitem[{{Hendler} {et~al.}(2020){Hendler}, {Pascucci}, {Pinilla}, {Tazzari}, {Carpenter}, {Malhotra}, \& {Testi}}]{Hendler2020}
{Hendler}, N., {Pascucci}, I., {Pinilla}, P., {et~al.} 2020, \apj, 895, 126, \dodoi{10.3847/1538-4357/ab70ba}

\bibitem[{{Hendler} {et~al.}(2018){Hendler}, {Pinilla}, {Pascucci}, {Pohl}, {Mulders}, {Henning}, {Dong}, {Clarke}, {Owen}, \& {Hollenbach}}]{Hendler2018}
{Hendler}, N.~P., {Pinilla}, P., {Pascucci}, I., {et~al.} 2018, \mnras, 475, L62, \dodoi{10.1093/mnrasl/slx184}

\bibitem[{{Henning} {et~al.}(2024){Henning}, {Kamp}, {Samland}, {Arabhavi}, {Kanwar}, {van Dishoeck}, {G{\"u}del}, {Lagage}, {Waelkens}, {Abergel}, {Absil}, {Barrado}, {Boccaletti}, {Bouwman}, {Caratti o Garatti}, {Geers}, {Glauser}, {Lahuis}, {Mueller}, {Nehm{\'e}}, {Olofsson}, {Pantin}, {Ray}, {Scheithauer}, {Vandenbussche}, {Waters}, {Wright}, {Argyriou}, {Christiaens}, {Franceschi}, {Gasman}, {Grant}, {Guadarrama}, {Jang}, {Morales-Calder{\'o}n}, {Pawellek}, {Perotti}, {Rodgers-Lee}, {Schreiber}, {Schwarz}, {Tabone}, {Temmink}, {Vlasblom}, {Colina}, {Greve}, \& {{\"O}stlin}}]{Henning2024}
{Henning}, T., {Kamp}, I., {Samland}, M., {et~al.} 2024, \pasp, 136, 054302, \dodoi{10.1088/1538-3873/ad3455}

\bibitem[{{Henning} {et~al.}(2017){Henning}, {Abergel}, {Absil}, {Barrado Navascues}, {Boccaletti}, {Bouwman}, {Caratti o Garatti}, {Geers}, {Glauser}, {Guedel}, {Kamp}, {Lagage}, {Lahuis}, {Mueller}, {Olofsson}, {Ray}, {Scheithauer}, {Van Dishoeck}, {Vandenbussche}, {Waelkens}, {Waters}, {nehme}, \& {pantin}}]{Henning2017}
{Henning}, T.~K., {Abergel}, A., {Absil}, O., {et~al.} 2017, {MIRI EC Protoplanetary and Debris Disks Survey}, JWST Proposal. Cycle 1, ID. \#1282

\bibitem[{{Hollenbach} \& {Gorti}(2009)}]{Hollenbach2009}
{Hollenbach}, D., \& {Gorti}, U. 2009, \apj, 703, 1203, \dodoi{10.1088/0004-637X/703/2/1203}

\bibitem[{{Hourihane} {et~al.}(2023){Hourihane}, {Fran{\c{c}}ois}, {Worley}, {Magrini}, {Gonneau}, {Casey}, {Gilmore}, {Randich}, {Sacco}, {Recio-Blanco}, {Korn}, {Allende Prieto}, {Smiljanic}, {Blomme}, {Bragaglia}, {Walton}, {Van Eck}, {Bensby}, {Lanzafame}, {Frasca}, {Franciosini}, {Damiani}, {Lind}, {Bergemann}, {Bonifacio}, {Hill}, {Lobel}, {Montes}, {Feuillet}, {Tautvai{\v{s}}ien{\.{e}}}, {Guiglion}, {Tabernero}, {Gonz{\'a}lez Hern{\'a}ndez}, {Gebran}, {Van der Swaelmen}, {Mikolaitis}, {Daflon}, {Merle}, {Morel}, {Lewis}, {Gonz{\'a}lez Solares}, {Murphy}, {Jeffries}, {Jackson}, {Feltzing}, {Prusti}, {Carraro}, {Biazzo}, {Prisinzano}, {Jofr{\'e}}, {Zaggia}, {Drazdauskas}, {Stonkut{\'e}}, {Marfil}, {Jim{\'e}nez-Esteban}, {Mahy}, {Guti{\'e}rrez Albarr{\'a}n}, {Berlanas}, {Santos}, {Morbidelli}, {Spina}, \& {Minkevi{\v{c}}i{\={u}}t{\.{e}}}}]{Hourihare2023}
{Hourihane}, A., {Fran{\c{c}}ois}, P., {Worley}, C.~C., {et~al.} 2023, \aap, 676, A129, \dodoi{10.1051/0004-6361/202345910}

\bibitem[{{Huang} {et~al.}(2018{\natexlab{a}}){Huang}, {Andrews}, {Dullemond}, {Isella}, {P{\'e}rez}, {Guzm{\'a}n}, {{\"O}berg}, {Zhu}, {Zhang}, {Bai}, {Benisty}, {Birnstiel}, {Carpenter}, {Hughes}, {Ricci}, {Weaver}, \& {Wilner}}]{Huang2018a}
{Huang}, J., {Andrews}, S.~M., {Dullemond}, C.~P., {et~al.} 2018{\natexlab{a}}, \apjl, 869, L42, \dodoi{10.3847/2041-8213/aaf740}

\bibitem[{{Huang} {et~al.}(2018{\natexlab{b}}){Huang}, {Andrews}, {P{\'e}rez}, {Zhu}, {Dullemond}, {Isella}, {Benisty}, {Bai}, {Birnstiel}, {Carpenter}, {Guzm{\'a}n}, {Hughes}, {{\"O}berg}, {Ricci}, {Wilner}, \& {Zhang}}]{Huang2018b}
{Huang}, J., {Andrews}, S.~M., {P{\'e}rez}, L.~M., {et~al.} 2018{\natexlab{b}}, \apjl, 869, L43, \dodoi{10.3847/2041-8213/aaf7a0}

\bibitem[{Hunter(2007)}]{Hunter:2007}
Hunter, J.~D. 2007, Computing in Science \& Engineering, 9, 90, \dodoi{10.1109/MCSE.2007.55}

\bibitem[{{J{\"o}nsson} {et~al.}(2020){J{\"o}nsson}, {Holtzman}, {Allende Prieto}, {Cunha}, {Garc{\'\i}a-Hern{\'a}ndez}, {Hasselquist}, {Masseron}, {Osorio}, {Shetrone}, {Smith}, {Stringfellow}, {Bizyaev}, {Edvardsson}, {Majewski}, {M{\'e}sz{\'a}ros}, {Souto}, {Zamora}, {Beaton}, {Bovy}, {Donor}, {Pinsonneault}, {Poovelil}, \& {Sobeck}}]{Jonsson2020}
{J{\"o}nsson}, H., {Holtzman}, J.~A., {Allende Prieto}, C., {et~al.} 2020, \aj, 160, 120, \dodoi{10.3847/1538-3881/aba592}

\bibitem[{{Keppler} {et~al.}(2019){Keppler}, {Teague}, {Bae}, {Benisty}, {Henning}, {van Boekel}, {Chapillon}, {Pinilla}, {Williams}, {Bertrang}, {Facchini}, {Flock}, {Ginski}, {Juhasz}, {Klahr}, {Liu}, {M{\"u}ller}, {P{\'e}rez}, {Pohl}, {Rosotti}, {Samland}, \& {Semenov}}]{Keppler2019}
{Keppler}, M., {Teague}, R., {Bae}, J., {et~al.} 2019, \aap, 625, A118, \dodoi{10.1051/0004-6361/201935034}

\bibitem[{{Kiman} {et~al.}(2019){Kiman}, {Schmidt}, {Angus}, {Cruz}, {Faherty}, \& {Rice}}]{Kiman2019}
{Kiman}, R., {Schmidt}, S.~J., {Angus}, R., {et~al.} 2019, \aj, 157, 231, \dodoi{10.3847/1538-3881/ab1753}

\bibitem[{{Kimmig} {et~al.}(2020){Kimmig}, {Dullemond}, \& {Kley}}]{Kimmig2020}
{Kimmig}, C.~N., {Dullemond}, C.~P., \& {Kley}, W. 2020, \aap, 633, A4, \dodoi{10.1051/0004-6361/201936412}

\bibitem[{{Kirwan} {et~al.}(2022){Kirwan}, {Murphy}, {Schneider}, {Whelan}, {Dougados}, \& {Eisl{\"o}ffel}}]{Kirwan2022}
{Kirwan}, A., {Murphy}, A., {Schneider}, P.~C., {et~al.} 2022, \aap, 663, A30, \dodoi{10.1051/0004-6361/202142862}

\bibitem[{{Klaassen} {et~al.}(2013){Klaassen}, {Juhasz}, {Mathews}, {Mottram}, {De Gregorio-Monsalvo}, {van Dishoeck}, {Takahashi}, {Akiyama}, {Chapillon}, {Espada}, {Hales}, {Hogerheijde}, {Rawlings}, {Schmalzl}, \& {Testi}}]{Klaassen2013}
{Klaassen}, P., {Juhasz}, A., {Mathews}, G., {et~al.} 2013, in Protostars and Planets VI Posters

\bibitem[{Kramida {et~al.}(2024)Kramida, Ralchenko, Reader, \& Team}]{NIST_ASD_2024}
Kramida, A., Ralchenko, Y., Reader, J., \& Team, N.~A. 2024, {NIST Atomic Spectra Database}, 5.12, \url{https://physics.nist.gov/asd}, \dodoi{10.18434/T4W30F}

\bibitem[{{Kraus} {et~al.}(2008){Kraus}, {Ireland}, {Martinache}, \& {Lloyd}}]{Kraus2008}
{Kraus}, A.~L., {Ireland}, M.~J., {Martinache}, F., \& {Lloyd}, J.~P. 2008, \apj, 679, 762, \dodoi{10.1086/587435}

\bibitem[{{Kurtovic} {et~al.}(2018){Kurtovic}, {P{\'e}rez}, {Benisty}, {Zhu}, {Zhang}, {Huang}, {Andrews}, {Dullemond}, {Isella}, {Bai}, {Carpenter}, {Guzm{\'a}n}, {Ricci}, \& {Wilner}}]{Kurtovic2018}
{Kurtovic}, N.~T., {P{\'e}rez}, L.~M., {Benisty}, M., {et~al.} 2018, \apjl, 869, L44, \dodoi{10.3847/2041-8213/aaf746}

\bibitem[{{Kurtovic} {et~al.}(2026){Kurtovic}, {Grant}, {Temmink}, {Sellek}, {van Dishoeck}, {Henning}, {Kamp}, {Christiaens}, {Banzatti}, {Gasman}, {Kaeufer}, {Stapper}, {Franceschi}, {G{\"u}del}, {Lagage}, {Vlasblom}, {Perotti}, {Schwarz}, \& {Somigliana}}]{Kurtovic2026}
{Kurtovic}, N.~T., {Grant}, S.~L., {Temmink}, M., {et~al.} 2026, \aap, 705, A97, \dodoi{10.1051/0004-6361/202554927}

\bibitem[{{Kutra} {et~al.}(2025){Kutra}, {Prato}, {Tofflemire}, {Akeson}, {Schaefer}, {Tang}, {Segura-Cox}, {Johns-Krull}, {Kraus}, {Andrews}, \& {Jensen}}]{Kutra2025}
{Kutra}, T., {Prato}, L., {Tofflemire}, B.~M., {et~al.} 2025, \aj, 169, 20, \dodoi{10.3847/1538-3881/ad900a}

\bibitem[{{Lee}(2020)}]{Lee2020}
{Lee}, C.-F. 2020, \aapr, 28, 1, \dodoi{10.1007/s00159-020-0123-7}

\bibitem[{{Lin} {et~al.}(2023){Lin}, {Ip}, {Hsiao}, {Chang}, {Song}, \& {Luo}}]{Lin2023}
{Lin}, C.-L., {Ip}, W.-H., {Hsiao}, Y., {et~al.} 2023, \aj, 166, 82, \dodoi{10.3847/1538-3881/ace322}

\bibitem[{{Long} {et~al.}(2019){Long}, {Herczeg}, {Harsono}, {Pinilla}, {Tazzari}, {Manara}, {Pascucci}, {Cabrit}, {Nisini}, {Johnstone}, {Edwards}, {Salyk}, {Menard}, {Lodato}, {Boehler}, {Mace}, {Liu}, {Mulders}, {Hendler}, {Ragusa}, {Fischer}, {Banzatti}, {Rigliaco}, {van de Plas}, {Dipierro}, {Gully-Santiago}, \& {Lopez-Valdivia}}]{Long2019}
{Long}, F., {Herczeg}, G.~J., {Harsono}, D., {et~al.} 2019, \apj, 882, 49, \dodoi{10.3847/1538-4357/ab2d2d}

\bibitem[{{Loomis} {et~al.}(2017){Loomis}, {{\"O}berg}, {Andrews}, \& {MacGregor}}]{Loomis2017}
{Loomis}, R.~A., {{\"O}berg}, K.~I., {Andrews}, S.~M., \& {MacGregor}, M.~A. 2017, \apj, 840, 23, \dodoi{10.3847/1538-4357/aa6c63}

\bibitem[{{Louvet} {et~al.}(2018){Louvet}, {Dougados}, {Cabrit}, {Mardones}, {M{\'e}nard}, {Tabone}, {Pinte}, \& {Dent}}]{Louvet2018}
{Louvet}, F., {Dougados}, C., {Cabrit}, S., {et~al.} 2018, \aap, 618, A120, \dodoi{10.1051/0004-6361/201731733}

\bibitem[{{Luhman}(2004)}]{Luhman2004}
{Luhman}, K.~L. 2004, \apj, 617, 1216, \dodoi{10.1086/425647}

\bibitem[{{Luhman} {et~al.}(2007){Luhman}, {Adame}, {D'Alessio}, {Calvet}, {McLeod}, {Bohac}, {Forrest}, {Hartmann}, {Sargent}, \& {Watson}}]{Luhman2007}
{Luhman}, K.~L., {Adame}, L., {D'Alessio}, P., {et~al.} 2007, \apj, 666, 1219, \dodoi{10.1086/520712}

\bibitem[{{Manara} {et~al.}(2023){Manara}, {Ansdell}, {Rosotti}, {Hughes}, {Armitage}, {Lodato}, \& {Williams}}]{Manara2023}
{Manara}, C.~F., {Ansdell}, M., {Rosotti}, G.~P., {et~al.} 2023, in Astronomical Society of the Pacific Conference Series, Vol. 534, Protostars and Planets VII, ed. S.~{Inutsuka}, Y.~{Aikawa}, T.~{Muto}, K.~{Tomida}, \& M.~{Tamura}, 539, \dodoi{10.48550/arXiv.2203.09930}

\bibitem[{{Manara} {et~al.}(2015){Manara}, {Testi}, {Natta}, \& {Alcal{\'a}}}]{Manara2015}
{Manara}, C.~F., {Testi}, L., {Natta}, A., \& {Alcal{\'a}}, J.~M. 2015, \aap, 579, A66, \dodoi{10.1051/0004-6361/201526169}

\bibitem[{{Manara} {et~al.}(2020){Manara}, {Natta}, {Rosotti}, {Alcal{\'a}}, {Nisini}, {Lodato}, {Testi}, {Pascucci}, {Hillenbrand}, {Carpenter}, {Scholz}, {Fedele}, {Frasca}, {Mulders}, {Rigliaco}, {Scardoni}, \& {Zari}}]{Manara2020}
{Manara}, C.~F., {Natta}, A., {Rosotti}, G.~P., {et~al.} 2020, \aap, 639, A58, \dodoi{10.1051/0004-6361/202037949}

\bibitem[{{Manset} {et~al.}(2009){Manset}, {Bastien}, {M{\'e}nard}, {Bertout}, {Le van Suu}, \& {Boivin}}]{Manset2009}
{Manset}, N., {Bastien}, P., {M{\'e}nard}, F., {et~al.} 2009, \aap, 499, 137, \dodoi{10.1051/0004-6361/200810945}

\bibitem[{{McCabe} {et~al.}(2011){McCabe}, {Duch{\^e}ne}, {Pinte}, {Stapelfeldt}, {Ghez}, \& {M{\'e}nard}}]{McCabe2011}
{McCabe}, C., {Duch{\^e}ne}, G., {Pinte}, C., {et~al.} 2011, \apj, 727, 90, \dodoi{10.1088/0004-637X/727/2/90}

\bibitem[{{McClure} {et~al.}(2017){McClure}, {Bailey}, {Beck}, {Boogert}, {Brown}, {Caselli}, {Chiar}, {Egami}, {Fraser}, {Garrod}, {Gordon}, {Ioppolo}, {Jimenez-Serra}, {Jorgensen}, {Kristensen}, {Linnartz}, {McCoustra}, {Murillo}, {Noble}, {Oberg}, {Palumbo}, {Pendleton}, {Pontoppidan}, {Van Dishoeck}, \& {Viti}}]{McClure2017}
{McClure}, M., {Bailey}, J., {Beck}, T., {et~al.} 2017, {IceAge: Chemical Evolution of Ices during Star Formation}, JWST Proposal ID 1309. Cycle 0 Early Release Science

\bibitem[{{McClure} {et~al.}(2021){McClure}, {Beck}, {Boogert}, {Cordiner}, {Dartois}, {Drozdovskaya}, {Espaillat}, {Harsono}, {Melnick}, {Palumbo}, {Pendleton}, {Pontoppidan}, {Terada}, \& {Van Dishoeck}}]{McClure2021}
{McClure}, M., {Beck}, T., {Boogert}, A.~C., {et~al.} 2021, {Mapping inclined disk astrochemical signatures (MIDAS)}, JWST Proposal. Cycle 1, ID. \#1751

\bibitem[{{McGinnis} {et~al.}(2018){McGinnis}, {Dougados}, {Alencar}, {Bouvier}, \& {Cabrit}}]{McGinnis2018}
{McGinnis}, P., {Dougados}, C., {Alencar}, S.~H.~P., {Bouvier}, J., \& {Cabrit}, S. 2018, \aap, 620, A87, \dodoi{10.1051/0004-6361/201731629}

\bibitem[{{Melo}(2003)}]{Melo2003}
{Melo}, C.~H.~F. 2003, \aap, 410, 269, \dodoi{10.1051/0004-6361:20031242}

\bibitem[{{Mendigut{\'\i}a} {et~al.}(2018){Mendigut{\'\i}a}, {Oudmaijer}, {Schneider}, {Hu{\'e}lamo}, {Baines}, {Brittain}, \& {Aberasturi}}]{Mendigutia2018}
{Mendigut{\'\i}a}, I., {Oudmaijer}, R.~D., {Schneider}, P.~C., {et~al.} 2018, \aap, 618, L9, \dodoi{10.1051/0004-6361/201834233}

\bibitem[{{Miley} {et~al.}(2024){Miley}, {Carpenter}, {Booth}, {Jennings}, {Haworth}, {Vioque}, {Andrews}, {Wilner}, {Benisty}, {Huang}, {Perez}, {Guzman}, {Ricci}, \& {Isella}}]{Miley2024}
{Miley}, J.~M., {Carpenter}, J., {Booth}, R., {et~al.} 2024, \aap, 682, A55, \dodoi{10.1051/0004-6361/202347135}

\bibitem[{{Miley} {et~al.}(2025){Miley}, {P{\'e}rez}, {Agurto-Gangas}, {Sierra}, {Trapman}, {Vioque}, {Kurtovic}, {Pinilla}, {Pascucci}, {Zhang}, {Anania}, {Carpenter}, {Cieza}, {Deng}, {Gonz{\'a}lez-Ruilova}, {Rosotti}, {Ruiz-Rodriguez}, \& {TorresVillanueva}}]{Miley2025}
{Miley}, J.~M., {P{\'e}rez}, L.~M., {Agurto-Gangas}, C., {et~al.} 2025, \apj, 989, 11, \dodoi{10.3847/1538-4357/add25c}

\bibitem[{{Mitchell} {et~al.}(1997){Mitchell}, {Sargent}, \& {Mannings}}]{Mitchell1997}
{Mitchell}, G.~F., {Sargent}, A.~I., \& {Mannings}, V. 1997, \apjl, 483, L127, \dodoi{10.1086/310750}

\bibitem[{{Mori} {et~al.}(2025){Mori}, {Bai}, \& {Tomida}}]{Mori2025}
{Mori}, S., {Bai}, X.-N., \& {Tomida}, K. 2025, \apj, 992, 85, \dodoi{10.3847/1538-4357/adf8d7}

\bibitem[{{Nakatani} {et~al.}(2026){Nakatani}, {Rosotti}, {Tabone}, \& {Sellek}}]{Nakatani2026}
{Nakatani}, R., {Rosotti}, G., {Tabone}, B., \& {Sellek}, A. 2026, \aap, 706, A295, \dodoi{10.1051/0004-6361/202556740}

\bibitem[{{Narang} {et~al.}(2025){Narang}, {Ohashi}, {Tobin}, {McClure}, {J{\o}rgensen}, {Sai (Insa Choi)}, \& {eDisk + IceAge Team}}]{Narang2025}
{Narang}, M., {Ohashi}, N., {Tobin}, J.~J., {et~al.} 2025, \aj, 169, 192, \dodoi{10.3847/1538-3881/adb1ba}

\bibitem[{{Narang} {et~al.}(2026{\natexlab{a}}){Narang}, {Pontoppidan}, {Salyk}, {Arulanantham}, {Blake}, {Banzatti}, {Najita}, {Pascucci}, {Huang}, {Krijt}, {Oberg}, {Rosotti}, {Kaeufer}, {Dahl}, {Cleeves}, {Zhang}, \& {Green}}]{Narang2026b}
{Narang}, M., {Pontoppidan}, K.~M., {Salyk}, C., {et~al.} 2026{\natexlab{a}}, arXiv e-prints, arXiv:2605.07016.
\newblock \doarXiv{2605.07016}

\bibitem[{{Narang} {et~al.}(2026{\natexlab{b}}){Narang}, {Tyagi}, {Ohashi}, {Manoj}, {Megeath}, {Tobin}, {Van Dishoeck}, {Evans}, {Watson}, {Caratti o Garatti}, {J{\o}rgensen}, {Gutermuth}, {Aso}, {Beuther}, {Looney}, {Neufeld}, {Anglada}, {Osorio}, {Rubinstein}, {Federman}, {Hartmann}, {Nazari}, {Karnath}, {Linz}, {Stanke}, {Bourke}, {Yang}, {Kuiper}, {Green}, {Klaassen}, {Zakri}, {Habel}, {Brunken}, {Muzerolle}, {Slavicinska}, {Stutz}, {Tychoniec}, {Wolk}, {Rocha}, \& {Fischer}}]{Narang2026a}
{Narang}, M., {Tyagi}, H., {Ohashi}, N., {et~al.} 2026{\natexlab{b}}, \apj, 1000, 184, \dodoi{10.3847/1538-4357/ae4354}

\bibitem[{{Natta} {et~al.}(2014){Natta}, {Testi}, {Alcal{\'a}}, {Rigliaco}, {Covino}, {Stelzer}, \& {D'Elia}}]{Natta2014}
{Natta}, A., {Testi}, L., {Alcal{\'a}}, J.~M., {et~al.} 2014, \aap, 569, A5, \dodoi{10.1051/0004-6361/201424136}

\bibitem[{{Natta} {et~al.}(2006){Natta}, {Testi}, \& {Randich}}]{Natta2006}
{Natta}, A., {Testi}, L., \& {Randich}, S. 2006, \aap, 452, 245, \dodoi{10.1051/0004-6361:20054706}

\bibitem[{{Navarro} {et~al.}(2025){Navarro}, {Nisini}, {Giannini}, {Kavanagh}, {Caratti o Garatti}, {Antoniucci}, {Arce}, {Bacciotti}, {Cabrit}, {Coffey}, {Dougados}, {Eisl{\"o}ffel}, {Hartigan}, {Crespo}, {Podio}, {van Dishoeck}, \& {Whelan}}]{Navarro2025}
{Navarro}, M.~G., {Nisini}, B., {Giannini}, T., {et~al.} 2025, \apj, 995, 199, \dodoi{10.3847/1538-4357/ae1f8f}

\bibitem[{{Neufeld} {et~al.}(1998){Neufeld}, {Melnick}, \& {Harwit}}]{Neufeld1998}
{Neufeld}, D.~A., {Melnick}, G.~J., \& {Harwit}, M. 1998, \apjl, 506, L75, \dodoi{10.1086/311636}

\bibitem[{{Neufeld} {et~al.}(2006){Neufeld}, {Melnick}, {Sonnentrucker}, {Bergin}, {Green}, {Kim}, {Watson}, {Forrest}, \& {Pipher}}]{Neufeld2006}
{Neufeld}, D.~A., {Melnick}, G.~J., {Sonnentrucker}, P., {et~al.} 2006, \apj, 649, 816, \dodoi{10.1086/506604}

\bibitem[{Newville {et~al.}(2025)Newville, Otten, Nelson, Stensitzki, Ingargiola, Allan, Fox, Carter, \& Rawlik}]{lmfit}
Newville, M., Otten, R., Nelson, A., {et~al.} 2025, LMFIT: Non-Linear Least-Squares Minimization and Curve-Fitting for Python, 1.3.4,  Zenodo, \dodoi{10.5281/zenodo.16175987}

\bibitem[{{Nisini} {et~al.}(2018){Nisini}, {Antoniucci}, {Alcal{\'a}}, {Giannini}, {Manara}, {Natta}, {Fedele}, \& {Biazzo}}]{Nisini2018}
{Nisini}, B., {Antoniucci}, S., {Alcal{\'a}}, J.~M., {et~al.} 2018, \aap, 609, A87, \dodoi{10.1051/0004-6361/201730834}

\bibitem[{{Nisini} {et~al.}(2005){Nisini}, {Antoniucci}, {Giannini}, \& {Lorenzetti}}]{Nisini2005}
{Nisini}, B., {Antoniucci}, S., {Giannini}, T., \& {Lorenzetti}, D. 2005, \aap, 429, 543, \dodoi{10.1051/0004-6361:20041409}

\bibitem[{{Nisini} {et~al.}(2015){Nisini}, {Santangelo}, {Giannini}, {Antoniucci}, {Cabrit}, {Codella}, {Davis}, {Eisl{\"o}ffel}, {Kristensen}, {Herczeg}, {Neufeld}, \& {van Dishoeck}}]{Nisini2015}
{Nisini}, B., {Santangelo}, G., {Giannini}, T., {et~al.} 2015, \apj, 801, 121, \dodoi{10.1088/0004-637X/801/2/121}

\bibitem[{{Nisini} {et~al.}(2024{\natexlab{a}}){Nisini}, {Navarro}, {Giannini}, {Antoniucci}, {Kavanagh}, {Hartigan}, {Bacciotti}, {Caratti o Garatti}, {Noriega-Crespo}, {van Dishoeck}, {Whelan}, {Arce}, {Cabrit}, {Coffey}, {Fedele}, {Eisl{\"o}ffel}, {Palumbo}, {Podio}, {Ray}, {Schultze}, {Urso}, {Alcal{\'a}}, {Bautista}, {Codella}, {Greene}, \& {Manara}}]{Nisini2024b}
{Nisini}, B., {Navarro}, M.~G., {Giannini}, T., {et~al.} 2024{\natexlab{a}}, \apj, 967, 168, \dodoi{10.3847/1538-4357/ad3d5a}

\bibitem[{{Nisini} {et~al.}(2024{\natexlab{b}}){Nisini}, {Gangi}, {Giannini}, {Antoniucci}, {Biazzo}, {Frasca}, {Alcal{\'a}}, {Manara}, \& {Weber}}]{Nisini2024a}
{Nisini}, B., {Gangi}, M., {Giannini}, T., {et~al.} 2024{\natexlab{b}}, \aap, 683, A116, \dodoi{10.1051/0004-6361/202346742}

\bibitem[{{Ogihara} {et~al.}(2018){Ogihara}, {Kokubo}, {Suzuki}, \& {Morbidelli}}]{Ogihara2018}
{Ogihara}, M., {Kokubo}, E., {Suzuki}, T.~K., \& {Morbidelli}, A. 2018, \aap, 615, A63, \dodoi{10.1051/0004-6361/201832720}

\bibitem[{{Okuzumi} \& {Inutsuka}(2015)}]{Okuzumi2015}
{Okuzumi}, S., \& {Inutsuka}, S.-i. 2015, \apj, 800, 47, \dodoi{10.1088/0004-637X/800/1/47}

\bibitem[{{Orihara} {et~al.}(2023){Orihara}, {Momose}, {Muto}, {Hashimoto}, {Liu}, {Tsukagoshi}, {Kudo}, {Takahashi}, {Yang}, {Hasegawa}, {Dong}, {Konishi}, \& {Akiyama}}]{Orihara2023}
{Orihara}, R., {Momose}, M., {Muto}, T., {et~al.} 2023, \pasj, 75, 424, \dodoi{10.1093/pasj/psad009}

\bibitem[{{Paneque-Carre{\~n}o} {et~al.}(2021){Paneque-Carre{\~n}o}, {P{\'e}rez}, {Benisty}, {Hall}, {Veronesi}, {Lodato}, {Sierra}, {Carpenter}, {Andrews}, {Bae}, {Henning}, {Kwon}, {Linz}, {Loinard}, {Pinte}, {Ricci}, {Tazzari}, {Testi}, \& {Wilner}}]{Paneque-Carreno2021}
{Paneque-Carre{\~n}o}, T., {P{\'e}rez}, L.~M., {Benisty}, M., {et~al.} 2021, \apj, 914, 88, \dodoi{10.3847/1538-4357/abf243}

\bibitem[{{Pascucci} {et~al.}(2021){Pascucci}, {Alexander}, {Ballabio}, {Clarke}, {Gaspar}, \& {Gorti}}]{Pascucci2021}
{Pascucci}, I., {Alexander}, R., {Ballabio}, G., {et~al.} 2021, {Caught in the act of dispersing their disks? MIRI MRS can tell}, JWST Proposal. Cycle 1, ID. \#2260

\bibitem[{{Pascucci} {et~al.}(2023){Pascucci}, {Cabrit}, {Edwards}, {Gorti}, {Gressel}, \& {Suzuki}}]{Pascucci2023}
{Pascucci}, I., {Cabrit}, S., {Edwards}, S., {et~al.} 2023, in Astronomical Society of the Pacific Conference Series, Vol. 534, Protostars and Planets VII, ed. S.~{Inutsuka}, Y.~{Aikawa}, T.~{Muto}, K.~{Tomida}, \& M.~{Tamura}, 567, \dodoi{10.48550/arXiv.2203.10068}

\bibitem[{{Pascucci} \& {Sterzik}(2009)}]{Pascucci2009}
{Pascucci}, I., \& {Sterzik}, M. 2009, \apj, 702, 724, \dodoi{10.1088/0004-637X/702/1/724}

\bibitem[{{Pascucci} {et~al.}(2020){Pascucci}, {Banzatti}, {Gorti}, {Fang}, {Pontoppidan}, {Alexander}, {Ballabio}, {Edwards}, {Salyk}, {Sacco}, {Flaccomio}, {Blake}, {Carmona}, {Hall}, {Kamp}, {K{\"a}ufl}, {Meeus}, {Meyer}, {Pauly}, {Steendam}, \& {Sterzik}}]{Pascucci2020}
{Pascucci}, I., {Banzatti}, A., {Gorti}, U., {et~al.} 2020, \apj, 903, 78, \dodoi{10.3847/1538-4357/abba3c}

\bibitem[{{Pascucci} {et~al.}(2025){Pascucci}, {Beck}, {Cabrit}, {Bajaj}, {Edwards}, {Louvet}, {Najita}, {Skinner}, {Gorti}, {Salyk}, {Brittain}, {Krijt}, {Muzerolle Page}, {Ruaud}, {Schwarz}, {Semenov}, {Duch{\^e}ne}, \& {Villenave}}]{Pascucci2025}
{Pascucci}, I., {Beck}, T.~L., {Cabrit}, S., {et~al.} 2025, Nature Astronomy, 9, 81, \dodoi{10.1038/s41550-024-02385-7}

\bibitem[{{Patapis} {et~al.}(2024){Patapis}, {Argyriou}, {Law}, {Glauser}, {Glasse}, {Labiano}, {{\'A}lvarez-M{\'a}rquez}, {Kavanagh}, {Gasman}, {Mueller}, {Larson}, {Vandenbussche}, {Lee}, {Klaassen}, {Guillard}, \& {Wright}}]{Patapis2024}
{Patapis}, P., {Argyriou}, I., {Law}, D.~R., {et~al.} 2024, \aap, 682, A53, \dodoi{10.1051/0004-6361/202347339}

\bibitem[{Pedregosa {et~al.}(2011)Pedregosa, Varoquaux, Gramfort, Michel, Thirion, Grisel, Blondel, Prettenhofer, Weiss, Dubourg, Vanderplas, Passos, Cournapeau, Brucher, Perrot, \& Duchesnay}]{scikit-learn}
Pedregosa, F., Varoquaux, G., Gramfort, A., {et~al.} 2011, Journal of Machine Learning Research, 12, 2825

\bibitem[{{Perotti} {et~al.}(2025){Perotti}, {Kurtovic}, {Henning}, {Olofsson}, {Arabhavi}, {Schwarz}, {Kanwar}, {van Boekel}, {Kamp}, {Pascucci}, {van Dishoeck}, {G{\"u}del}, {Lagage}, {Barrado}, {Garatti}, {Glauser}, {Lahuis}, {Christiaens}, {Franceschi}, {Gasman}, {Grant}, {Jang}, {Kaeufer}, {Morales-Calder{\'o}n}, {Temmink}, \& {Vlasblom}}]{Perotti2025}
{Perotti}, G., {Kurtovic}, N.~T., {Henning}, T., {et~al.} 2025, arXiv e-prints, arXiv:2504.11424, \dodoi{10.48550/arXiv.2504.11424}

\bibitem[{{Pety} {et~al.}(2006){Pety}, {Gueth}, {Guilloteau}, \& {Dutrey}}]{Pety2006}
{Pety}, J., {Gueth}, F., {Guilloteau}, S., \& {Dutrey}, A. 2006, \aap, 458, 841, \dodoi{10.1051/0004-6361:20065814}

\bibitem[{{Pineda} {et~al.}(2023){Pineda}, {Arzoumanian}, {Andre}, {Friesen}, {Zavagno}, {Clarke}, {Inoue}, {Chen}, {Lee}, {Soler}, \& {Kuffmeier}}]{Pineda2023}
{Pineda}, J.~E., {Arzoumanian}, D., {Andre}, P., {et~al.} 2023, in Astronomical Society of the Pacific Conference Series, Vol. 534, Protostars and Planets VII, ed. S.~{Inutsuka}, Y.~{Aikawa}, T.~{Muto}, K.~{Tomida}, \& M.~{Tamura}, 233, \dodoi{10.48550/arXiv.2205.03935}

\bibitem[{{Podio} {et~al.}(2021){Podio}, {Tabone}, {Codella}, {Gueth}, {Maury}, {Cabrit}, {Lefloch}, {Maret}, {Belloche}, {Andr{\'e}}, {Anderl}, {Gaudel}, \& {Testi}}]{Podio2021}
{Podio}, L., {Tabone}, B., {Codella}, C., {et~al.} 2021, \aap, 648, A45, \dodoi{10.1051/0004-6361/202038429}

\bibitem[{{Pontoppidan} {et~al.}(2010){Pontoppidan}, {Salyk}, {Blake}, {Meijerink}, {Carr}, \& {Najita}}]{Pontoppidan2010}
{Pontoppidan}, K.~M., {Salyk}, C., {Blake}, G.~A., {et~al.} 2010, \apj, 720, 887, \dodoi{10.1088/0004-637X/720/1/887}

\bibitem[{{Raul} {et~al.}(2026){Raul}, {Zhang}, {Waggoner}, {Xie}, {Tallon}, {Banzatti}, {Salyk}, {Pontoppidan}, {Pascucci}, {Arulanantham}, {Vioque}, {Empey}, {Manara}, {Blake}, {Pinilla}, {Long}, {Yao}, {Kanwar}, {Bajaj}, {Jos{\'e} Colmenares}, {Kaeufer}, {Tabone}, {Bergin}, {Cieza}, {Narang}, {Miley}, {Krijt}, \& {Rosotti}}]{Raul2026}
{Raul}, E., {Zhang}, K., {Waggoner}, A., {et~al.} 2026, arXiv e-prints, arXiv:2606.27476, \dodoi{10.48550/arXiv.2606.27476}

\bibitem[{{Reipurth} {et~al.}(2004){Reipurth}, {Rodr{\'\i}guez}, {Anglada}, \& {Bally}}]{Reipurth2004}
{Reipurth}, B., {Rodr{\'\i}guez}, L.~F., {Anglada}, G., \& {Bally}, J. 2004, \aj, 127, 1736, \dodoi{10.1086/381062}

\bibitem[{{Riaud} {et~al.}(2006){Riaud}, {Mawet}, {Absil}, {Boccaletti}, {Baudoz}, {Herwats}, \& {Surdej}}]{Riaud2006}
{Riaud}, P., {Mawet}, D., {Absil}, O., {et~al.} 2006, \aap, 458, 317, \dodoi{10.1051/0004-6361:20065232}

\bibitem[{{Ricci} {et~al.}(2010){Ricci}, {Testi}, {Natta}, {Neri}, {Cabrit}, \& {Herczeg}}]{Ricci2010}
{Ricci}, L., {Testi}, L., {Natta}, A., {et~al.} 2010, \aap, 512, A15, \dodoi{10.1051/0004-6361/200913403}

\bibitem[{{Rigliaco} {et~al.}(2015){Rigliaco}, {Pascucci}, {Duchene}, {Edwards}, {Ardila}, {Grady}, {Mendigut{\'\i}a}, {Montesinos}, {Mulders}, {Najita}, {Carpenter}, {Furlan}, {Gorti}, {Meijerink}, \& {Meyer}}]{Rigliaco2015}
{Rigliaco}, E., {Pascucci}, I., {Duchene}, G., {et~al.} 2015, \apj, 801, 31, \dodoi{10.1088/0004-637X/801/1/31}

\bibitem[{{Rigliaco} {et~al.}(2016){Rigliaco}, {Wilking}, {Meyer}, {Jeffries}, {Cottaar}, {Frasca}, {Wright}, {Bayo}, {Bonito}, {Damiani}, {Jackson}, {Jim{\'e}nez-Esteban}, {Kalari}, {Klutsch}, {Lanzafame}, {Sacco}, {Gilmore}, {Randich}, {Alfaro}, {Bragaglia}, {Costado}, {Franciosini}, {Lardo}, {Monaco}, {Morbidelli}, {Prisinzano}, {Sousa}, \& {Zaggia}}]{Rigliaco+2016}
{Rigliaco}, E., {Wilking}, B., {Meyer}, M.~R., {et~al.} 2016, \aap, 588, A123, \dodoi{10.1051/0004-6361/201527253}

\bibitem[{{Rodriguez} {et~al.}(2018){Rodriguez}, {Loomis}, {Cabrit}, {Haworth}, {Facchini}, {Dougados}, {Booth}, {Jensen}, {Clarke}, {Stassun}, {Dent}, \& {Pety}}]{Rodriguez2018}
{Rodriguez}, J.~E., {Loomis}, R., {Cabrit}, S., {et~al.} 2018, \apj, 859, 150, \dodoi{10.3847/1538-4357/aac08f}

\bibitem[{{Rosenthal} {et~al.}(2000){Rosenthal}, {Bertoldi}, \& {Drapatz}}]{Rosenthal2000}
{Rosenthal}, D., {Bertoldi}, F., \& {Drapatz}, S. 2000, \aap, 356, 705, \dodoi{10.48550/arXiv.astro-ph/0002456}

\bibitem[{{Ruiz-Rodriguez} {et~al.}(2025){Ruiz-Rodriguez}, {Gonz{\'a}lez-Ruilova}, {Cieza}, {Zhang}, {Trapman}, {Sierra}, {Pinilla}, {Pascucci}, {P{\'e}rez}, {Deng}, {Agurto-Gangas}, {Carpenter}, {Tabone}, {Rosotti}, {Anania}, {Miley}, {Schwarz}, {Kuznetsova}, {Vioque}, \& {Kurtovic}}]{Ruiz-Rodriguez2025}
{Ruiz-Rodriguez}, D.~A., {Gonz{\'a}lez-Ruilova}, C., {Cieza}, L.~A., {et~al.} 2025, \apj, 989, 2, \dodoi{10.3847/1538-4357/add2ec}

\bibitem[{{Sacco} {et~al.}(2012){Sacco}, {Flaccomio}, {Pascucci}, {Lahuis}, {Ercolano}, {Kastner}, {Micela}, {Stelzer}, \& {Sterzik}}]{Sacco2012}
{Sacco}, G.~G., {Flaccomio}, E., {Pascucci}, I., {et~al.} 2012, \apj, 747, 142, \dodoi{10.1088/0004-637X/747/2/142}

\bibitem[{{Salyk} {et~al.}(2021){Salyk}, {Pontoppidan}, {Andrews}, {Banzatti}, {Bergin}, {Blake}, {Bosman}, {Carr}, {Krijt}, {Najita}, {Pascucci}, {Pauly}, {Pinilla}, \& {Zhang}}]{Salyk2021}
{Salyk}, C., {Pontoppidan}, K.~M., {Andrews}, S.~M., {et~al.} 2021, {A DSHARP-MIRI Treasury survey of Chemistry in Planet-forming Regions}, JWST Proposal. Cycle 1, ID. \#1584

\bibitem[{{Sanchis} {et~al.}(2020){Sanchis}, {Testi}, {Natta}, {Manara}, {Ercolano}, {Preibisch}, {Henning}, {Facchini}, {Miotello}, {de Gregorio-Monsalvo}, {Lopez}, {Mu{\v{z}}i{\'c}}, {Pascucci}, {Santamar{\'\i}a-Miranda}, {Scholz}, {Tazzari}, {van Terwisga}, \& {Williams}}]{Sanchis2020}
{Sanchis}, E., {Testi}, L., {Natta}, A., {et~al.} 2020, \aap, 633, A114, \dodoi{10.1051/0004-6361/201936913}

\bibitem[{{Schaefer} {et~al.}(2009){Schaefer}, {Dutrey}, {Guilloteau}, {Simon}, \& {White}}]{Schaefer2009}
{Schaefer}, G.~H., {Dutrey}, A., {Guilloteau}, S., {Simon}, M., \& {White}, R.~J. 2009, \apj, 701, 698, \dodoi{10.1088/0004-637X/701/1/698}

\bibitem[{{Schwarz} {et~al.}(2025){Schwarz}, {Samland}, {Olofsson}, {Henning}, {Sellek}, {G{\"u}del}, {Tabone}, {Kamp}, {Lagage}, {van Dishoeck}, {Caratti o Garatti}, {Glauser}, {Ray}, {Arabhavi}, {Christiaens}, {Franceschi}, {Gasman}, {Grant}, {Kanwar}, {Kaeufer}, {Kurtovic}, {Perotti}, {Temmink}, \& {Vlasblom}}]{Schwarz2025}
{Schwarz}, K.~R., {Samland}, M., {Olofsson}, G., {et~al.} 2025, \apj, 991, 232, \dodoi{10.3847/1538-4357/ae0603}

\bibitem[{{Sellek} {et~al.}(2024{\natexlab{a}}){Sellek}, {Grassi}, {Picogna}, {Rab}, {Clarke}, \& {Ercolano}}]{Sellek2024b}
{Sellek}, A.~D., {Grassi}, T., {Picogna}, G., {et~al.} 2024{\natexlab{a}}, \aap, 690, A296, \dodoi{10.1051/0004-6361/202450171}

\bibitem[{{Sellek} {et~al.}(2024{\natexlab{b}}){Sellek}, {Bajaj}, {Pascucci}, {Clarke}, {Alexander}, {Xie}, {Ballabio}, {Deng}, {Gorti}, {Gaspar}, \& {Morrison}}]{Sellek2024}
{Sellek}, A.~D., {Bajaj}, N.~S., {Pascucci}, I., {et~al.} 2024{\natexlab{b}}, \aj, 167, 223, \dodoi{10.3847/1538-3881/ad34ae}

\bibitem[{{Shang} {et~al.}(2020){Shang}, {Krasnopolsky}, {Liu}, \& {Wang}}]{Shang2020}
{Shang}, H., {Krasnopolsky}, R., {Liu}, C.-F., \& {Wang}, L.-Y. 2020, \apj, 905, 116, \dodoi{10.3847/1538-4357/abbdb0}

\bibitem[{{Shang} {et~al.}(2023){Shang}, {Liu}, {Krasnopolsky}, \& {Wang}}]{Shang2023}
{Shang}, H., {Liu}, C.-F., {Krasnopolsky}, R., \& {Wang}, L.-Y. 2023, \apj, 944, 230, \dodoi{10.3847/1538-4357/aca763}

\bibitem[{{Simon} {et~al.}(2016){Simon}, {Pascucci}, {Edwards}, {Feng}, {Gorti}, {Hollenbach}, {Rigliaco}, \& {Keane}}]{Simon2016}
{Simon}, M.~N., {Pascucci}, I., {Edwards}, S., {et~al.} 2016, \apj, 831, 169, \dodoi{10.3847/0004-637X/831/2/169}

\bibitem[{{Sturm} {et~al.}(2023){Sturm}, {McClure}, {Law}, {Harsono}, {Bergner}, {Dartois}, {Drozdovskaya}, {Ioppolo}, {{\"O}berg}, {Palumbo}, {Pendleton}, {Rocha}, {Terada}, \& {Urso}}]{Sturm2023}
{Sturm}, J.~A., {McClure}, M.~K., {Law}, C.~J., {et~al.} 2023, \aap, 677, A17, \dodoi{10.1051/0004-6361/202346052}

\bibitem[{{Sullivan} {et~al.}(2019){Sullivan}, {Wilking}, {Greene}, {Lisalda}, {Gibb}, \& {Ejeta}}]{Sullivan2019}
{Sullivan}, T., {Wilking}, B.~A., {Greene}, T.~P., {et~al.} 2019, \aj, 158, 41, \dodoi{10.3847/1538-3881/ab24c0}

\bibitem[{{Tabone} {et~al.}(2020){Tabone}, {Godard}, {Pineau des For{\^e}ts}, {Cabrit}, \& {van Dishoeck}}]{Tabone2020}
{Tabone}, B., {Godard}, B., {Pineau des For{\^e}ts}, G., {Cabrit}, S., \& {van Dishoeck}, E.~F. 2020, \aap, 636, A60, \dodoi{10.1051/0004-6361/201937383}

\bibitem[{{Tabone} {et~al.}(2022){Tabone}, {Rosotti}, {Cridland}, {Armitage}, \& {Lodato}}]{Tabone2022}
{Tabone}, B., {Rosotti}, G.~P., {Cridland}, A.~J., {Armitage}, P.~J., \& {Lodato}, G. 2022, \mnras, 512, 2290, \dodoi{10.1093/mnras/stab3442}

\bibitem[{{Tabone} {et~al.}(2017){Tabone}, {Cabrit}, {Bianchi}, {Ferreira}, {Pineau des For{\^e}ts}, {Codella}, {Gusdorf}, {Gueth}, {Podio}, \& {Chapillon}}]{Tabone2017}
{Tabone}, B., {Cabrit}, S., {Bianchi}, E., {et~al.} 2017, \aap, 607, L6, \dodoi{10.1051/0004-6361/201731691}

\bibitem[{{Takasao} {et~al.}(2022){Takasao}, {Tomida}, {Iwasaki}, \& {Suzuki}}]{Takasao2022}
{Takasao}, S., {Tomida}, K., {Iwasaki}, K., \& {Suzuki}, T.~K. 2022, \apj, 941, 73, \dodoi{10.3847/1538-4357/ac9eb1}

\bibitem[{{Thanathibodee} {et~al.}(2020){Thanathibodee}, {Molina}, {Calvet}, {Serna}, {Bae}, {Reynolds}, {Hern{\'a}ndez}, {Muzerolle}, \& {Hern{\'a}ndez}}]{Thanathibodee2020}
{Thanathibodee}, T., {Molina}, B., {Calvet}, N., {et~al.} 2020, \apj, 892, 81, \dodoi{10.3847/1538-4357/ab77c1}

\bibitem[{{The pandas development Team}(2025)}]{pandas}
{The pandas development Team}. 2025, {pandas-dev/pandas: Pandas}, v2.3.1,  Zenodo, \dodoi{10.5281/zenodo.3509134}

\bibitem[{{Tu} {et~al.}(2025){Tu}, {Li}, {Zhu}, {Hsu}, \& {Hu}}]{Tu2025}
{Tu}, Y., {Li}, Z.-Y., {Zhu}, Z., {Hsu}, C.-Y., \& {Hu}, X. 2025, \apj, 988, 107, \dodoi{10.3847/1538-4357/addf3c}

\bibitem[{{Tychoniec} {et~al.}(2024){Tychoniec}, {van Gelder}, {van Dishoeck}, {Francis}, {Rocha}, {Caratti o Garatti}, {Beuther}, {Gieser}, {Justtanont}, {Linnartz}, {Le Gouellec}, {Perotti}, {Devaraj}, {Tabone}, {Ray}, {Brunken}, {Chen}, {Kavanagh}, {Klaassen}, {Slavicinska}, {G{\"u}del}, \& {{\"O}stlin}}]{Tychoniec2024}
{Tychoniec}, {\L}., {van Gelder}, M.~L., {van Dishoeck}, E.~F., {et~al.} 2024, \aap, 687, A36, \dodoi{10.1051/0004-6361/202348889}

\bibitem[{{van Boekel} {et~al.}(2009){van Boekel}, {G{\"u}del}, {Henning}, {Lahuis}, \& {Pantin}}]{vanBoekel2009}
{van Boekel}, R., {G{\"u}del}, M., {Henning}, T., {Lahuis}, F., \& {Pantin}, E. 2009, \aap, 497, 137, \dodoi{10.1051/0004-6361/200811440}

\bibitem[{{Van Dishoeck} {et~al.}(2017){Van Dishoeck}, {Beuther}, {Caratti o Garatti}, {Greene}, {Justtanont}, {Klaassen}, {Ray}, {Ressler}, {Tychoniec}, \& {Waelkens}}]{vanDishoeck2017}
{Van Dishoeck}, E.~F., {Beuther}, H., {Caratti o Garatti}, A., {et~al.} 2017, {MIRI EC Protostars Survey}, JWST Proposal. Cycle 1, ID. \#1290

\bibitem[{{van Dishoeck} {et~al.}(2025){van Dishoeck}, {Tychoniec}, {Rocha}, {Slavicinska}, {Francis}, {van Gelder}, {Ray}, {Beuther}, {Caratti o Garatti}, {Brunken}, {Chen}, {Devaraj}, {Geers}, {Gieser}, {Greene}, {Justtanont}, {Le Gouellec}, {Kavanagh}, {Klaassen}, {Janssen}, {Navarro}, {Nazari}, {Notsu}, {Perotti}, {Ressler}, {Reyes}, {Sellek}, {Tabone}, {Tap}, {Theijssen}, {Colina}, {G{\"u}del}, {Henning}, {Lagage}, {{\"O}stlin}, {Vandenbussche}, \& {Wright}}]{vanDishoeck2025}
{van Dishoeck}, E.~F., {Tychoniec}, {\L}., {Rocha}, W.~R.~M., {et~al.} 2025, \aap, 699, A361, \dodoi{10.1051/0004-6361/202554444}

\bibitem[{{van't Hoff} {et~al.}(2023){van't Hoff}, {McClure}, {Sturm}, {Tabone}, {Tychoniec}, \& {Van Dishoeck}}]{vanthoff2023}
{van't Hoff}, M., {McClure}, M., {Sturm}, A., {et~al.} 2023, {The Butterfly Effect: Determining the distribution of ices across a young disk to constrain planet formation}, JWST Proposal. Cycle 2, ID. \#4201

\bibitem[{{Villenave} {et~al.}(2020){Villenave}, {M{\'e}nard}, {Dent}, {Duch{\^e}ne}, {Stapelfeldt}, {Benisty}, {Boehler}, {van der Plas}, {Pinte}, {Telkamp}, {Wolff}, {Flores}, {Lesur}, {Louvet}, {Riols}, {Dougados}, {Williams}, \& {Padgett}}]{Villenave2020}
{Villenave}, M., {M{\'e}nard}, F., {Dent}, W.~R.~F., {et~al.} 2020, \aap, 642, A164, \dodoi{10.1051/0004-6361/202038087}

\bibitem[{Virtanen {et~al.}(2020)Virtanen, Gommers, Oliphant, Haberland, Reddy, Cournapeau, Burovski, Peterson, Weckesser, Bright, {van der Walt}, Brett, Wilson, Millman, Mayorov, Nelson, Jones, Kern, Larson, Carey, Polat, Feng, Moore, {VanderPlas}, Laxalde, Perktold, Cimrman, Henriksen, Quintero, Harris, Archibald, Ribeiro, Pedregosa, {van Mulbregt}, \& {SciPy 1.0 Contributors}}]{2020SciPy-NMeth}
Virtanen, P., Gommers, R., Oliphant, T.~E., {et~al.} 2020, Nature Methods, 17, 261, \dodoi{10.1038/s41592-019-0686-2}

\bibitem[{{Wassell} {et~al.}(2006){Wassell}, {Grady}, {Woodgate}, {Kimble}, \& {Bruhweiler}}]{Wassell2006}
{Wassell}, E.~J., {Grady}, C.~A., {Woodgate}, B., {Kimble}, R.~A., \& {Bruhweiler}, F.~C. 2006, \apj, 650, 985, \dodoi{10.1086/507268}

\bibitem[{{Weber} {et~al.}(2020){Weber}, {Ercolano}, {Picogna}, {Hartmann}, \& {Rodenkirch}}]{Weber2020}
{Weber}, M.~L., {Ercolano}, B., {Picogna}, G., {Hartmann}, L., \& {Rodenkirch}, P.~J. 2020, \mnras, 496, 223, \dodoi{10.1093/mnras/staa1549}

\bibitem[{{White} \& {Hillenbrand}(2004)}]{White2004}
{White}, R.~J., \& {Hillenbrand}, L.~A. 2004, \apj, 616, 998, \dodoi{10.1086/425115}

\bibitem[{{Williams} \& {Cieza}(2011)}]{Williams2011}
{Williams}, J.~P., \& {Cieza}, L.~A. 2011, \araa, 49, 67, \dodoi{10.1146/annurev-astro-081710-102548}

\bibitem[{Wilson(1927)}]{Wilson1927}
Wilson, E.~B. 1927, Journal of the American Statistical Association, 22, 209.
\newblock \url{http://www.jstor.org/stable/2276774}

\bibitem[{{Wolff} {et~al.}(2017){Wolff}, {Perrin}, {Stapelfeldt}, {Duch{\^e}ne}, {M{\'e}nard}, {Padgett}, {Pinte}, {Pueyo}, \& {Fischer}}]{Wolff2017}
{Wolff}, S.~G., {Perrin}, M.~D., {Stapelfeldt}, K., {et~al.} 2017, \apj, 851, 56, \dodoi{10.3847/1538-4357/aa9981}

\bibitem[{{Wright} \& {Eastman}(2014)}]{Wright2014}
{Wright}, J.~T., \& {Eastman}, J.~D. 2014, \pasp, 126, 838, \dodoi{10.1086/678541}

\bibitem[{{Xie} {et~al.}(2021){Xie}, {Haffert}, {de Boer}, {Kenworthy}, {Brinchmann}, {Girard}, {Snellen}, \& {Keller}}]{Xie2021}
{Xie}, C., {Haffert}, S.~Y., {de Boer}, J., {et~al.} 2021, \aap, 650, L6, \dodoi{10.1051/0004-6361/202140602}

\bibitem[{{Xie} {et~al.}(2026){Xie}, {Pascucci}, {Long}, {Gorti}, {Banzatti}, {Booth}, {Pontoppidan}, {Molyarova}, {Carpenter}, {Fang}, {Liu}, {Raul}, {Zhang}, {Ertel}, {Stone}, {Empey}, {Manara}, {Pinilla}, {Salyk}, {Tabone}, {Vioque}, {Cieza}, {Rosotti}, {Miley}, {Blake}, \& {Waggoner}}]{Xie2026}
{Xie}, C., {Pascucci}, I., {Long}, F., {et~al.} 2026, arXiv e-prints, arXiv:2606.27477, \dodoi{10.48550/arXiv.2606.27477}

\bibitem[{{Zhang} {et~al.}(2023{\natexlab{a}}){Zhang}, {Anderson}, {Beatty}, {Bergin}, {Blake}, {Carpenter}, {Cieza}, {Hogerheijde}, {Miley}, {Pascucci}, {Perez}, {Pinilla}, {Pontoppidan}, {Rosotti}, {Salyk}, {Schwarz}, {Tabone}, {Trapman}, \& {Vioque}}]{Zhang2023}
{Zhang}, K., {Anderson}, D., {Beatty}, T.~G., {et~al.} 2023{\natexlab{a}}, {Building on ALMA: a JWST legacy survey of the chemical evolution of planet-forming disks}, JWST Proposal. Cycle 2, ID. \#3034

\bibitem[{{Zhang} {et~al.}(2023{\natexlab{b}}){Zhang}, {Ginski}, {Huang}, {Zurlo}, {Beust}, {Bae}, {Benisty}, {Garufi}, {Hogerheijde}, {van Holstein}, {Kenworthy}, {Langlois}, {Manara}, {Pinilla}, {Rab}, {Ribas}, {Rosotti}, \& {Williams}}]{ZhangY2023}
{Zhang}, Y., {Ginski}, C., {Huang}, J., {et~al.} 2023{\natexlab{b}}, \aap, 672, A145, \dodoi{10.1051/0004-6361/202245577}

\bibitem[{{Zsidi} {et~al.}(2022){Zsidi}, {Fiorellino}, {K{\'o}sp{\'a}l}, {{\'A}brah{\'a}m}, {B{\'o}di}, {Hussain}, {Manara}, \& {P{\'a}l}}]{Zsidi2022}
{Zsidi}, G., {Fiorellino}, E., {K{\'o}sp{\'a}l}, {\'A}., {et~al.} 2022, \apj, 941, 177, \dodoi{10.3847/1538-4357/ac7229}

\end{thebibliography}

\end{document}